\documentclass[11pt,a4paper]{article}
\pdfoutput=1
\usepackage{jheppub}
\usepackage{amsmath,amsfonts,amssymb, mathtools, mathrsfs,upgreek}
\usepackage{hyperref, csquotes}
\usepackage{graphicx, color}
\usepackage{tikz,pgf,tikz-cd,float}
\usepackage{physics,tensor,subcaption,multirow}
\usepackage[scr=boondox]{mathalfa}
\usepackage{enumitem}
\usepackage{todonotes}
\newcommand{\diff}{\text{d}}
\newcommand{\R}{\mathbb{R}}
\newcommand{\SUT}{\mathrm{SU}(2)}
\newcommand{\SUO}{\mathrm{SU}(1,1)}
\newcommand{\e}{\mathrm{e}}

\usepackage{braket}
\usepackage{dsfont}

\usepackage{tikzit}

\tikzstyle{Node}=[fill=black, draw=black, shape=circle, scale=0.3px]
\tikzstyle{coh}=[fill=white, draw=black, shape=circle, scale=0.6px, line width=1px]
\tikzstyle{coh_big}=[fill=white, draw=black, shape=circle, scale=0.6px]
\tikzstyle{coh_black}=[fill=black, draw=black, shape=circle, scale=0.6px, line width=1px]
\tikzstyle{coh_blue}=[fill=blue, draw=blue, shape=circle, tikzit fill=blue, scale=0.5px]
\tikzstyle{coh_wb}=[fill=white, draw=blue, shape=circle, scale=0.6px, line width=1px]
\tikzstyle{square  coh}=[fill=black, draw=black, shape=regular polygon, scale=0.4px, line width=1px, regular polygon sides=3]
\tikzstyle{square coh white}=[fill=white, draw=black, shape=regular polygon, scale=0.4px, line width=1px, regular polygon sides=3]
\tikzstyle{orange}=[fill={rgb,255: red,255; green,128; blue,0}, draw=none, shape=circle, scale=0.4px]
\tikzstyle{blue}=[fill=blue, draw=none, shape=circle, scale=0.4px]

\tikzstyle{line}=[-, fill=none, line width=1px]
\tikzstyle{blockline}=[-, fill=black, line width=3px]
\tikzstyle{boost}=[-, fill={rgb,255: red,128; green,128; blue,128}, draw={rgb,255: red,128; green,128; blue,128}, tikzit fill={rgb,255: red,128; green,128; blue,128}, tikzit draw={rgb,255: red,128; green,128; blue,128}, line width=5px]
\tikzstyle{specialsu2}=[-, line width=5px, fill=black, draw={rgb,255: red,0; green,0; blue,189}, tikzit fill=white, tikzit draw=black]
\tikzstyle{boxdash}=[-, line width=5px, fill=black, dash pattern=on 2pt off 2pt]
\tikzstyle{arrow}=[line width=1.1px, ->]
\tikzstyle{arrowdotted}=[line width=1.1px, ->, dash pattern=on 2pt off 1.5pt]
\tikzstyle{dashed}=[-, line width=1px, dash pattern=on 2pt off 1pt, fill=none]
\tikzstyle{thindash}=[-, line width=0.5px, dash pattern=on 1pt off 3pt, draw={rgb,255: red,158; green,158; blue,158}]
\tikzstyle{grayfill}=[-, fill={rgb,255: red,234; green,234; blue,234}, draw=black, line width=1px]
\tikzstyle{thingray}=[-, line width=0.8px, draw={rgb,255: red,158; green,158; blue,158}]
\tikzstyle{arrowgray}=[draw={rgb,255: red,158; green,158; blue,158}, ->, line width=1px]
\tikzstyle{line_blue}=[-, line width=1.5px, draw=blue, tikzit draw=blue]
\tikzstyle{blue_dashed}=[-, draw=blue, tikzit draw=blue, line width=1.5px, dash pattern=on 2pt off 1.5pt]
\tikzstyle{orange_dashed}=[-, draw={rgb,255: red,255; green,128; blue,0}, tikzit draw={rgb,255: red,255; green,128; blue,0}, line width=1.5px, dash pattern=on 2pt off 1.5pt]
\tikzstyle{new edge style 0}=[-]
\tikzstyle{line_orange}=[-, draw={rgb,255: red,255; green,128; blue,0}, tikzit draw={rgb,255: red,255; green,128; blue,0}, line width=1.5px]
\tikzstyle{black fill}=[-, fill=black, draw=none]

\usetikzlibrary{shapes.geometric}
\usepackage{manfnt}
\newcommand*\cube{\mbox{\mancube}}
\renewcommand\bra[1]{{\langle{#1}|}}
\makeatletter
\renewcommand\ket[1]{%
  \@ifnextchar\bra{\k@t{#1}\!}{\k@t{#1}}%
}
\newcommand\k@t[1]{{|{#1}\rangle}}
\makeatother
\usetikzlibrary{patterns}
\usetikzlibrary{matrix,decorations.pathreplacing}
\pgfkeys{tikz/mymatrixenv/.style={decoration={brace},every left delimiter/.style={xshift=8pt},every right delimiter/.style={xshift=-8pt}}}
\pgfkeys{tikz/mymatrix/.style={matrix of math nodes,nodes in empty cells,left delimiter={(},right delimiter={)},inner sep=1pt,outer sep=4pt,column sep=8pt,row sep=8pt,nodes={minimum width=20pt,minimum height=10pt,anchor=center,inner sep=0,outer sep=0}}}
\pgfkeys{tikz/mymatrixbrace/.style={decorate, thick, decoration={raise=3pt}}}

\DeclareMathOperator*{\sumint}{%
\mathchoice%
  {\ooalign{$\displaystyle\sum$\cr\hidewidth$\displaystyle\int$\hidewidth\cr}}
  {\ooalign{\raisebox{.14\height}{\scalebox{.7}{$\textstyle\sum$}}\cr\hidewidth$\textstyle\int$\hidewidth\cr}}
  {\ooalign{\raisebox{.2\height}{\scalebox{.6}{$\scriptstyle\sum$}}\cr$\scriptstyle\int$\cr}}
  {\ooalign{\raisebox{.2\height}{\scalebox{.6}{$\scriptstyle\sum$}}\cr$\scriptstyle\int$\cr}}
}

\newlength{\boxsize}
\begin{document}

\title{(2+1) Lorentzian quantum cosmology from spin-foams: opportunities and obstacles for semi-classicality}

\author[*,a,b,c]{Alexander F. Jercher}
\emailAdd{alexander.jercher@uni-jena.de}

\author[a]{José Diogo Simão,}
\emailAdd{j.d.simao@uni-jena.de}

\author[a]{Sebastian Steinhaus}
\emailAdd{sebastian.steinhaus@uni-jena.de}

\affiliation[a]{Theoretisch-Physikalisches Institut, Friedrich-Schiller-Universit\"{a}t Jena\\ Max-Wien-Platz 1, 07743 Jena, Germany, EU}
\affiliation[b]{Arnold Sommerfeld Center for Theoretical Physics,\\ Ludwig-Maximilians-Universit\"at München \\ Theresienstrasse 37, 80333 M\"unchen, Germany, EU}
\affiliation[c]{Munich Center for Quantum Science and Technology (MCQST),\\ Schellingstr. 4, 80799 M\"unchen, Germany, EU}
\affiliation[*]{corresponding author}

\date{\today}

\begin{abstract}
{
We construct an effective cosmological spin-foam model for a (2+1) dimensional spatially flat universe, discretized on a 
cubical lattice, containing both space- and time-like regions. Our starting point is the recently proposed coherent state spin-foam model for (2+1) Lorentzian quantum gravity. The full amplitude is assumed to factorize into single vertex amplitudes with boundary data corresponding to Lorentzian 3-frusta. A stationary phase approximation is performed at each vertex individually, where the inverse square root of the Hessian determinant serves as a measure for the effective path integral. Additionally, a massive scalar field is coupled to the geometry, and we show that its mass renders the partition function convergent. For a single 3-frustum with time-like struts, we compute the expectation value of the bulk strut length and show that it generically agrees with the classical solutions and that it is a discontinuous function of the scalar field mass. Allowing the struts to be space-like introduces causality violations, which drive the expectation values away from the classical solutions due to the lack of an exponential suppression of these configurations. This is a direct consequence of the semi-classical amplitude only containing the real part of deficit angles, in contrast with the Lorentzian Regge action used in effective spin-foams. We give an outlook on how to evaluate the partition function on an extended discretization including a bulk spatial slice. This serves as a foundation for future investigations of physically interesting scenarios such as a quantum bounce or the viability of massive scalar field clocks. Our results demonstrate that the effective path integral in the causally regular sector serves as a viable quantum cosmology model, but that the agreement of expectation values with classical solutions is tightly bound to the path integral measure. 
}
\end{abstract}

\maketitle

\section{Introduction}\label{sec:Introduction}

Quantum cosmology represents an ideal testing ground for theories of quantum gravity (QG). 
The close-to-homogeneous Universe exhibits a large degree of symmetry which, imposed on the considered QG model, renders explicit analytical and numerical computations feasible. 
Furthermore, quantum fluctuations in the early Universe could possibly lead to observable signatures~e.g.~in the cosmic microwave background~\cite{Ashtekar:2021kfp} or in gravitational waves~\cite{Calcagni:2019kzo}. 

A promising path integral approach to QG are spin-foam models~\cite{Perez:2013uz}, which associate transition amplitudes to states describing the boundary of a given spacetime region. 
To regularize the path integral, the bulk of spacetime is discretized on a 2-complex, most commonly chosen to be dual to a triangulation. 
Closely related to spin-foam models are canonical loop quantum gravity (LQG)~\cite{Ashtekar:2021kfp} and group field theories (GFTs)~\cite{Oriti:2006ts,Freidel:2005qe}, all of which have in common the feature that on the microscopic quantum level spacetime is described in terms of discrete degrees of freedom encoded in group theoretic data. 
Recovering continuum spacetime from such an underlying description is one of the most pressing challenges shared among these approaches.

Identifying macroscopic cosmological observables from the microscopic degrees of freedom involves either a coarse-graining, as in the GFT condensate cosmology approach~\cite{Marchetti:2020umh,Oriti:2016qtz,Gielen:2016dss,Jercher:2021bie}, or a truncation in the form of a symmetry reduction as in loop quantum cosmology~\cite{Agullo:2016tjh} and approaches to cosmology in spin-foams~\cite{Bahr:2017bn,Bianchi:2010ej,Dittrich:2023rcr,Han:2024ydv}. 

First developments towards spin-foam cosmology within the full EPRL spin-foam model~\cite{Engle:2007em,Engle:2008fj} appeared in~\cite{Bianchi:2010ej,Bianchi:2011ym,Rennert:2013pfa,Sarno:2018ses}, followed by investigations on the Hartle-Hawking no-boundary state~\cite{Gozzini:2019nbo,Frisoni:2022urv,Frisoni:2023lvb}. 
Most recently~\cite{Han:2024ydv}, the complex critical point method~\cite{Han:2021kll,Han:2023cen,Han:2024lti} has been applied to investigate cosmological transition amplitudes for triangulations of a single hypercube and two connected hypercubes within the Conrady-Hnybida (CH) extension~\cite{Conrady:2010kc,Conrady:2010vx} of the EPRL model.

Another approach to cosmology has been introduced in~\cite{Bahr:2017bn} in the context of reduced spin-foam models. 
In this setting, coherent intertwiners are restricted to adhere to the cosmological symmetries of spatial homogeneity and isotropy. 
For spatially flat geometries this leads to so-called frusta geometries, the spectral dimension of which has been investigated in~\cite{Jercher:2023rno}. 
Other examples of symmetry-reduced spin-foam models are given by cuboid~\cite{Bahr:2015gxa,Bahr:2016dl,Bahr:2017klw} and parallelepiped models~\cite{Assanioussi:2020fml}. 
The numerically challenging nature of symmetry reduced quantum amplitudes for large spins~\cite{Allen:2022unb} motivated resorting to semi-classical amplitudes for explicit computations.

A similar strategy for simplifying the amplitudes and thus speeding up computations is pursued in effective spin-foams~\cite{Asante:2021zzh,Asante:2020iwm,Asante:2020qpa}. 
In this approach, amplitudes are assumed to be given by the exponentiated area Regge action~\cite{Barrett:1997tx,Asante:2018wqy} supplemented with constraints that encode the transition from area to length variables and thus Regge calculus~\cite{Regge:1961ct}. 
Effective spin-foams have been applied to cosmology in~\cite{Dittrich:2023rcr,Dittrich:2021gww,Asante:2021phx}, considering spatially spherical geometries with a positive cosmological constant and discretizations without spatial hypersurfaces in the bulk. 
Despite the simple setting, a number of intriguing effects can be observed in the Lorentzian effective spin-foam path integral. First, expectation values of the strut length (corresponding to the lapse function in the continuum) have been shown to agree with the classical solutions of the Regge equations in a large regime of parameters and boundary states. 
Second, and most importantly, a subset of configurations that is being summed over exhibits so-called causality violations, characterized by more or less than two light cones located at space-like sub-simplices. 
It has been demonstrated in~\cite{Dittrich:2023rcr} that the amplitudes of effective spin-foams exhibit an exponential suppression or enhancement of these configurations due to complex-valued deficit angles. 

Identifying a cosmological sub-sector of Lorentzian spin-foam models and investigating the causal structure therein suggests employing causally extended spin-foam models, such as the aforementioned CH-extension of the Lorentzian EPRL model or the complete Barrett-Crane model~\cite{Jercher:2022mky}, and studying their semi-classical behavior. 
For the latter, such an analysis is still missing. 
In contrast, a stationary phase approximation has been applied to the EPRL-CH model yielding a cosine of the Regge action for space-like interfaces between tetrahedra of arbitrary causal character.
However, for time-like interfaces, a stationary phase approximation cannot be straightforwardly applied and thus a closed asymptotic formula remains unknown. 
That is because 1) the critical points are non-isolated~\cite{Liu:2018gfc,Simao:2021qno}, 2) $\SUO$ coherent states in the continuous series either require an asymptotic approximation~\cite{Liu:2018gfc} or a regularization~\cite{Simao:2024don} and 3) the integrand exhibits a branch cut at the critical points~\cite{Simao:2021qno}. 
The results of~\cite{Dittrich:2023rcr,Jercher:2023csk} in the context of cosmology suggest that it is precisely the time-like interfaces which are relevant to identify a causally regular sub-sector of Lorentzian spin-foams.

In order to address problems 1) and 2), a coherent state model has been developed in the simpler setting of (2+1)-dimensional Lorentzian spin-foams~\cite{Simao:2024don}, where the problematic configurations involve space-like edges. 
By introducing a regularization of the space-like $\SUO$ coherent states and supplementing them with an ad-hoc Gaussian constraint that ensures the correct gluing condition, a closed semi-classical formula can be attained for the full set of causal configurations. 
Investigations in~\cite{Jercher:2024kig} have shown that for faces either all time-like or all space-like, the typical cosine-like asymptotics is recovered. 
In every other case only a single critical point exists, leading to a single factor of the exponentiated Regge action. With regard to previous iterations of the Lorentzian Ponzano-Regge model, formulations in the magnetic basis had already been proposed as extensions of the $\SUT$ $\{6j\}$ symbol to the discrete~\cite{Davids:1998bp} and the continuous series~\cite{Garcia-Islas:2003ges}, thus restricting to either entirely time-like or entirely space-like boundaries, respectively. 
The definition in~\cite{Freidel:2000uq,Freidel:2005bb} contains also mixed causal characters, but it is only given formally. 
In contrast, the coherent model of~\cite{Simao:2024don} not only offers a clear geometric interpretation via the coherent states, but also offers an explicit definition of a Lorentzian (2+1) model for the full set of causal configurations.

When neglecting perturbations, the homogeneous and isotropic continuum metric of cosmology effectively reduces to a one-dimensional field, and it is therefore insensitive to the dimensionality of space (except for numerical factors entering the dynamical equations). 
Furthermore, propagating degrees of freedom only become relevant when including perturbations. 
In the absence of an explicit expression for the semiclassical limit of the well-known (3+1)-dimensional EPRL-CH model for all causal configurations, the (2+1)-dimensional coherent model proposed in~\cite{Simao:2024don} is well-suited as a first feasible model for investigating the homogeneous sector of quantum cosmology. 

The strategy and main results of this article is summarized in the following three paragraphs. 
We construct an effective cosmological amplitude via a sequence of modifications of the full underlying (2+1) coherent spin-foam model discretized on a 
cubical lattice. 
These modifications include a separation of the vertex amplitudes and an application of the stationary phase approximation at each vertex individually, similar to the hybrid algorithm idea~\cite{Asante:2022lnp}. 
The boundary data of the semi-classical vertices is symmetry reduced to that of 3-frusta which form a discretization of spatially flat, homogeneous and isotropic geometries. 
Causality violating configurations are obtained if struts are space-like. Crucially, for these configurations, the semi-classical vertex amplitudes involve only the real part of otherwise complex-valued deficit angles, such that causality violating configurations are not exponentially suppressed. 
The stationary phase approximation equips the amplitudes with measure factors given by inverse square roots of Hessian determinants the properties of which we investigate. 
To obtain non-trivial dynamics, and to introduce degrees of freedom that serve as a relational clock, we minimally couple a massive scalar field to the system. 
Given this setup, we discuss the role of causality violations in the effective cosmological model and provide a comparison to the effective spin-foam construction.

The effective cosmological spin-foam model allows to explicitly study intriguing aspects of the Lorentzian cosmological path integral for a single building block, such as the influence of the scalar field mass as well as expectation values of the strut length. 
Restricting to the causally regular sector, we find freezing oscillations of the effective amplitudes for vanishing scalar field mass. This leads to a divergence of the path integral, which is already to be expected from the continuum. 
Introducing a non-vanishing scalar field mass as a regulator of the aforementioned divergence, we study the strut length expectation value for varying boundary data and mass parameters. 
The real part of the expectation value shows good agreement with semi-classical solutions in a wide range of parameters, while the imaginary part curiously tends towards $-\frac{1}{2}$. 
Explicit computations show that the effective path integral is discontinuous in the scalar field mass $\mu$ at the point $\mu = 0$ which is reminiscent of Proca theory and massive gravity. 
Including causality violating configurations in the path integral generically leads to substantial deviations because of the lack of exponential suppression. 
In comparison, the effective spin-foam computation shows that causality violations are suppressed and thus only negligibly contribute to expectation values. 

Lastly, we take first steps towards including a spatial slice in the bulk, and outline a recipe for computing the associated partition function as well as geometric and matter expectation values. 
We find an intricate relation between semi-classical behavior and the measure factor of the path integral, highlighted by the results of a toy model computation. 

In short, our article is organized as follows: in Sec.~\ref{sec:model}, we introduce the (2+1) coherent model, as well as its semi-classical limit for cuboidal combinatorics and for an arbitrary choice of causal characters. 
Thereafter, in Sec.~\ref{sec:A proposal for an effective cosmological amplitude}, we perform a sequence of simplifications and approximations to adapt the model to spatially flat, homogeneous and isotropic cosmologies, and minimally couple a massive scalar field to the cosmological system. 
In Sec.~\ref{sec:Numerical evaluation: Strut in the bulk}, we investigate the effective path integral numerically by studying the issue of freezing oscillations, while computing the strut length expectation as a function of varying boundary data and scalar field mass. 
We furthermore quantify the influence of causality violations on these expectation values, and present a comparison of our results to that of the effective spin-foam approach. 
In Sec.~\ref{sec:1slice}, we include a spatial slice in the bulk, sketch the strategy to compute the partition function for the effective model and introduce a toy model which yields expectation values close to classicality.
\section{A cuboidal spin-foam amplitude for Lorentzian gravity}\label{sec:model}

In this work, we employ the Ponzano-Regge model for (2+1)-dimensional Lorentzian quantum gravity. First formulated in terms of the magnetic basis~\cite{Barrett:1993db,Freidel:2000uq,Freidel:2002hx}, a coherent-state representation of the (2+1) model has been developed recently in~\cite{Simao:2024don}. This formulation elucidates the geometric interpretation of the model, allowing for the restriction to the cosmological sector we introduce later on. Although the model has originally been put forth in a simplicial setting, its generalization to other spacetime discretizations is straightforward. So as to make our presentation as self-contained as possible, we review here the relevant construction for a cuboidal cell complex. We remind the reader that the model is based upon the theory of unitary irreducible representations of $\mathrm{SU}(1,1)$; a general overview of the relevant concepts, notation and references can be found in Appendix~\ref{app:su11}.

\subsection{The vertex, the amplitude and the partition function}\label{sec:The vertex, the amplitude and the partition function}
Consider then a cuboidal lattice $\mathcal{X}$, i.e. a 3-cell complex where the adjacency matrix of each 2-cell follows that of a regular cube. A generic geometric realization of each 3-cell corresponds to a cuboid, and that of a 2-cell to a quadrilateral. Each 2-cell $\Box$ is identified by a single integer $a$, and each 1-cell $\diagup$ by a pair $ab$ of adjacent 2-cells. Given $\mathcal{X}$, a \textit{history} $\psi$ is an assignment of data to $\mathcal{X}$, as follows:
\begin{enumerate}
    \item To each 1-cell $\diagup_{ab}$ one assigns a spin $s_{ab}\in \mathbb{R}^+$ or $k_{ab}\in -\frac{\mathbb{N}}{2}$ of the continuous $\mathcal{C}^0_s$ or discrete $\mathcal{D}^q_k$ series of unitary representations of $\mathrm{SU}(1,1)$, respectively. The discrete spin is moreover complemented by a sign $\tau_{ab}:=-q_{ab}$, selecting between the positive or negative family of the series. An edge with discrete spin is termed time-like, and space-like otherwise\footnote{Throughout this work we consider the Minkowski metric signature $(+,-,-)$.}. Semi-classically, the spin is in correspondence with the length of the edges of $\mathcal{X}$;
    \item To each ordered pair $(\Box_a, \Box_b)$ of adjacent boundary 2-cells one assigns an ordered pair $(n_{ab},n_{ba})$ of $\mathrm{SU}(1,1)$ group elements, together with an orientation $\mathscr{o}_{ab}=\pm$.
\end{enumerate}
The spin-foam model prescribes an \emph{amplitude map} $A: (\mathcal{X},\psi) \rightarrow \mathbb{C}$, i.e. it assigns a complex number to each spin-foam history $\psi$ on a given lattice $\mathcal{X}$. Its fundamental building block is the \textit{vertex amplitude} $A_v: \psi \rightarrow \mathbb{C}$, obtained by evaluating the amplitude map on a single 3-cell lattice $\cube$. The precise form of the vertex amplitude is most clearly expressed in a diagrammatic representation; for a generic $\psi$, such a diagram has the form
\begin{equation}
\label{cube_vertex}
    A_v(\psi)= \scalebox{0.4}{\cubevertex}\,,
\end{equation}
the meaning of which we discuss next.

The diagram is composed of boxes $\raisebox{2pt}{\scalebox{0.5}{\boxtikz}}_a$ and links $\raisebox{2pt}{\scalebox{0.5}{\link}}_{ab}$, the latter indexed by the two boxes which intersect each link. A history $\psi$ induces a coloring of the diagram, according to the rules:
\begin{enumerate}
    \item Each link $\raisebox{2pt}{\scalebox{0.5}{\link}}_{ab}$ is in correspondence with an edge $\diagup_{ab}$, and inherits its data. If the associated spin is of the discrete series, the endpoints are colored white $\raisebox{2pt}{\scalebox{0.5}{\link}}$, and the link is said to be time-like as it corresponds to a time-like edge. If the spin is of the continuous series, it is colored black $\raisebox{2pt}{\scalebox{0.5}{\linkblack}}$, and the link is said to be space-like as it corresponds to a space-like edge;
    \item Each box $\raisebox{2pt}{\scalebox{0.5}{\boxtikz}}_a$ is associated with a 2-cell $\Box_a$. For a single 3-cell lattice all adjacent 2-cells $(\Box_a, \Box_b)$ are boundary, so each link $\raisebox{2pt}{\scalebox{0.5}{\link}}_{ab}$ inherits a pair $(n_{ab},n_{ba})$ and an orientation $\mathscr{o}_{ab}$. Accordingly, to the end-point $\raisebox{2pt}{\scalebox{0.5}{\ept}}_{ab}$ close to the box $\raisebox{2pt}{\scalebox{0.5}{\boxtikz}}_a$ there corresponds the element $n_{ab}\in\mathrm{SU}(1,1)$, and the orientation prescribes an ordering to the link $\overset{\rightarrow}{\raisebox{2pt}{\scalebox{0.5}{\link}}}_{ab}$. Semi-classically, the $n_{ab}$ are in correspondence with the edge vectors of the cuboid;
\end{enumerate}
Note that in Eq.~\eqref{cube_vertex} we have omitted most of the $\psi$ data for simplicity. The diagram is now evaluated as follows: for a time-like link, colored with a spin $k_{ab}$ of the discrete series, we set
\begin{equation}
\begin{gathered}
\label{brakettl}
\!{}^{n_{ab}}\overset{\rightarrow}{\scalebox{0.65}{\braketblack}}{}^{n_{ba}}= d_{k_{ab}} \braket{\tau_{ab}|g^\dagger_a \sigma_3 g_b |\tau_{ba}}^{2k_{ab}}\,, \quad d_k:=-2k-1\,, \\
\ket{+_{ab}}:=n_{ab}\left(\begin{smallmatrix} 1 \\ 0 \end{smallmatrix}\right)\,, \quad \ket{-_{ab}}:=n_{ab}\left(\begin{smallmatrix} 0 \\ 1 \end{smallmatrix}\right)\,,
\end{gathered}
\end{equation}
where $g_a,g_b \in \mathrm{SU}(1,1)$ are associated to the two boxes, and remain unspecified. For a space-like link, colored with a spin $s_{ab}$ of the continuous series, we take
\begin{equation}
\begin{gathered}
\label{braketsl}
\!{}^{n_{ab}}\overset{\rightarrow}{\scalebox{0.65}{\brakettikz}}{}^{n_{ba}}=   d_{s_{ab}}  \mathcal{C}_{n_{ab},n_{ba}} \braket{l^+_{ab}|g_a^\dagger \sigma_3 g_b|l^-_{ba}}^{-1+2is_{ab}}\,, \quad d_s:=-\frac{\gamma}{\pi} s \tanh \pi s\,, \\
 \mathcal{C}_{n_{ab},n_{ba}} := e^{s_{ab} \braket{l^+_{ab}|g_a^\dagger \sigma_3 g_b |l^+_{ba}}^2}\,, \quad \ket{l^\pm_{ab}}:=\frac{n_{ab}}{\sqrt{2}}  \left(\begin{smallmatrix} 1 \\ \pm 1 \end{smallmatrix}\right)\,.
\end{gathered}
\end{equation}
Here $\gamma$ denotes the Euler-Mascheroni constant. The states $\ket{\tau_{ab}}$ and $\ket{l^\pm_{ab}}$ are $\mathrm{SU}(1,1)$ Perelomov coherent states~\cite{Perelomov:1986tf} in the defining representation. As discussed in~\cite{Simao:2024don}, the term $ \mathcal{C}_{n_{ab},n_{ba}}$ corresponds to a Gaussian constraint that has been introduced by hand. It ensures a well-behaved semi-classical limit by implementing the otherwise absent gluing between edges $n_{ab}$ and $n_{ba}$. The vertex amplitude is now obtained from Eq.~\eqref{cube_vertex} by taking the product of all links, and integrating over the boxes; note that the same box crosses more than one link, and consequently different links may share the same box variable $g_a$. Explicitly, for a certain choice of orientation,
\begin{multline}
     A_{v}(\psi)=\int_{\mathrm{SU}(1,1)^6} \prod_a \mathrm{d} g_a\, \delta(g_6) \prod_{ab\; \tikz\draw[black,fill=white] (0,0) circle (.3ex); } d_{k_{ab}} \braket{\tau_{ab}|g^\dagger_a \sigma_3 g_b  |\tau_{ba}}^{2k_{ab}} \\\cdot \prod_{ab \; \tikz\draw[black,fill=black] (0,0) circle (.3ex);}    d_{s_{ab}}  \mathcal{C}_{n_{ab},n_{ba}} \braket{l^+_{ab}|g_a^\dagger \sigma_3 g_b|l^-_{ba}}^{-1+2is_{ab}}\,.
\end{multline}
By introducing the function $\delta(g_6)$ to $\mathcal{A}_v$, we regularize the amplitude, which would otherwise diverge due to a redundant non-compact group integration~\cite{Engle:2008ev}. This regularization procedure is commonly referred to as gauge-fixing, and we will use this terminology for the remainder.\footnote{This gauge-fixing is, however, not related to the equivalently termed procedure in gauge theories.}

Having explained the structure of the vertex amplitude, the full amplitude for a general lattice $\mathcal{X}$ follows straightforwardly: to each cuboid one associates a diagram as in Eq.~\eqref{cube_vertex}. If two cuboids are glued at a quadrilateral, the respective diagrams are connected at the corresponding ends. If the gluing is such that a closed loop arises, the link is assigned a value
\begin{equation}
\label{eq:looptl}
 \scalebox{0.5}{\looptikz}_k   = \sum_{q=\pm} (-2k-1)\mathrm{Tr}\left[D^{k(q)}(g)\right]\
\end{equation}
or 
\begin{equation}
\label{eq:loopsl}
  \scalebox{0.5}{\looptikz}_s  =\sum_{\delta=0,\frac{1}{2}} s\tanh^{1-4\delta} (\pi s) \, \mathrm{Tr}\left[D^{j(\delta)}(g)\right]\,,
\end{equation}
depending on the history $\psi$. Note that the unitary representations of $\mathrm{SU}(1,1)$ are infinite-dimensional, so that the previous two equations must be understood in the sense of their regularization. An example configuration $\mathcal{X}$ inducing such a loop is given by four cuboids sharing an edge, for which a corresponding diagram would take the form
\begin{equation}
     A(\cube^4, \psi)= \scalebox{0.4}{\cubefour}\,,
\end{equation}
where the loop associated to the internal edge is dashed for clarity. On such an extended diagram one must still gauge-fix all redundant group integrations.

Finally, the spin-foam \emph{partition function} is given by a sum over histories $\psi$ in agreement with a fixed choice of boundary data. By boundary data $\partial \psi$ we mean the data assignments of $\psi$ restricted to the boundary edges $\diagup_{ab}\in \partial \mathcal{X}$. Thus, for given $\mathcal{X}$ and boundary data $\phi$,
\begin{equation}
\label{eq:ptfunc}
    Z(\mathcal{X}, \phi):= \sumint_{\psi\,|\, \partial \psi=\phi} A(\mathcal{X},\psi)\,.
\end{equation}
Recovering the previous example of the $\cube^4$ lattice, the partition function would read
\begin{equation}
    Z(\cube^4, \phi):= \sum_{k\,  \mapsto  \, \raisebox{1pt}{\scalebox{0.3}{\looptikz}}} \int_{s\,  \mapsto  \, \raisebox{1pt}{\scalebox{0.3}{\looptikz}}}\mathrm{d}s  \;\scalebox{0.4}{\cubefour}\,,
\end{equation}
and include a sum and integral over assignments of spins $k$ and $s$ to the bulk looped link. Notice that the causal character of bulk edges is being summed over. This can be traced back to the fact that the Plancherel-decomposition of functions on $\SUO$ contains contributions of both the continuous and the discrete series, as discussed in Appendix~\ref{app:su11}.

At this stage we must remark on discretization invariance of the theory. It is well known since Ponzano's original paper \cite{Ponzano:1968wi} that the simplicial spin-foam amplitude for 3-dimensional Riemannian gravity exhibits topological invariance only once the amplitude is regularized with an upper cut-off, and complemented with an additional factor for each bulk vertex of the triangulation; one can take the cut-off to infinity as the last computational step, and in this manner the amplitude can be shown to be invariant under the 4-1 and 3-2 Pachner moves, which completely characterize all homeomorphisms. The need for such a procedure is related to the unbounded sums over representation traces as in Eqs.~\eqref{eq:looptl} and~\eqref{eq:loopsl}. A similar argument can be made for the present Lorentzian theory in its simplicial version.
However, once the model is generalized to higher polyhedra (e.g. cuboids) one looses the straightforward characterization of homeomorphisms provided by the Pachner moves, and it is conceivable that the correction introduced in the simplicial case does not guarantee topological invariance. This obstacle had already been identified in~\cite{Bahr:2015gxa} for 4 dimensions, where the authors opted to remain agnostic on the right correction, choosing instead to parametrize different choices of factors associated to loops. Our choice is to preserve the definition in Eq.~\eqref{eq:ptfunc} as-is, and later introduce a modification in Sec.~\ref{sec:A proposal for an effective cosmological amplitude}.

\subsection{The semi-classical limit of the vertex}\label{sec:The semi-classical limit of the vertex}

Let us return to the vertex amplitude $A_{v}$, and consider a history with particular color (i.e. causal character) assignments
\begin{equation}
\label{vertex_color}
    A_{v}(\psi)= \scalebox{0.4}{\cubevertex}\,,
\end{equation}
being here two opposing 2-cells with space-like edges connected by time-like edges. The semi-classical limit of the vertex amplitude was already studied in~\cite{Simao:2024don} when all edges are space-like. This constitutes the most challenging case due to the lack of a gluing condition and the necessity for a regularization of $\SUO$ coherent states; here we extend the discussion to this mixed case. The general strategy is to write the link functions in terms of complex exponentials weighted by the spins, such that stationary phase methods can be applied \cite[Th. 7.7.5]{Hormander2003}. Defining then
\begin{equation}
    \!{}^{n_{ab}}\overset{\rightarrow}{\scalebox{0.65}{\braketblack}}{}^{n_{ba}}= d_{k_{ab}} e^{S_{ab}^{\mathrm{tl}}}\,, \quad \!{}^{n_{ab}}\overset{\rightarrow}{\scalebox{0.65}{\brakettikz}}{}^{n_{ba}}=   \braket{l^+_{ab}|g_a^\dagger \sigma_3 g_b|l^-_{ba}}^{-1} d_{s_{ab}}  e^{S_{ab}^{\mathrm{sl}}}
\end{equation}
the model assigns the actions
\begin{equation}
\begin{gathered}
S_{ab}^{\mathrm{tl}}= 2k_{ab} \ln  \braket{\tau_{ab}|g_a^\dagger \sigma_3 g_b | \tau_{ba}}\,, \\
S_{ab}^{\mathrm{sl}}=2is_{ab} \ln \braket{l^+_{ab}|g_a^\dagger \sigma_3 g_b | l^-_{ab}} + s_{ab} \braket{l^+_{ab}|g_a^\dagger \sigma_3 g_b | l^+_{ab}}^2\,
\end{gathered}
\end{equation}
to the spin-foam amplitude. The critical point equations $\sum_{ab} \delta_{g_a} S_{ab}=0$ and $\mathfrak{Re}\{ S_{ab}\}=0$ (the value at which it is maximal) can be shown to imply, for space-like $ab$, 
\begin{equation}
\begin{gathered}
\label{gluesl}
g_b \ket{l^+_{ba}}=\vartheta_{ab} g_a \ket{l^+_{ab}} \\
g_b \ket{l^-_{ba}}=\vartheta_{ab}^{-1} g_a \ket{l^-_{ab}}
\end{gathered}\,, \quad \vartheta_{ab}\in \mathbb{R}^+\,,
\end{equation}
while for time-like $ab$
\begin{equation}
\begin{gathered}
\label{gluetl}
g_b \ket{\tau_{ba}}=\varrho_{ab} g_a \ket{\tau_{ab}} \\
g_b \ket{-\tau_{ba}}=\overline{\varrho}_{ab} g_a \ket{-\tau_{ab}}
\end{gathered}\,, \quad \varrho_{ab} \in e^{i \mathbb{R}}\,,
\end{equation}
together with a closure relation 
\begin{equation}
\label{eq:closure}
   \forall a\,, \quad  \sum_b^{ab\, \mathrm{tl}} - \mathscr{o}_{ab} k_{ab}  v_{ab} +  \sum_b^{ab\, \mathrm{sl}} \mathscr{o}_{ab} s_{ab} v_{ab} =0\,.
\end{equation}
In the equation above $v_{ab}$ stands for a 3-vector, whose definition depends on the coloring of the link $ab$. For a space-like link,
\begin{equation}
\label{geo_vec_sl}
    v_{ab}:= \pi(n_{ab}) \hat{e}_2\in H^{\mathrm{sl}}\subset \mathbb{R}^{1,2} \,, \quad \braket{l^+_{ab}|\sigma_3 \varsigma^I |l^-_{ab}}=i v_{ab}^I\,,
\end{equation}
while for a time-like link 
\begin{equation}
\label{geo_vec_tl}
    v_{ab}:= \tau_{ab} \pi(n_{ab}) \hat{e}_0\in H^\tau \subset \mathbb{R}^{1,2} \,, \quad \tau_{ab}\braket{\tau_{ab}|\sigma_3 \varsigma^I |\tau_{ab}}=v_{ab}^I\,.
\end{equation}
The symbol $\varsigma^I=(\sigma_3, -i\sigma_2, i\sigma_1)^I$ denotes a tuple of Pauli matrices, and $\mathscr{o}_{ab}$ is a sign which depends on the choice of orientation of the link $ab$: it is positive when the orientation is incoming at $n_{ab}$ and negative otherwise. The geometrical vectors are obtained from $\mathrm{SU}(1,1)$ group elements via the spin homomorphism
\begin{equation}
    \begin{gathered}
        \pi: \; \mathrm{SU}(1,1)\; \rightarrow \; \mathrm{SO}_0(1,2) \\
        g \sigma_\mu g^\dagger = \pi(g)^\nu_{\;\;\mu} \sigma_\nu\,, \quad\mu,\nu=0,1,2\,,
    \end{gathered}
\end{equation}
projecting down to the connected identity component of the Lorentz group. Note that Eq.~\eqref{eq:closure} implies that the critical points are characterized by (possibly skewed) 2-polygons at each of the six boxes, whose lengths are determined by the spin assignments. The data $n_{ab}\in \mathrm{SU}(1,1)$, in turn, are in correspondence with the edge vectors.

Further clarity can be achieved by following the algorithm of \cite{Dona:2017dvf, Dona:2020yao}, which allows for determining critical points of spin-foam amplitudes based on non-simplicial polytopes. The idea is to consider sets of three quadrilaterals $\Box_a,\Box_b,\Box_c$ which are pair-wise adjacent, so as to determine $g_a,g_b,g_c$ at criticality; one then finds two other quadrilaterals adjacent to one of the former, and reiterates the method until all critical points are identified. There are only two types of such sets in Eq.~\eqref{vertex_color}: either all quadrilaterals meet at space-like edges (a case which reduces to what was studied in \cite{Simao:2024don}), or two quadrilaterals share a time-like edge. Consider then the latter, and assume the time-like edge is shared between  $\Box_a$ and $\Box_c$. The sets of equations \eqref{gluesl} and \eqref{gluetl} imply
\begin{equation}
\label{system}
    \begin{cases}
        g_a^{-1} g_b n_{ba} n_{ab}^{-1} = e^{-i \theta_{ab} v_{ab} \cdot \varsigma^\dagger} \\
        g_b^{-1} g_c n_{cb} n_{bc}^{-1} = e^{-i \theta_{bc} v_{bc} \cdot \varsigma^\dagger} \\
        g_a^{-1} g_c n_{ca} n_{ac}^{-1} = e^{-i \rho_{ac} v_{ac} \cdot \varsigma^\dagger} 
    \end{cases} \,,
\end{equation}
where we have introduced $\theta_{ab}:=\ln \vartheta_{ab}$ and $\rho_{ac}:=i\ln \varrho_{ac}$.\footnote{Recall that $|v_{ab}|^2=|v_{bc}|^2=-1$ and $|v_{ac}^2|=1$.} As in \cite{Dona:2017dvf, Dona:2020yao}, we make the explicit gauge choice of setting $n_{ab}=n_{ba}$ for all boundary data to simplify one set of solutions of Eqs.~\eqref{gluesl}--\eqref{eq:closure}. In complete analogy to the fully space-like case of \cite{Simao:2024don}, straightforward - if tedious - algebra then yields the angle formulas
\begin{equation}
\label{eq:angle1}
  \rho_{ac}=0 \quad \vee \quad  \tan \rho_{ac} = \frac{v_{cb} \cdot v_{ab}\times v_{ac}}{(v_{ac}\times v_{cb})\cdot (v_{ab}\times v_{ac})}\,,
\end{equation}
\begin{equation}
\label{eq:angle2}
  \theta_{ab}=0 \quad \vee \quad  \tanh \theta_{ab} = \frac{v_{cb} \cdot v_{ac}\times v_{ab}}{(v_{ab}\times v_{cb})\cdot (v_{ac}\times v_{ab})}\,,
\end{equation}
\begin{equation}
\label{eq:angle3}
  \theta_{cb}=0 \quad \vee \quad  \tanh \theta_{cb} = \frac{v_{ac} \cdot v_{ab}\times v_{cb}}{(v_{ac}\times v_{cb})\cdot (v_{ab}\times v_{cb})}\,,
\end{equation}
involving Minkowski vector $\times$ and scalar $\cdot$ products. The angle associated to a space-like edge lies in the corresponding orthogonal plane which is isomorphic to $\R^{1,1}$. As a result, the angle $\theta_{ab}$ is Lorentzian and thus defined via a tangent hyperbolic. In contrast, the angle associated to a time-like edge lies in the corresponding orthogonal plane which is isomorphic to $\R^2$. Thus, the angle formula for $\rho_{ac}$ contains a trigonometric tangent. Note moreover that the equations yield two sectors of solutions: if some angle is zero then all remaining ones must vanish as well as per Eq.~\eqref{system}; this propagates to every other set of three quadrilaterals in the cuboid, as can be seen from equations analogous to Eq.~\eqref{system} for all other sets, and all critical group elements are identified with the identity $g_a = \mathds{1}$. As to the second sector, once all angles $\theta,\rho$ are determined (provided the relevant equations admit solutions, see \cite{Jercher:2024kig}), one may resort to equations of the type \eqref{system} to determine all critical $g_a$.

Putting everything together, the asymptotic amplitude for an arbitrary assignment of causal characters reads
\begin{equation}
\label{as1}
    A_{v} = e^{\frac{7i\pi}{4}}\frac{2^{\Delta_{\mathrm{tl}}-10}}{(2\pi)^{5/2}}\prod_{ab}^{\mathrm{sl}} d_{s_{ab}} \prod_{ab}^{\mathrm{tl}}d_{k_{ab}}  \left(\frac{1}{\sqrt{\det H_{\mathds{1}}}} + \Theta \frac{e^{2 i\sum_{ab}^{\mathrm{sl}} s_{ab} \theta_{ab}  + 2i \sum_{ab}^{\mathrm{tl}} (-k_{ab}) \rho_{ab}}}{\sqrt{\det H_{\vartheta}} \prod_{ab}^{\mathrm{sl}} \vartheta_{ab} } \right) + \mathcal{O}\left(j^{\frac{11}{2}}\right) \,,
\end{equation}
where $\Theta=0,1$ is a binary toggle for the second sector of solutions: if the boundary data is such that the quadrilaterals are either all time-like or all space-like, then $\Theta=1$. The numerical factors heading the equation are obtained from 1) the Haar measure, 2) Hörmander's theorem \cite[Th. 7.7.5]{Hormander2003} and 3) a spin redundancy ($g_a \mapsto -g_a$) of factor $2$ depending on the number $\Delta_{\mathrm{tl}}$ of non-gauge-fixed squares with entirely time-like edges. The matrices appearing in the asymptotic formula are the Hessian matrices of the total spin-foam action $S=\sum_{ab}^\mathrm{tl} S_{ab}^{\mathrm{tl}}+\sum_{ab}^\mathrm{sl} S_{ab}^{\mathrm{sl}}$ evaluated at the two critical points; we denote by $H_{\mathds{1}}$ the Hessian  at all $g_a=\pm \mathds{1}$, and by $H_\vartheta$ the Hessian at the non-trivial critical point. The Hessian matrices have the symmetric form
\begin{equation}
     H(\{g_a\})=\begin{pmatrix}
        
       H^{11} & H^{12} & H^{13} & H^{14} & H^{15}  \\
     &  H^{22} & H^{23} & H^{24} & H^{25}    \\
     & & H^{33} & H^{34} & H^{35}   \\
     & & &H^{44} & H^{45}  \\
     & & & & H^{55} 
     \end{pmatrix}\,,
\end{equation}
where each component $H^{ab}$ is a $3\times 3$ matrix, the form of which depends on the causal character assigned to the edge $\diagup_{ab}$, and where $g_6=\mathds{1}$ has been gauge-fixed. The diagonal blocks read
\begin{equation}\label{eq:diag Hess}
H_{IJ}^{aa}=-\sum_b^{ab\, \mathrm{tl}} \frac{k_{ab}}{2}\left[\eta_{IJ} -v_{ab,I}^{(a)} v_{ab,J}^{(a)} \right]  -\sum_b^{ab\, \mathrm{sl}}  \frac{i s_{ab}}{2}\left[\eta_{IJ} +v_{ab,I}^{(a)} v_{ab,J}^{(a)}-i\vartheta_{ab}^2 m_{ab,I}^{(a)} m_{ab,J}^{(a)} \right]\,,
\end{equation}
while for the off-diagonal $a\neq b$ blocks we have
\begin{equation}
    {}^{\mathrm{tl}}H_{IJ}^{ab}=\frac{k_{ab}}{2}\left[\eta_{IJ} -v_{ab,I}^{(a)} v_{ab,J}^{(a)} -i \epsilon_{IJK} \mathscr{o}_{ab} \eta^{KL} v_{ab,L}^{(a)}\right]\,,
\end{equation}
\begin{equation}\label{eq:off-diag sl Hess}
       {}^{\mathrm{sl}}H_{IJ}^{ab}=\frac{i s_{ab}}{2}\left[\eta_{IJ} +v_{ab,I}^{(a)} v_{ab,J}^{(a)} + \epsilon_{IJK}  \mathscr{o}_{ab} \eta^{KL} v_{ab,L}^{(a)}-i\vartheta_{ab}^2 m_{ab,I}^{(a)} m_{ab,J}^{(a)} \right]\,.
\end{equation}
In the equations above we have denoted $v_{ab,I}^{(a)}:= \left[\pi(g_a) v_{ab}\right]_I$ for simplicity, and introduced the future null vectors
\begin{equation}
m_{ab}=\pi(n_{ab})(\hat{e}_0-\hat{e}_1) \in C^+ \subset \mathbb{R}^{1,2}\,, \quad \braket{l^+_{ab}|\sigma_3 \varsigma^I|l^+_{ab}}=m_{ab}^I\,, \quad ab \;\text{sl}\,,
\end{equation}
constructed from the boundary data associated to space-like edges.

The exponential function of Eq.~\eqref{as1} is a global phase. On the other hand, note from Eqs.~\eqref{geo_vec_sl} and~\eqref{geo_vec_tl} that the geometrical vectors determined by the boundary data $\partial \psi$ are oblivious to a phase change at each coherent state. A precedent has therefore appeared in the literature \cite{Barrett:2009gg, Barrett:2009mw, Kaminski:2017eew, Dona:2017dvf} where the phase of each coherent state is fixed (given the global information of the vertex amplitude) in order to bring the asymptotic amplitude into a more symmetric form. Proceeding as such, under a concrete choice of phase 
\begin{equation}
    \begin{cases}
        \ket{\tau_{ab}} \mapsto e^{i k_{ab} \rho_{ab}} \ket{\tau_{ab}} \\
        \ket{l^-_{ab}} \mapsto e^{-i s_{ab} \theta_{ab}} \ket{l^-_{ab}}\,,
    \end{cases}
\end{equation}
the first term of Eq.~\eqref{as1} becomes
\begin{multline}\label{eq:vertex semi-classics}
    A_{v}^{\mathrm{asy}} = e^{\frac{7i\pi}{4}}\frac{2^{\Delta_{\mathrm{tl}}-10}}{(2\pi)^{5/2}}\prod_{ab}^{\mathrm{sl}} d_{s_{ab}} \prod_{ab}^{\mathrm{tl}}d_{k_{ab}}  \Biggl(\frac{e^{ -i\sum_{ab}^{\mathrm{sl}} s_{ab} \theta_{ab}  - i \sum_{ab}^{\mathrm{tl}} (-k_{ab}) \rho_{ab}}}{\sqrt{\det H_{\mathds{1}}}}  \\
    + \Theta \frac{e^{ i\sum_{ab}^{\mathrm{sl}} s_{ab} \theta_{ab}  + i \sum_{ab}^{\mathrm{tl}} (-k_{ab}) \rho_{ab}}}{\sqrt{\det H_{\vartheta}} \prod_{ab}^{\mathrm{sl}} \vartheta_{ab} } \Biggr) \,.
\end{multline}
The above expression captures the well-known result that spin-foam asymptotics tend to reproduce the cosine of the boundary Regge action \cite{Barrett:2009gg, Barrett:2009mw, Kaminski:2017eew, Liu:2018gfc, Simao:2021qno, Dona:2017dvf, Dona:2020yao}. Here, we note the additional property of the current model that the second term of the cosine may be absent depending on the causal structure, which has been shown in~\cite{Jercher:2024kig}. Since the arguments therein are local and only assume that the edges of the 3-cells are 3-valent these results transfer to the cuboidal case considered here.
\section{A proposal for an effective cosmological amplitude}\label{sec:A proposal for an effective cosmological amplitude}

We are now interested in building a model for spin-foam cosmology from the fundamental theory outlined earlier. Our approach will be an effective one, in that we shall make a sequence of assumptions and simplifications such as a symmetry reduction to spatially flat discrete geometries. The goal is to construct a model that is numerically feasible and thus allows for the explicit computation of the partition function and expectation values.

\subsection{Geometrical structure of the model}\label{sec:Geometrical structure of the model}

To start with, we conceive of spacetime as a homogeneous and isotropic Lorentzian manifold, in line with the usual cosmological assumptions. We further require for its topology the product $\mathcal{T} \simeq \mathbb{R}\times T^2$, with each toroidal $T^2$ leaf taken as space-like and flat. 

\begin{figure}
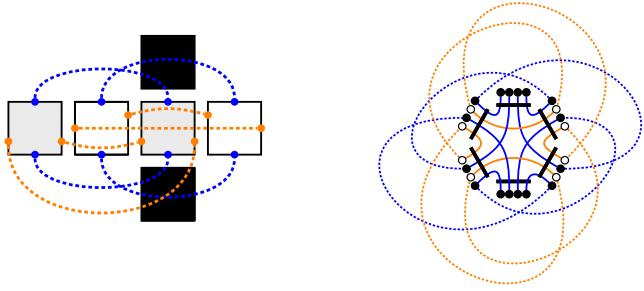

    \centering
     \scalebox{0.7}{\unfold} \scalebox{0.45}{\eldrich}
    \vspace{-1cm} \caption{Left: \enquote{unfolded} Lorentzian cuboid with spacelike top and bottom faces (black) and four faces of arbitrary signature (gray and white). Edge identifications necessary to fold the boundary back into a cuboid are not explicitly drawn and understood implicitly. To obtain a spatial $T^2$ toroidal topology, similarly-colored white and gray faces are identified, as are their respective edges. The orange-colored edge identifications lead to a single bulk edge. Blue-colored edge identifications readily induce a toroidal topology of the black faces. Right: representation of edge identifications of the left diagram in the amplitude diagram. One can verify that the orange line consists of a single loop. There are four open blue lines, two at the top and two at the bottom spacelike face. These correspond to the initial and final tori each of which is characterized by two radii.} 
    \label{torus}
\end{figure}

Looking to discretize the above continuum picture, we consider a homeomorphic polyhedral complex $\mathcal{X}_{\mathcal{V}}$ of $\mathcal{V}$ frusta-like 3-cells, organized in a linear chain along the temporal direction. Such configurations have proven ideal to capture classical discrete cosmological dynamics with the correct continuum time limit in Euclidean~\cite{Bahr:2017bn} and Lorentzian~\cite{Jercher:2023csk} signature, respectively. The cosmological principle is implemented in this discrete setting by demanding the 2-cells to be squares, and thus characterized by a single edge-length to which we refer to in the following as spatial edge length. Due to the topological requirements, a single intrinsically flat 2-cell suffices to capture the entire spatial geometry of each leaf. Consecutive squares are connected by four edges we term \emph{struts}, which are identified amongst themselves, carrying the same length. The appropriate strut identifications are depicted in Fig.~\ref{torus}.
We allow for varying spatial edge lengths at different instances of time (i.e. at different frusta), standing in analogy to the change of scale factor in time. 

\begin{figure}
    \centering
    \includegraphics[width=0.3\linewidth]{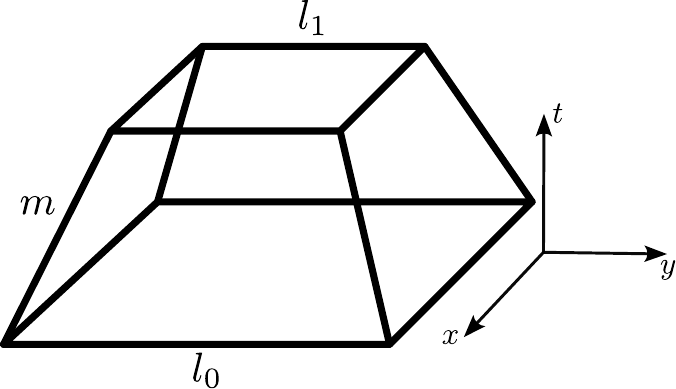}
    \caption{A Lorentzian 3-frustum where the absolute value of the edge lengths is given by $l_0$ and $l_1$ for the spatial edges and $m$ for the struts.}
    \label{fig:3frustum}
\end{figure}

As a result of this construction, spacetime is reduced to 3-dimensional Lorentzian frusta\footnote{Fusta geometries can be realized on a triangulation by sub-division into Lorentzian tetrahedra and imposing flatness conditions on deficit angles located at sub-dividing edges. For a detailed description in (3+1) dimensions, see~\cite{Jercher:2023csk}.} bounded by two (generically different) squares and four trapezoids; see Fig.~\ref{fig:3frustum} for a visualization. In this restricted setting, a single 3-dimensional building block is fully characterized by three geometric variables $(l_0,l_1,m)$\footnote{We remind the reader that in three dimensions the geometric data of spin-foams already correspond to edge length. Consequently, no obstructions arise from a transition between area and length variables~\cite{Barrett:1997tx,Asante:2024rrd} which led to the introduction of the so-called gluing constraints in 4-dimensional effective spin-foams~\cite{Asante:2020qpa,Asante:2021zzh}. Irrespective of the dimension, in the symmetry reduced setting considered here, there is a one-to-one correspondence between area and length variables such that the constraints trivialize, as discussed in~\cite{Dittrich:2023rcr,Jercher:2023csk,Asante:2021phx}.}: the spatial edge lengths $l_0$ and $l_1$ of the two square slices labelled by $0$ and $1$, and the length $m$ of the struts connecting the two slices. 
Notice that these edge lengths correspond to absolute values, i.e. $l_0,l_1,m > 0$. While the spatial edges are always space-like by assumption, the struts are allowed to be either space-like or time-like. For the remainder of this work, we denote the \textit{signed} volumes in boldface, so signed squared strut length are denoted as $\vb*{m}^2$. For a time-like strut, $m = \sqrt{\vb*{m}^2}$, and for a space-like strut, $m = \sqrt{-\vb*{m}^2}$.

\subsection{Spin-foam amplitude simplifications}\label{sec:Spin-foam amplitude simplifications}

Having characterized the geometry of our setting, we now discuss the assignment of a spin-foam amplitude to a particular geometrical configuration. Given the 3-complex $\mathcal{X}_{\mathcal{V}}$ described above and some a priori unrestricted history $\psi$, the model of Sec.~\ref{sec:model} would prescribe a spin-foam amplitude of the form
\begin{equation}
     A(\mathcal{X}_{\mathcal{V}}, \psi)= \scalebox{0.4}{\cubethree}\,,
\end{equation}
where all open lines are taken to be joined to adjacent diagrams; there is one hexagonal diagram per 3-cell in $\mathcal{X}_{\mathcal{V}}$. An equivalent form of the amplitude can be obtained by making use of a completeness relation of coherent states on the Hilbert spaces of the relevant representations,
\begin{equation}
    \label{eq:restl}
        \mathds{1}_{k(q)} = (-2k-1) \int \mathrm{d} g \; D^{k (q)} (g) \ket{k, -qk}\bra{k, -qk} D^{ k (q) \dagger}(g) =: d_k\int \mathrm{d}g \raisebox{1pt}{\scalebox{0.7}{\restime}}\,,
\end{equation}
\begin{equation}
\label{eq:ressl}
    \mathds{1}_{j(\delta)} = s \tanh (\pi s)^{1-4\delta}\int \mathrm{d} g \; D^{j (\delta)} (g) \ket{j, ij, 0}\bra{j, \overline{ij}, 0} D^{ j (\delta) \dagger}(g) =:  d_j\int \mathrm{d}g \raisebox{1pt}{\scalebox{0.7}{\resspace}}\,,
\end{equation}
which follow from the orthonormality of $\mathrm{SU}(1,1)$ Wigner matrices in $L^2(SU(1,1))$, as per the discussion in Appendix~\ref{app:su11}. Consequently, any adjacent diagrams can be rewritten as 
\begin{equation}
\label{glue}
     \scalebox{0.4}{\cubetwo} = \prod_{i=1}^4 d_{j_i}\int\mathrm{d}g_i \scalebox{0.4}{\cubetwores},
\end{equation}
having made in this example an explicit assumption on the causal character of the joined lines - namely that they are time-like, and that the appropriate resolution is $d_k \int \mathrm{d}g \raisebox{1pt}{\scalebox{0.7}{\restime}}$. Furthermore, recall that each coherent state appearing in Eqs.~\eqref{eq:restl} and~\eqref{eq:ressl} admits a geometrical interpretation: note that \cite{Simao:2024don}
\begin{equation}
    \braket{j, \overline{ij}, 0| D^{ j (\delta) \dagger}(g)\, D^{ j (\delta)}( \varsigma^I)\,  D^{ j (\delta)}( g) | j, ij, 0 } = - \frac{\gamma}{\pi} \braket{l^+|g^\dagger \sigma_3 \varsigma^I g| l^-}^{2j}\,, 
\end{equation}
\begin{equation}
    \braket{k, -qk| D^{ k (q) \dagger}(g)\, D^{ k (q)}( \varsigma^I)\,  D^{ k (q)}( g) | k, -qk } =  \braket{-q|g^\dagger \sigma_3 \varsigma^I g|-q}^{2k}\,, 
\end{equation}
and Eqs.~\eqref{geo_vec_sl} and~\eqref{geo_vec_tl} show that the right-hand side above relates to vector components of 3-vectors in either the space-like or time-like hyperboloids, respectively. One is thus justified in thinking of Eq.~\eqref{glue} as a gluing identity between vertices, where the integrations range over all possible geometric vectors assigned to the boundary of each diagram. Notice that also at the four identified struts in Fig.~\ref{torus}, a resolution of the identity is inserted.

The next step in our simplification leverages the geometrical interpretation of the coherent states, and enforces a particular boundary geometry. The process consists in performing a symmetry reduction, by fixing the integration domain at each gluing to a particular group element - and consequently a particular coherent state - in correspondence with the frustum geometry of Fig.~\ref{fig:3frustum}. In other words, we insist on complementing each history $\psi \mapsto \psi'$ with coherent states at the boundary of each diagram, and operate the reduction
\begin{equation}
    A(\mathcal{X}_{\mathcal{V}},\psi) =  \prod_{i=1}^4 d_{k_i}\int \mathrm{d}g_i \scalebox{0.4}{\cubetwores}\; \mapsto\;  \hat{A}_1(\mathcal{X}_{\mathcal{V}},\psi')
    :=  \scalebox{0.4}{\cubetwosplit}\,,
\end{equation}
arriving at a first modified amplitude $\hat{A}_1$.\footnote{Notice that we also dropped the Plancherel factors $d_j$ and $d_k$, corresponding to a modification of the face amplitude of the model. In~\cite{Bahr:2015gxa,Bahr:2017bn}, different choices of face amplitudes were parametrized by an exponent $d_j^\alpha$ and cuboid and frustum intertwiners have been introduced with a normalization factor modifying the edge amplitude $\mathcal{A}_e$. Here, we effectively set $\mathcal{A}_e = 1$ and introduce Plancherel factors only for closed loops. For highly oscillating amplitudes in one variable as considered in Secs.~\ref{sec:Numerical evaluation: Strut in the bulk}, modified edge and face amplitudes are expected to have a negligible influence on the qualitative behavior of expectation values. They can however influence the numerical stability of series accelerations and numerical integration. The choices made here are going to prove numerically feasible.} The structure of this new amplitude is such that it factorizes into a product of frusta amplitudes, each of which is of a similar structure to that of the original vertex amplitude $A_v$ of Eq.~\eqref{cube_vertex}. The amplitudes $\hat{A}_1$ and $A_v$ are however not strictly equivalent. The time-like coherent states of the identity resolution satisfy
\begin{equation}
 \!{}^{n_{ab}}\overset{\rightarrow}{\scalebox{0.65}{\braketsquare}}{}^{n_{ba}} = \!{}^{n_{ab}}\overset{\rightarrow}{\scalebox{0.65}{\braketblack}}{}^{n_{ba}} / d_{k_{ab}}
\end{equation}
as per the definitions in Eqs.~\eqref{brakettl} and~\eqref{eq:restl}. In particular, the pairings of Eq.~\eqref{brakettl} do not come with an additional constraint $\mathcal{C}$, as gluing is implicitly ensured~\cite{Simao:2024don}. The same is not true for space-like pairings: the closure constraint $\mathcal{C}_{n_{ab},n_{ba}}$ of Eq.~\eqref{braketsl} is missing, and it is only that
\begin{equation}
 \!{}^{n_{ab}}\overset{\rightarrow}{\scalebox{0.65}{\braketsquareblack}}{}^{n_{ba}} = \!{}^{n_{ab}}\overset{\rightarrow}{\scalebox{0.65}{\brakettikz}}{}^{n_{ba}} / d_{j_{ab}} \,.
\end{equation}
Following~\cite{Simao:2024don}, the constraint $\mathcal{C}_{n_{ab},n_{ba}}$ must be added by hand in order for the vertex amplitude to have a well-behaved asymptotic formula. Hence, we perform another modification to the amplitude by including the constraint on every vertex, 
\begin{equation}
    \hat{A}_1(\mathcal{X}_{\mathcal{V}},\psi') =  \scalebox{0.4}{\cubetwosplit}\; \mapsto\;  \hat{A}_2 ( \mathcal{X}_{\mathcal{V}},\psi')
    :=  \scalebox{0.4}{\cubetwosplitorig}\,,
\end{equation}
such that, finally, the amplitude $\hat{A}_2 (\mathcal{X}_{\mathcal{V}},\psi')$ amounts to a simple product of the vertex amplitude $A_v$ over every frustum in $\mathcal{X}_{\mathcal{V}}$, i.e. 
\begin{equation}
    \hat{A}_2 (\mathcal{X}_{\mathcal{V}},\psi') = \prod_{{\cube} \in \mathcal{X}_{\mathcal{V}}} A_{v}(\psi'|_{\cube}) \prod_{\mathrm{bulk }\diagup \in \mathcal{X}_{\mathcal{V}}} A_f(\psi'|_\diagup)\,,
\end{equation}
where $A_f$ is the face amplitude associated to bulk edges $\diagup \in \mathcal{X}_{\mathcal{V}}$. It corresponds to the Plancherel measure appearing in Eqs.~\eqref{brakettl} and~\eqref{braketsl}, depending on the history, as
\begin{equation}\label{eq:face amplitudes}
    A_f = \begin{cases}
        -2k-1\,, \quad \diagup \text{ is t.l.} \\
        s \tanh(\pi s )\,, \quad \diagup \text{ is s.l.}
    \end{cases}\,.
\end{equation}

As it stands, the amplitude $\hat{A}_2$ does not yet single out a frustum geometry. The interpretation of the spin-foam vertex amplitude in terms of a geometrical polyhedron follows from the semi-classical limit, where only histories derived from convex polyhedra dominate. Yet another simplification step in our sequence is therefore to replace the amplitude $\hat{A}_2$ with the semi-classical formula at each frustum, evaluating it on the geometrical data of Fig.~\ref{fig:3frustum}. That is, we define
\begin{equation}
\label{eq:a3}
    \hat{A}_2(\psi', \mathcal{X}_{\mathcal{V}}) \; \mapsto \; \hat{A}_3 (\psi', \mathcal{X}_{\mathcal{V}}) = \prod_{{\cube} \in \mathcal{X}_{\mathcal{V}}} A^{\mathrm{asy}}_{v}(\psi'|_{\cube}) \prod_{\mathrm{bulk }\diagup \in \mathcal{X}_{\mathcal{V}}} A_f(\psi'|_\diagup)\,,
\end{equation}
with $A^{\mathrm{asy}}_{v}$ prescribed by Eq.~\eqref{eq:vertex semi-classics}. 

\subsection{Boundary data for symmetry reduced Lorentzian 3-frusta}
\label{sec:bdata}

The effective model summarized in Eq.~\eqref{eq:a3} is valid for arbitrary boundary data $\partial \psi$ associated to a boundary graph of cuboidal combinatorics. In order to specify to the setting of 3-dimensional spatially flat Lorentzian cosmology with toroidal topology, as laid out in Section \ref{sec:Geometrical structure of the model}, we now prescribe the necessary boundary data. 

One first needs to appropriately identify the geometric data of a 3-frustum $(l_0,l_1,m)$ with the boundary data $\partial\psi = \{t_{ab},n_{ab}\}$, consisting of spins $t_{ab}$ (standing collectively for spins of both continuous $j$ and discrete $k$ types) and group elements $n_{ab}\in\SUO$. The spins $t_{ab}$, which lie in the discrete (continuous) series, and are associated to the time-like (space-like) 
edges $(ab)$, can be semi-classically identified with the geometrical edge length.\footnote{The Casimir spectrum of the continuous series of $\mathrm{SU}(1,1)$ representations is given by $s^2+\frac{1}{4}$ with $s\in\R$ and thus exhibits a length gap. Different identifications of space-like geometrical edge length and $s$ are possible, and we choose here $l_1 = s+\frac{1}{2}$ with a real off-set to map the length gap. As we are going to detail in Sec.~\ref{sec:Evaluation of the partition function}, this choice yields an effective amplitude finite in the bulk spatial edge length.} Thus, for edges $(ab)$ of squares, we identify the spins of the continuous series $s_{ab}$ with the length $l$ as $l = s_{ab}+\frac{1}{2}$. For time-like struts, we identify $-k_{ab}$ with the strut length $m$, while for space-like struts, we set $m=s_{ab}+\frac{1}{2}$. 

As a consequence of the symmetry reduction, the normalized geometric edge vectors $v_{ab}^{(a)}$, which determine the group elements $n_{ab}$ up to a phase, are a function of the edge lengths $(l_0,l_1,m)$. Choosing an embedding of the 3-frustum into $\R^{1,2}\ni(t,x,y)$ where the squares lie in constant-$t$ planes with edges parallel to the $x$ and $y$ directions, the edge vectors of squares are given by $(0,\pm 1, 0)$ or $(0,0,\pm 1)$. Strut edge vectors take the form
\begin{equation}\label{eq:strut vec}
v_{ab} = \frac{1}{m}\left(\epsilon^0_{ab}\sqrt{\frac{(l_0-l_1)^2}{2}+\vb*{m}^2},\epsilon^1_{ab}\frac{l_0-l_1}{2},\epsilon^2_{ab}\frac{l_0 - l_1}{2}\right).
\end{equation}
Here, the signs $\epsilon^I_{ab} =\pm$ are chosen such that the vectors $v_{ab}$ and $v_{ba}$ are parallel and closure holds as in Eq.~\eqref{eq:closure}. The $0$-component of the strut vectors correspond to the height of the 3-frustum which follows immediately from the chosen embedding.

Provided this boundary data, the geometry of a single 3-frustum is fully characterized. In particular, the signed squared volumes of different sub-cells can be determined as a function of $l_0,l_1$ and $m$, which we summarize in the following. The squared height of $3$-frusta is given by
\begin{equation}
    \vb*{H}^2 = \vb*{m}^2+\frac{(l_0-l_1)^2}{2}\,.
\end{equation}
Embeddability of a $3$-frustum in $\R^{1,2}$ requires $\vb*{H}^2 > 0$ and thus poses the condition $\vb*{m}^2 > -\frac{(l_0-l_1)^2}{2}$. Configurations that violate this bound correspond to Euclidean $3$-frusta. Notice furthermore, that this condition sets a bound on space-like struts with $\vb*{m}^2<0$, while time-like struts remain unbounded relative to $l_0$ and $l_1$. To see explicitly that the sign of $\vb*{m}^2$ determines the causal character of the struts, we compute the Minkowski norm of $v_{ab}$ in Eq.~\eqref{eq:strut vec}, yielding $v_{ab}^2 = \vb*{m}^2/m^2 = \mathrm{sign}(\vb*{m}^2)$. Hence, $\vb*{m}^2 > 0$ $(<0)$ corresponds to a time-like (space-like) strut. The squared height of trapezoids is given by
\begin{equation}
\vb*{h}^2 = \vb*{H}^2-\frac{(l_0-l_1)^2}{4} = \vb*{m}^2 +\frac{(l_0-l_1)^2}{4}\,,
\end{equation}
from which the squared area of trapezoids follows as
\begin{equation}
\vb*{v}^2 = \left(\frac{l_0+l_1}{2}\right)^2 \vb*{h}^2 = \left(\frac{l_0+l_1}{2}\right)^2\left[\vb*{m}^2 +\frac{(l_0-l_1)^2}{4}\right]\,.
\end{equation}
A trapezoid is therefore space-like if $-\frac{(l_0-l_1)^2}{2} < \vb*{m}^2 < -\frac{(l_0-l_1)^2}{4}$, and time-like otherwise. The squared 3-volume of a 3-frustum is given by
\begin{equation}\label{eq:3-volume}
\vb*{V}^2 =\left(\frac{l_0^2+l_0 l_1+l_1^2}{3}\right)^2\vb*{H}^2 = \left(\frac{l_0^2+l_0 l_1+l_1^2}{3}\right)^2\left[\vb*{m}^2+\frac{(l_0-l_1)^2}{2}\right]\,,
\end{equation}
which is positive if the 3-frustum is Lorentzian, i.e. if it can be embedded into $\R^{1,2}$. Finally, due to the topological identifications described in Figure \ref{torus}, note that a single vertex amplitude is sufficient to capture the whole space-like slab.

\begin{figure}
    \centering
   \includegraphics[width=0.7\linewidth]{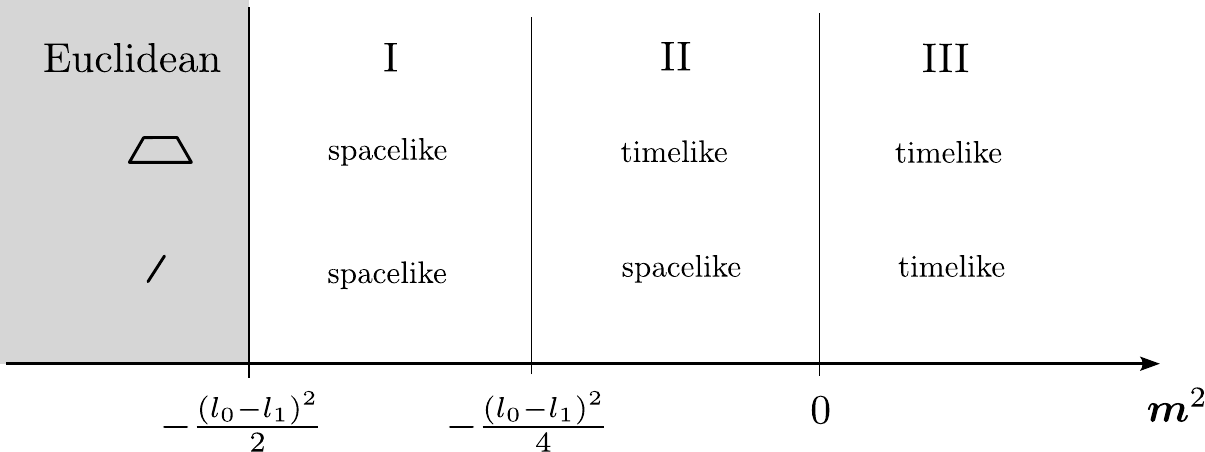}
   \caption{A categorization of the boundary data of a Lorentzian 3-frusta according to the causal character of its sub-cells. In Sector I, struts and trapezoids are space-like, in Sector II struts are space-like and trapezoids are time-like. In Sector III, struts and trapezoids are time-like. The special case of a Lorentzian cuboid with $l_0 = l_1$ can be realized only with time-like struts.}
    \label{fig:sectors}
\end{figure}

Following~\cite{Dittrich:2021gww,Jercher:2023csk,Dittrich:2023rcr,Asante:2021phx}, the different causal characters of the trapezoids and struts contained in Lorentzian $3$-frusta are conveniently captured by three sectors of the theory, depicted in Fig.~\ref{fig:sectors}. In Sector I, defined by a signed strut length in the range $-\frac{(l_0-l_1)^2}{2} < \vb*{m}^2 < -\frac{(l_0-l_1)^2}{4}$, the struts and the trapezoids are space-like. In Sector II, the signed strut length lies in the range $-\frac{(l_0-l_1)^2}{4} < \vb*{m}^2 <0$ such that the strut is still space-like while the trapezoid is time-like. At the interface of Sector I and II, i.e. at $\vb*{m}^2 = -\frac{(l_0-l_1)^2}{4}$, the trapezoid is light-like and constitutes still a critical point of the spin-foam partition function. In Sector III, defined by $\vb*{m}^2 > 0$, the trapezoid and the struts are time-like. This range of configurations also captures the Lorentzian cuboid with squares $l_0 = l_1$ and height $m$. A cuboid with all edges space-like can only be embedded in Euclidean space.  

\subsection{Asymptotic vertex amplitude and measure factors}\label{sec:Asymptotic vertex amplitude and measure factors}

The semi-classical amplitude can be evaluated for the particular choice of boundary data discussed in this section, amounting to 
%
\begin{equation}\label{eq:vertex semi-classics 2}
 A_v^{\mathrm{asy}} = \left(\upmu_{\mathds{1}}(l_0,l_1,m)\e^{-i\,\mathfrak{Re}\{S_{\mathrm{R}}\}}+\Theta\,\upmu_{\vartheta}(l_0,l_1,m)\e^{i\,\mathfrak{Re}\{S_{\mathrm{R}}\}}\right) \,,
\end{equation}
where $\Theta$ is one in Sector I and vanishes in II and III. The phases $\mathfrak{Re} \{S_{\mathrm{R}}\}$ contain the Lorentzian Regge action, taking over the three sectors the values
\begin{align}
S_{\textsc{i}} &= 4\abs{l_0-l_1}\left[i\frac{\pi}{2}-\cosh^{-1}\left(\frac{l_0-l_1}{\sqrt{4m^2-(l_0-l_1)^2}}\right)\right]-4m\left[i\frac{\pi}{2}-\cosh^{-1}\left(\frac{(l_0-l_1)^2}{4m^2-(l_0-l_1)^2}\right)\right]\,,\label{eq:S_I}\\[7pt]
S_{\textsc{ii}} &= 4(l_0-l_1)\sinh^{-1}\left(\frac{l_1-l_0}{\sqrt{4m^2-(l_0-l_1)^2}}\right)+4m\left[i\frac{\pi}{2}+\cosh^{-1}\left(\frac{(l_0-l_1)^2}{4m^2-(l_0-l_1)^2}\right)\right]\,,\label{eq:S_II}\\[7pt]
S_{\textsc{iii}} &= 4(l_0-l_1)\sinh^{-1}\left(\frac{l_1-l_0}{\sqrt{4m^2-(l_0-l_1)^2}}\right)+4m\left[\frac{\pi}{2}-\cos^{-1}\left(\frac{(l_0-l_1)^2}{4m^2-(l_0-l_1)^2}\right)\right]\,,\label{eq:S_III}
\end{align}
according to Eq.~\eqref{eq:vertex semi-classics} and the angle formulas in Eqs.~\eqref{eq:angle1}--\eqref{eq:angle3}. Importantly, only the real part of the Lorentzian Regge action enters $A_v^{\mathrm{asy}}$ which is a result of the stationary phase approximation of the full spin-foam quantum amplitude; we shall have more to say on this topic in Secs.~\ref{sec:causality violations},~\ref{sec:including I and II} and~\ref{sec:comparison to ESF}. 

\begin{figure}
    \centering
    \begin{subfigure}{0.5\textwidth}
    \includegraphics[width=\linewidth]{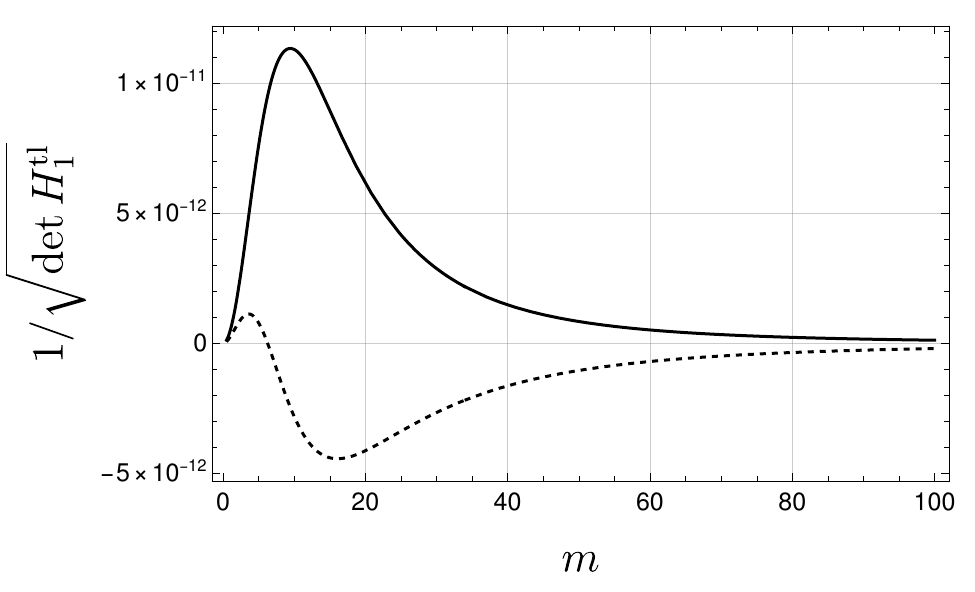}
    \end{subfigure}%
    \begin{subfigure}{0.5\textwidth}
    \includegraphics[width=\linewidth]{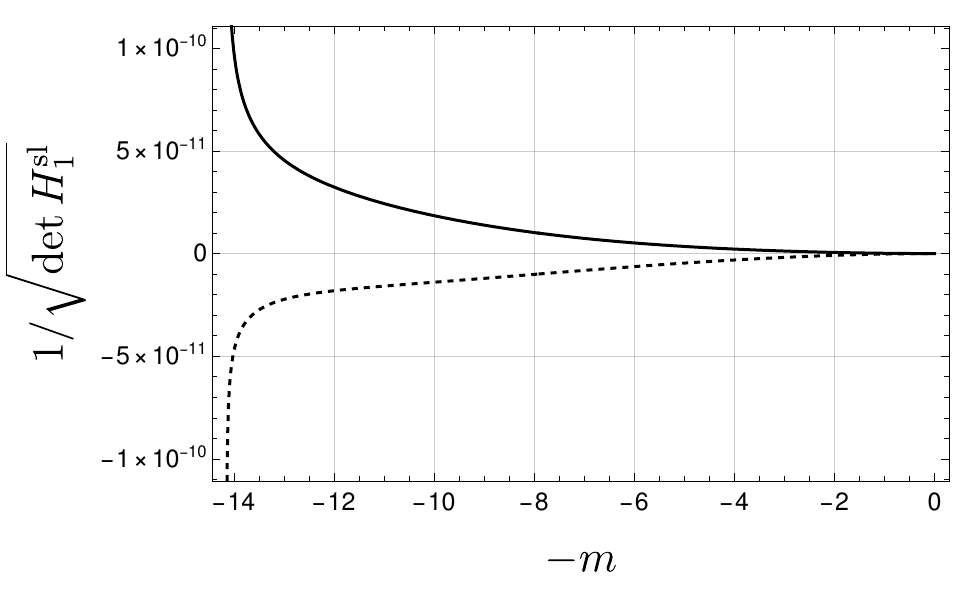}
    \end{subfigure}\\
    \begin{subfigure}{0.5\textwidth}
    \includegraphics[width=\linewidth]{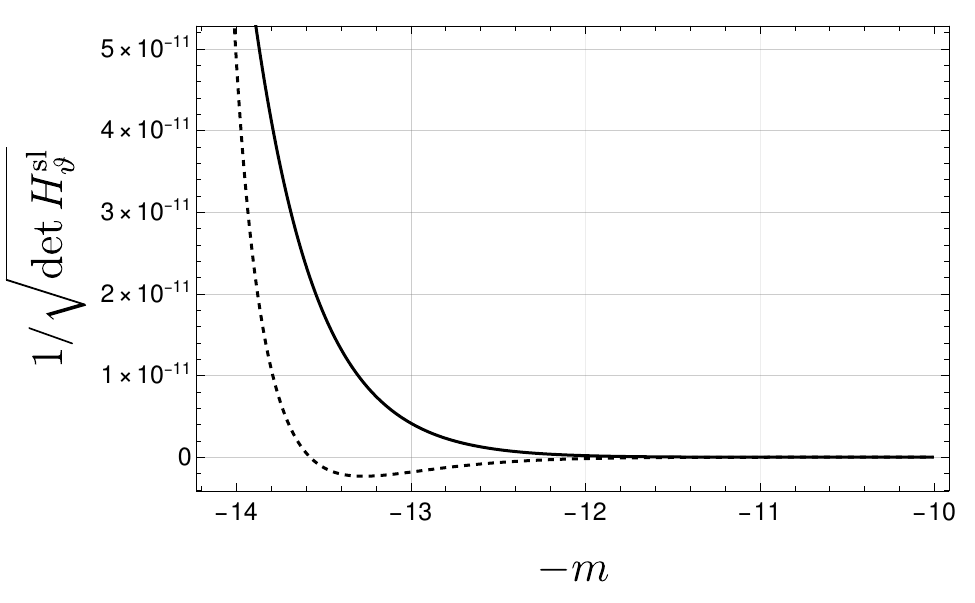}
    \end{subfigure}
    \caption{Real (solid) and imaginary (dashed) part of inverse square root of the Hessian determinants which enter the measure factors $\upmu_{\mathds{1},\vartheta}$. Plotted as a function of the strut length $m$ for spatial edge length $l_0 = 10$ and $l_1 = 30$ chosen for demonstration purposes. Top left: inverse square root of $\det H_{\mathds{1}}$ in Sector III for a time-like strut. Top right: inverse square root of $\det H_{\mathds{1}}$ in Sectors I and II (which are separated at $m = 10$.), depicted as a function of $-m$ to emphasize that the strut is space-like. Notice that $(\det H_{\mathds{1}}^{\mathrm{sl}})^{-1/2}$ is finite at $m = 10\sqrt{2}$. Bottom: inverse square root of $\det H_{\vartheta}$ in Sector I at the non-identity solution, depicted as a function of $-m$ to emphasize that the strut is space-like. Plot ranges only from $-10\sqrt{2}$ to $-10$ since the non-identity solution only exists in Sector I, see Sec.~\ref{sec:The semi-classical limit of the vertex} for a discussion. Note that also $(\det H_{\vartheta})^{-1/2}$ is finite at $m = 10\sqrt{2}$.}
    \label{fig:measures}
\end{figure}

The functions $\upmu_{\mathds{1},\vartheta}$ constitute measure factors arising from the spin-foam asymptotics and consist of the inverse square root of the Hessian and factors of $\vartheta$, given as exponentials of dihedrals angles, at the non-identity solution of the critical points,
\begin{equation}
\upmu_{\mathds{1}} = e^{\frac{7i\pi}{4}}\frac{2^{-10}}{(2\pi)^{5/2}} 
\frac{1}{\sqrt{\det H_{\mathds{1}}}}\,,\qquad \upmu_{\vartheta} = e^{\frac{7i\pi}{4}}\frac{2^{-10}}{(2\pi)^{5/2}}
\frac{1}{\vartheta_m^4 \sqrt{\det H_{\vartheta}} }\,,
\end{equation}
where $\vartheta_m$ is associated to the struts. An  exemplary plot of the measure factors is given in Fig.~\ref{fig:measures}. Since $\vartheta = 1$ at the identity solution $g_a = \mathds{1}$, the $15\times 15$-matrix $H_\mathds{1}$ simplifies significantly with many entries being zero. As a consequence, the determinant $\det H_{\mathds{1}}$ can be computed explicitly, with its functional form given by
\begin{equation}
\det H_{\mathds{1}} = \frac{1}{m^4}(l_0l_1)^3\sum_{\substack{n_0,n_1,n_m,n_s \\ n_0+n_1+n_m+n_s = 13}}c^{\mathds{1}}_{n_0,n_1,n_m,n_s}l_0^{n_0} l_1^{n_1} m^{n_m} \left(\sqrt{\frac{(l_0-l_1)^2}{2}+\vb*{m}^2}\right)^{n_s}\,,
\end{equation}
where $c^{\mathds{1}}_{n_0,n_1,n_m,n_s}$ are constant complex coefficients. At the non-identity solution $\vartheta\neq 1$, the Hessian $H_\vartheta$ has less zero entries compared to $H_\mathds{1}$ which furthermore take a more involved form. As a result, we have not been able to find an analytical formula for $\det H_{\vartheta}$. Still, one can give the functional form also for $\det H_{\vartheta}$, which is now
\begin{equation}
\det H_\vartheta = \frac{1}{m^4}(l_0l_1)^3\sum_{\substack{n_0,n_1,n_m,n_s \\ n_0+n_1+n_m+n_s = 13}}c^\vartheta_{n_0,n_1,n_m,n_s}l_0^{n_0} l_1^{n_1} m^{n_m} \left(\sqrt{\frac{(l_0-l_1)^2}{2}+\vb*{m}^2}\right)^{n_s}\,.
\end{equation}
In contrast to the case before, the $c^\vartheta_{n_0,n_1,n_m,n_s}$ are now not constants anymore but complex-valued scale-invariant functions of the variables $l_0,l_1$ and $m$, that is
\begin{equation}
c^\vartheta_{n_0,n_1,n_m,n_s}(\lambda l_0, \lambda l_1, \lambda m) = c^\vartheta_{n_0,n_1,n_m,n_s}(l_0,l_1, m) ,\qquad \forall\,\lambda\in\mathbb{C}^*\,.
\end{equation}
This behavior can already be anticipated at the level of the Hessian matrices associated to space-like edges in Eqs.~\eqref{eq:diag Hess} and~\eqref{eq:off-diag sl Hess}. While at the identity solution $\vartheta=1$, the non-identity solution introduces to the Hessian involved functions of the spins, given by $\vartheta\neq 1$ and the reflected edge vectors. As the $\vartheta$ are defined as the exponential of dihedral angles, they are found to be scale-invariant. Similarly, the normalized edge vectors are scale-invariant. Note that for both sets of solutions, the $\vartheta$-contributions arise from the Gaussian constraint introduced in Eq.~\eqref{braketsl} for space-like pairings. 

Both Hessian determinants satisfy the correct scaling behavior $\det H(\lambda l_0,\lambda l_1,\lambda m)= \lambda^{15}\det H(l_0,l_1,m)$ which is consistent with the stationary phase approximation conducted in Sec.~\ref{sec:The semi-classical limit of the vertex}.

Deriving a measure from the asymptotics of spin-foam models is a strategy that has been used already extensively in the context of symmetry reduced spin-foam models~\cite{Bahr:2015gxa,Bahr:2016hwc,Bahr:2017klw,Steinhaus:2018aav,Bahr:2018gwf,Allen:2022unb,Bahr:2017bn,Jercher:2023rno}. Therein, the measure plays an important role for the behavior of the coarse-graining flow and the spectral dimension. Although asymptotics provide a reasonable motivation for these measure factors, different choices are in principle conceivable. For instance in effective spin-foams $\upmu = 1$ is a common choice~\cite{Asante:2021zzh,Asante:2020qpa,Asante:2020iwm}. Another approach, followed in~\cite{Dittrich:2023rcr,Asante:2021phx}, is to derive an effective spin-foam measure from continuum quantum cosmology~\cite{Feldbrugge:2017kzv}; we will discuss this explicitly in Sec.~\ref{sec:comparison to ESF}. Discretization independence has moreover been used as a guiding principle for the choice of measure, see~\cite{Dittrich:2011vz,Borissova:2024pfq,Borissova:2024txs,Dittrich:2014rha}. 

\subsection{Minimally coupled massive scalar field}\label{sec:Minimally coupled massive scalar field}

Coupling matter to the cosmological system serves two purposes: rendering the dynamics non-trivial in the absence of spatial curvature and a cosmological constant, and introducing degrees of freedom that serve as a relational clock. The manner in which we do so is discussed in the following.

In a strict sense, matter should be included by coupling classical $BF$ theory, already fully describing gravity in (2+1) dimensions, to the relevant matter fields, subsequently proceeding with spin-foam quantization in order to arrive at a spin-foam model of both gravitational and matter degrees of freedom. Such an approach is however difficult to implement
(chiefly, because the $B$-field integration cannot be straightforwardly performed), having only been attained, to our knowledge, in the context of 3-dimensional Riemannian gravity coupled to fermions \cite{Fairbairn:2006dn} or Yang-Mills \cite{Speziale:2007mt}. Alternative approaches include justifying reasonable Ansätze for the partition function of the coupled system \cite{Mikovic:2001xi, Bianchi:2010bn, Han:2011as}, inserting matter as topological defects~\cite{Freidel:2005bb,Freidel:2005me,Livine:2024iyk}, modifying the spin-foam amplitudes appropriately by analogy with LQG coupled to scalar fields \cite{Kisielowski:2018oiv}, or by conceiving of each spin-foam history as providing a geometry where the discrete matter action is defined \cite{Oriti:2002bn, Mikovic:2002uq, Ali:2022vhn, Jercher:2023rno}. 
In this work we follow the latter perspective, in that we \textit{ad-hoc} add a factor of $\e^{iS_{\mathrm{m}}}$ to the amplitude $\hat{A}_3$ in Eq.~\eqref{eq:a3}, with $S_{\mathrm{m}}$ the discretized matter action on the background determined by the geometry of the 3-frusta. While such an approach is evidently limited\footnote{It ignores for instance possible modifications of the amplitude measure, perhaps to be expected in a more fundamental coupling \cite{Fairbairn:2006dn, Speziale:2007mt}. Furthermore, additional ambiguities are introduced via different possible discretizations of the matter action and different couplings to gravity.}, the hope is that it is sufficient to describe gravity-matter interactions in an effective manner \cite{Ali:2022vhn}.

The particular matter content we consider is a spatially homogeneous minimally coupled massive scalar field discretized on vertices of $\mathcal{X}$, $\phi(t)\rightarrow\phi_n$, where $n$ labels the spatial slices. Here, ``minimal coupling'' refers to the fact that no additional coupling with the curvature tensor is being considered such as a term $R\phi$ with $R$ the Ricci scalar. We consider a massive scalar field since 1) it is more general than the massless case and 2) the mass acts as a regulator for the partition function as demonstrated in Secs.~\ref{sec:freezing oscillations} and~\ref{sec:mass term regularization}.

The scalar field action discretized on a single 3-frustum is given by
\begin{equation}\label{eq:S_phi}
S_\phi = w_0(l_0,l_1,\vb*{m}_0)(\phi_0-\phi_1)^2-M_0(l_0,l_1,\vb*{m}_0)(\phi_0^2+\phi_1^2)\,,
\end{equation}
with 
\begin{equation}\label{eq:w and M}
w_n(l_n,l_{n+1},m_n) = \frac{(l_n+l_{n+1})^2}{8\sqrt{\frac{(l_n-l_{n+1})^2}{2}+\vb*{m}_n^2}}\,,\qquad M_n(l_n,l_{n+1},m_n) = \frac{\mu^2}{4}V(l_n,l_{n+1},\vb*{m}_n)\,.
\end{equation}
Here, $\mu$ is the scalar field mass (not to be confused with the measure factors $\upmu_{\mathds{1},\vartheta}$) and $V(l_0,l_1,\vb*{m}_0)$ is the 3-volume depending on the causal character of the strut, given in Eq.~\eqref{eq:3-volume}. Together with the Regge action, the continuum Friedmann and scalar field equations are obtained in the classical continuum limit~\cite{Jercher:2023csk}. 

For a discrete spacetime with one space-like bulk slice and two boundary slices, the discrete scalar field equations are solved by
\begin{equation}\label{eq:phi sol}
\phi_1 = \frac{\phi_0 w_0+\phi_2w_1}{w_0+w_1-M_0-M_1}\,,
\end{equation}
which can be iterated for arbitrarily many slices. 

The combined bare vertex amplitude for the coupled system of geometry and matter for a single 3-frustum is defined from the matter-less amplitude as 
\begin{equation}
A_v^{\mathrm{asy}}(l_0,l_1,m) \; \mapsto \; A^\phi_v(l_0,l_1,m;\phi_0,\phi_1) = \left(\upmu_{\mathds{1}}\e^{-i(\mathfrak{Re}\{S_{\mathrm{R}}\}+S_\phi)}+\Theta\,\upmu_{\vartheta}\e^{i(\mathfrak{Re}\{S_{\mathrm{R}}\}+S_\phi)}\right)\,,
\end{equation}
with the measure factors $\upmu_{\mathds{1},\vartheta}$, $\Theta$ and the Regge action as in Eq.~\eqref{eq:vertex semi-classics 2}. This allows us to define the final amplitude of our effective model: complementing $\psi' \mapsto \psi ''$ with additional scalar field data, a last manipulation results in the vertex amplitude 
\begin{equation}\label{eq:final amplitude}
\hat{A}_3(\psi',\mathcal{X}_{\mathcal{V}}) \; \mapsto \; \hat{A}(\psi'', \mathcal{X}_{\mathcal{V}})  = \prod_{{\cube} \in \mathcal{X}_{\mathcal{V}}} A^\phi_{v}(\psi''|_{\cube}) \prod_{\mathrm{bulk }\diagup \in \mathcal{X}_{\mathcal{V}}} A_f(\psi''|_\diagup) \,,
\end{equation}
with the face amplitudes defined in Eq.~\eqref{eq:face amplitudes}. As we will demonstrate in Secs.~\ref{sec:Numerical evaluation: Strut in the bulk} and~\ref{sec:1slice}, the scalar field does indeed render the dynamics of the effective model non-trivial in that its solutions are non-stationary and thus curved. Therefore, the first purpose of the matter coupling is met.

As mentioned previously, the second reason for coupling the scalar field is to obtain a relational clock: in the absence of a \textit{fixed} background manifold structure, evolution in time and space needs to be understood in a relational sense~\cite{Rovelli:1990ph,Hoehn:2019fsy,Rovelli:2001bz,Dittrich:2005kc,Goeller:2022rsx,Giesel:2012tj}. A particularly simple matter reference frame that can be straightforwardly coupled is the massless scalar field. In spatially flat continuum cosmology, such a field is strictly monotonic in unphysical background time and thus ideal to deparametrize the cosmological system. Examples of quantum cosmology approaches where this strategy is routinely followed are loop quantum cosmology~\cite{Bojowald:2008wn,Banerjee:2012fn,Ashtekar:2021kfp}, Wheeler-de Witt cosmology~\cite{Kiefer2004} and GFT condensate cosmology~\cite{Oriti:2016qtz}.

In the present setting we explicitly consider the scalar field to be massive, while allowing its mass to be arbitrarily small for the reasons detailed in Sec.~\ref{sec:freezing oscillations}. Note that in the continuum picture a finite mass $\mu\neq 0$ spoils the global monotonicity property, such that there is no global inversion of scalar field values in time - such an inversion is only possible locally. Recent works in the context of Hamiltonian quantum mechanics and quantum cosmology~\cite{Bojowald:2021uqo,Martinez:2023fsd} show however that a massive scalar field still defines a relational clock globally, so long as the clock values are supplemented with a cycle count - much in the same way time is read from a watch.\footnote{Global monotonic time therefore consists of a pair $(\phi,c)$, with $\phi$ the scalar field value and $c$ a cycle integer. Cycles are maximal intervals of monotonicity of $\phi$, forming an ordered set $C = \{c_i\,\vert\, c_i < c_j\text{ for } i<j\}$. The causal ordering of events associated to different cycles is given by the ordering of $C$, irrespective of the scalar field value. Within the same cycle, an event $E_1$ with clock reading $(\phi_1,c_i)$ and $i$ odd (even) is said to happen after (before) event $E_2$ with clock reading $(\phi_2,c_i)$ if $\phi_1 > \phi_2$.} Although it will not be necessary for our purposes to account for multiple cycles, we expect our follow-up investigations to require a careful analysis of this notion of clock time. 

\subsection{Effective cosmological partition function}\label{sec:Effective cosmological partition function}

In this section we set up an effective cosmological partition function for a spatially flat discrete cosmological spacetime minimally coupled to a massive scalar field from the effective amplitude defined in the previous sections. The form of the partition function and the numerical costs for evaluating it depend on the number of space-like slices of the lattice $\mathcal{X}_\mathcal{V}$. For $\mathcal{N} = \mathcal{V}+1$ space-like slices, there are $\mathcal{N}-1$ bulk struts and $\mathcal{N}-2$ bulk spatial edges and scalar field values making up in total $3\mathcal{N}-5$ bulk variables. The partition function is then given by
%
\begin{equation}\label{eq:Z tot}
\begin{aligned}
    & Z_{\mathcal{X}_{\mathcal{V}}}(l_0,l_{\mathcal{N}-1},\phi_0,\phi_{\mathcal{N}-1}) =\sumint_{\{l,m,\phi\}} \hat{A}(l_n,l_{n+1},m_n,\phi_n,\phi_{n+1}) \\
    =& \sumint_{\{l,m,\phi\}} \prod_{n=0}^{\mathcal{N}-2}A_v^\phi (l_n,l_{n+1},m_n,\phi_n,\phi_{n+1}) \prod_{n=0}^{\mathcal{N}-1} A_f(m_n)\prod_{n=1}^{\mathcal{N}-2}A_f(l_n)\,,
\end{aligned}
\end{equation}
with fixed boundary data $(l_0,l_{\mathcal{N}-1},\phi_0,\phi_{\mathcal{N}-1})$. Notice that the sum/integration over all bulk variables includes in particular a sum over time-like strut lengths in Sector III and integrations over space-like strut lengths in Sectors I and II. The sum in III is unbounded from above while the integrations in I and II are bounded. 

At this point we must remark that although 3-dimensional $BF$ theory is topological (and $Z_{BF}$ is consequently  discretization independent~\cite{Ponzano:1968wi}), the effective partition function we propose generically depends on the lattice $\mathcal{X}_{\mathcal{V}}$. This is first and foremost a result of the simplification steps of Sec.~\ref{sec:Geometrical structure of the model}. At the level of quantum Regge calculus, also the measure factors $\upmu_{\mathds{1},\vartheta}$ and the matter coupling render the partition function discretization dependent. In particular, it has been shown in~\cite{Jercher:2023csk} that the presence of a scalar field breaks the discretization independence of solutions to the Regge equations.

The partition function in Eq.~\eqref{eq:Z tot} can be used to define expectation values of functions $\mathcal{O}(\{l,m,\phi\})$ of the bulk geometric and matter variables,
\begin{equation}\label{eq:general exp val}
\langle \mathcal{O}(\{l,m,\phi\})\rangle = \frac{1}{Z}\sumint_{\{l,m,\phi\}} \mathcal{O}(\{l,m,\phi\})\hat{A}(l_n,l_{n+1},m_n,\phi_n,\phi_{n+1})\,.
\end{equation}
Following~\cite{Jercher:2023csk}, for real variables classical solutions and their correct continuum limit are found exclusively in Sector III. In contrast, configurations of Sectors I and II are entirely off-shell and contain an irregular causal structure as discussed in the section hereafter. Restricting the partition function to causally regular configurations therefore amounts to a restriction to Sector III, 
\begin{equation}
    Z^{\textsc{iii}}_{\mathcal{X}_{\mathcal{V}}} = \sumint_{\{l,m,\phi\}}  \hat{A}^{\textsc{iii}}(l_n,l_{n+1},m_n,\phi_n,\phi_{n+1})\,.
\end{equation}
Expectation values are computed accordingly and are denoted as $\langle\,\cdot\,\rangle_{\textsc{iii}}$. In Sec.~\ref{sec:including I and II}, we investigate the influence of Sectors I and II on expectation values of the squared strut length, i.e. we present a comparison of $\langle m^2\rangle$ and $\langle m^2\rangle_{\textsc{iii}}$. 

This concludes the construction and setup of the effective cosmological spin-foam partition function from the 3$d$ Lorentzian coherent spin-foam model. In previous related works, reduced spin-foam models~\cite{Bahr:2015gxa,Bahr:2016hwc,Bahr:2017klw,Steinhaus:2018aav,Bahr:2018gwf,Allen:2022unb,Bahr:2017bn,Jercher:2023rno} and effective spin-foams~\cite{Asante:2020iwm,Asante:2020qpa,Asante:2021zzh,Dittrich:2023rcr,Dittrich:2024awu} have already proven computationally more feasible than full spin-foam models as they avoid the costly calculation of quantum vertex amplitudes. As we are going to demonstrate in Secs.~\ref{sec:Numerical evaluation: Strut in the bulk} and~\ref{sec:1slice}, also the present model allows performing explicit computations of the partition function and expectation values. Accessing new regimes of computability however poses conceptual questions such as the interpretation of the numbers that the partition function puts out, the meaning of generically complex expectation values or whether the set of transition amplitudes for all possible boundary states is sufficient to constitute a quantum and not merely a statistical theory. We comment on these points in the following paragraphs.

In terms of the \textit{general boundary formulation} introduced by Oeckl~\cite{Oeckl:2003vu,Oeckl:2005bv,Oeckl:2006rs,Oeckl:2011qd}, the effective cosmological partition function can be formally understood as an amplitude map from the space of boundary states to the complex numbers for a fixed lattice $\mathcal{X}_{\mathcal{V}}$. Here, the boundary configurations on $\partial\mathcal{X}_{\mathcal{V}}$ are parametrized by the spatial edge length $l_0,l_{\mathcal{N}-1}$ and the scalar field values $\phi_0,\phi_{\mathcal{N}-1}$. The partition function then enters the \textit{transition probability}
\begin{equation}
\mathbb{P}_{\mathcal{X}_{\mathcal{V}}}\left((l_0,\phi_0)\rightarrow (l_{\mathcal{N}-1},\phi_{\mathcal{N}-1})\right) = \frac{\abs{Z_{\mathcal{X}_{\mathcal{V}}}(l_0,\phi_0,l_{\mathcal{N}-1},\phi_{\mathcal{N}-1)}}^2}{\int\dd{l}\dd{\phi}\dd{l'}\dd{\phi'}\abs{Z_{\mathcal{X}_{\mathcal{V}}}(l,\phi,l',\phi')}^2}\,,
\end{equation}
which is normalized by summing/integrating $\abs{Z_{\mathcal{X}_{\mathcal{V}}}}^2$ over all possible boundary data. Of course, the denominator may not be finite, but the ratio of probabilities might still be a meaningful quantity as the infinite normalization factor would drop out. Defined in this way, the transition probability will depend on the lattice $\mathcal{X}_{\mathcal{V}}$ as the partition function does depend on it. Defining probabilities that are independent of $\mathcal{X}_{\mathcal{V}}$ could either be achieved by 1) renormalizing the partition function~\cite{Steinhaus:2020lgb,Asante:2022dnj}, i.e. computing $Z_{\mathcal{X}_{\mathcal{V}}}$ for different $\mathcal{X}_{\mathcal{V}}$ and eventually determining a fixed-point $(\mathcal{X}_{\mathcal{V}})_*$ at which discretization independence is obtained or 2) summing over discretizations $\mathcal{X}_{\mathcal{V}}$ as done for instance in group field theories~\cite{Freidel:2005jy,Oriti:2011jm} and as suggested by the structure of the LQG Hilbert space~\cite{Rovelli:2004wb}.

Following~\cite{Oeckl:2011qd}, also expectation values of observables can be defined within the general boundary formulation. In the present setting, where the boundary consists of two disjoint space-like hypersurfaces, the expectation value defined in~\cite{Oeckl:2011qd} in fact reduces to the definition we have given in Eq.~\eqref{eq:general exp val}. Since complex amplitudes are utilized to define the partition function, also the expectation values of real observables are generically complex. In~\cite{Dittrich:2023rcr}, where a partition function similar to Eq.~\eqref{eq:Z tot} was used, the expectation value of the strut length has a non-vanishing imaginary part. The authors consider this as representing the quantum nature of the model, in accordance with~\cite{Oeckl:2011qd}. As the numerical evaluations of Secs.~\ref{sec:Numerical evaluation: Strut in the bulk} and~\ref{sec:1slice} will show, also our model generically yields imaginary parts of expectation values (albeit tending to a constant half imaginary unit, the physical interpretation of which is left to future research).

Lastly, we consider it as an open conceptual question whether the set of spin-foam amplitudes for all boundary data is sufficient to provide a proper quantum theory. That is because in the language of quantum field theory, the partition function serves to compute the expectation value of only the time-ordered product of operators. It is argued in~\cite{Oeckl:2011qd} that the non-commutative operator product can be retrieved from a time-ordered product of operators. Also in the context of quantum measure theory~\cite{Sorkin:1994dt,Craig:2006ny,Frauca:2016eup,Dowker:2010ng}, it is argued that complex amplitudes, allowing for interference effects, are sufficient to define a quantum theory. To which extent these arguments apply to spin-foams is unclear.

\subsection{On semi-classics and causality violations}\label{sec:causality violations}

The effective cosmological vertex amplitude derived in Sec.~\ref{sec:Asymptotic vertex amplitude and measure factors} shows a particularity on which we elaborate in this section: in the semi-classical limit of the spin-foam vertex only the real part of the Lorentzian Regge action is recovered. This has important consequences for semi-classical configurations with an irregular light cone structure, as we detail now.

Following the definition of~\cite{Sorkin:2019llw}, the Lorentzian dihedral angle of a $d$-cell $\sigma$ located at a $(d-2)$-dimensional space-like hinge $h$, denoted here as $\psi_{\sigma,h}^{\mathrm{L},\pm}$, is generically complex valued. More precisely, the normal vectors of two $(d-1)$-cells meeting at the space-like hinge $h$ lie in 2-dimensional Minkowski space $\R^{1,1}$. Then, for every light ray that lies within the convex wedge spanned by the two normal vectors, the angle $\psi_{\sigma,h}^{\mathrm{L},\pm}$ takes up a contribution of $\pm i\frac{\pi}{2}$.\footnote{Notice that the choice of sign \enquote{$\pm$} is a matter of convention. In the context of Lorentzian quantum Regge calculus~\cite{Asante:2021phx} and Lorentzian effective spin-foams~\cite{Asante:2021zzh,Dittrich:2023rcr} the choice of sign bears physical consequences for the exponential suppression or enhancement of Yarmulke and trouser-like singularities.} Lorentzian deficit angles located at space-like hinges are then defined as~\cite{Asante:2021phx,Asante:2021zzh} 
\begin{equation}
\delta_h^{\mathrm{L},\pm} = \mp i 2\pi - \sum_{\sigma\supset h}\psi_{\sigma,h}^{\mathrm{L,\pm}}\,.
\end{equation}
As a consequence, the Lorentzian deficit angle $\delta_h^{\mathrm{L},\pm}$ is real if and only if there are two light cones (or four light rays) located at the hinge $h$. Configurations with less or more than exactly two light cones are considered as causally irregular and therefore referred to as \textit{hinge causality violating}~\cite{Asante:2021phx}. For causality violations defined at lower-dimensional sub-cells we refer to~\cite{Asante:2021phx} for a definition and to~\cite{Jercher:2023csk} for an exemplary explicit investigation. 

The non-vanishing imaginary part of Lorentzian deficit angles for hinge causality violating configurations has important consequences for the partition function of effective spin-foams and Lorentzian quantum Regge calculus. That is because the amplitude $\e^{iS_{\mathrm{R}}}$ yields an exponential suppression or enhancement if $\mathfrak{Im}\{S_R\}\neq 0$. In fact, it is argued in~\cite{Dittrich:2023rcr,Asante:2021zzh} that this behavior provides a physical mechanism for suppressing causality violating configurations instead of ad-hoc excluding such configurations from the partition function.

Let us now investigate hinge causality violations in the context of the present effective cosmological model. As Eqs.~\eqref{eq:S_I}--\eqref{eq:S_III} show, the Lorentzian Regge action attains an imaginary part in Sectors I and II, i.e. when the strut is space-like. Thus, these sectors are considered causality violating which is in analogy to the 4-dimensional case investigated in~\cite{Jercher:2023csk}. However, as stated at the beginning of this section, the bare vertex amplitude in Eq.~\eqref{eq:vertex semi-classics 2} only contains Regge exponentials $\e^{\pm i\mathfrak{Re}\{S_R\}}$. Consequently, the mechanism of exponential suppression or enhancement of causality violating configurations does not figure in the effective cosmological amplitude constructed in Sec.~\ref{sec:Geometrical structure of the model}. 

The mismatch of Lorentzian Regge calculus and the effective model derived from spin-foam asymptotics could have different reasons; we elaborate on a few: 1) This behavior is a particularity of working in three dimensions. Indeed, for the 3$d$ coherent model, a deficit angle with non-vanishing imaginary part would violate the critical point equations derived in Sec.~\ref{sec:The semi-classical limit of the vertex}. As discussed in~\cite{Jercher:2024kig}, this is particular to the 3-dimensional model where coherent states are associated to edge vectors of $3$-cells. In the asymptotic analysis of known $4$d spin-foam models, where coherent states are associated to normal vectors of triangles, a similar phenomenon does not occur~\cite{Simao:2021qno,Kaminski:2017eew,Liu:2018gfc}. 2) The mismatch follows from performing the stationary phase approximation of the spin-foam amplitude at each vertex individually. It is expected that the product of the individual semi-classical vertex amplitudes does not correspond to the semi-classical amplitude of the total complex~\cite{Asante:2022lnp}. Thus, it is conceivable that either the spin-foam asymptotics on extended complexes yields the full Lorentzian Regge action with imaginary parts or that causality violating configurations are not critical points. These points are particularly important since causal regularity is defined for the gluing of multiple building blocks and cannot be inferred from a single building block in the absence of symmetry assumptions. 3) The mismatch reflects an inherent property of Lorentzian spin-foams and is not merely an artifact of the simplifications performed here.  Tentatively, this would imply that for causality violating configurations, full spin-foam amplitudes are not related to the amplitudes of Lorentzian effective spin-foams via an asymptotic limit. Future investigations on the semi-classical limit of Lorentzian spin-foam amplitudes for extended complexes will hopefully give a definite answer to this question, selecting one of the three explanations above or even reveal a different mechanism.  

Without the exponential suppression of causality violations, such configurations can be dealt with by excluding them by hand and thus only considering $Z_{\textsc{iii}}$. Since this restriction can be straightforwardly performed in the current setup, this is the strategy we employ in Sec.~\ref{sec:1slice}. However, beyond the symmetry restricted setting here, an ad-hoc exclusion of causality violations might be computationally unfeasible. This argument has been put forward in~\cite{Asante:2021phx} supporting the importance of including causality violations in the partition function. If one does so, then there are two further 
possible suppression mechanisms besides the one from effective spin-foams. First, there are no classical solutions in Sectors I and II, i.e. the Regge action does not exhibit stationary points. Consequently, the oscillations could lead to destructive interference rendering causality violations negligible. Second, the measure factors $\mu_{\mathds{1},\vartheta}$ could lead to a suppression of such configurations relative to Sector III. We explicitly investigate this question in Sec.~\ref{sec:including I and II}.

\section{Numerical evaluation: strut in the bulk}\label{sec:Numerical evaluation: Strut in the bulk}

In this section, we numerically evaluate the effective spin-foam partition function for a single 3-frustum with one bulk strut. To that end, we first introduce Wynn's algorithm for sequence convergence acceleration~\cite{Wynn1956,Weniger2003}. Thereafter, in Sec.~\ref{sec:freezing oscillations}, we demonstrate issues that arise for vertex amplitudes with freezing oscillations and show in Sec.~\ref{sec:mass term regularization} that a scalar field mass resolves these obstacles. In Sec.~\ref{sec:including I and II} we include Sectors I and II for the computation of the bulk strut and in Sec.~\ref{sec:comparison to ESF} we compare our results to those obtained in effective spin-foams~\cite{Dittrich:2023rcr}.

\subsection{Wynn's algorithm for sequence convergence acceleration}\label{sec:Wynn}

Before we explicitly address the numerical evaluation of the partition function, let us digress briefly to introduce the Shanks transform and Wynn's $\epsilon$-algorithm which have been applied to effective spin-foams\footnote{A closely related series convergence acceleration, known as Aitken's $\Delta^2$-process~\cite{Weniger2003} has been applied to infinite bulk variable summations in~\cite{Dona:2023myv}.}  in~\cite{Dittrich:2023rcr}. In the subsequent sections, we will frequently encounter unbounded sums of the form
\begin{equation}
\mathfrak{S} = \lim\limits_{n\rightarrow\infty}\mathfrak{S}_n = \lim\limits_{n\rightarrow\infty}\sum_{j = 1}^{n}a_j
\end{equation}
for some complex sequence $\{a_j\}$. Assuming the partial sum $\mathfrak{S}_n$ to be known and of the form~\cite{Weniger2003}
\begin{equation}
\mathfrak{S}_n = \mathfrak{S} + \sum_{j=0}^{k-1}c_j\lambda_j^n\,,
\end{equation}
with $c_j$ coefficients and the $1 > \abs{\lambda_0} > \dots > \abs{\lambda_{k-1}}$ referred to as transients, there are $2k+1$ unknowns given by the limiting value $\mathfrak{S}$, the $c_j$ and the $\lambda_j$. In order to solve for $\mathfrak{S}$, one utilizes $2k+1$ consecutive sequence values $\mathfrak{S}_n,\mathfrak{S}_{n+1},\dots,\mathfrak{S}_{n+2k}$ to obtain the $k$-th Shanks transform~\cite{Schmidt1941,Shanks1955} $e_k(\mathfrak{S}_n)$ as a ratio of determinants. For sufficiently large $n$, the Shanks transform $e_k(\mathfrak{S}_n)$ approximates $\mathfrak{S}$ faster than taking the limit $n\rightarrow \infty$ of the partial sums $\mathfrak{S}_n$.

In~\cite{Wynn1956}, Wynn introduced a non-linear recursive relation to efficiently compute the Shanks transform, commonly referred to as $\epsilon$-algorithm. Given the partial sums $\mathfrak{S}_1,\dots,\mathfrak{S}_n$, define~\cite{Weniger2003}
\begin{equation}
\epsilon_{-1}^{(n)} = 0,\qquad \epsilon_0^{(n)} = \mathfrak{S}_n\,.
\end{equation}
Then, $\epsilon$'s of higher $k$ are obtained via the relation
\begin{equation}
\epsilon_{k+1}^{(n)} = \epsilon_{k-1}^{(n+1)} + \frac{1}{\epsilon_k^{(n+1)}-\epsilon_k^{(n)}}\,.
\end{equation}
Wynn has shown~\cite{Wynn1956} that $\epsilon^{(n)}_{2k} = e_k(\mathfrak{S}_n)$ thus providing a fast algorithm for computing the $k$-th Shanks transform of a sequence $\mathfrak{S}_n$.\footnote{A Mathematica algorithm for convergence acceleration via Wynn's method can be found in the repository \href{https://resources.wolframcloud.com/FunctionRepository/resources/SequenceLimit/}{SequenceLimit}. An implementation in \textsc{Julia} can be found in \href{https://github.com/Jercheal/3d-cosmology/blob/master/wynn.jl}{https://github.com/Jercheal/3d-cosmology}.}


Note that Wynn's $\epsilon$-algorithm can be applied to sequences of more than one variable, which we discuss explicitly in Sec.~\ref{sec:Summation of bulk strut lengths}. 

\subsection{Freezing oscillations}\label{sec:freezing oscillations}

\begin{figure}
    \centering
    \includegraphics[width=0.6\linewidth]{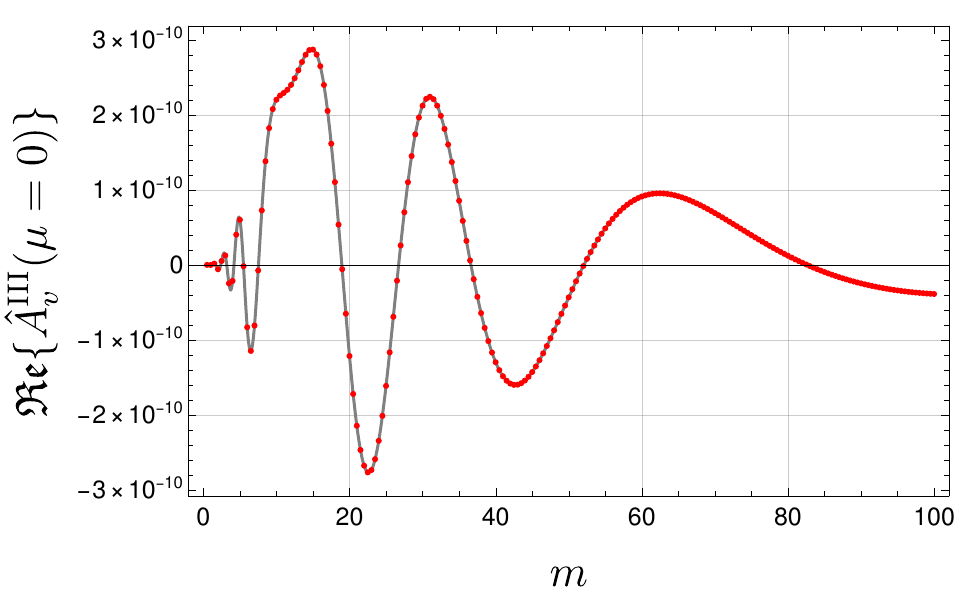}
    \caption{The dressed vertex amplitude $\hat{A}^{\textsc{iii}}$, coupled to a scalar field with vanishing mass, $\mu = 0$, evaluated on discrete values $m\in\mathbb{N}/2$ (red dots) and a continuum of values (gray graph). A saddle point can be spotted at $m\approx 12$ and the oscillations are freezing out asymptotically for large strut length. The boundary data, given by $l_0 = 10$, $l_1 = 30$, $\phi_0 = 2$ and $\phi_1 = 4.1$ has been chosen for demonstrational purposes.}
    \label{fig:freezing}
\end{figure}

The vacuum partition function for a single 3-frustum with boundary data $(l_0,l_1)$ is defined by
\begin{equation}\label{eq:Z single frustum}
Z(l_0,l_1) = \sum_{m\in\mathbb{{N}}/2}\hat{A}_3^{\textsc{iii}}(l_0,l_1,m)+\int\limits_{1/2}^{m_{\mathrm{int}}}\dd{m}\hat{A}_3^{\textsc{ii}}(l_0,l_1,m)+\int\limits_{m_{\mathrm{int}}}^{m_{\mathrm{max}}}\dd{m}\hat{A}_3^{\textsc{i}}(l_0,l_1,m)\,,
\end{equation}
with $\hat{A}_3$ defined in Eq.~\eqref{eq:a3}. The integrations in I and II can in principle be computed straightforwardly with numerical integration methods.\footnote{We remind the reader that the lower bound of the  space-like strut length arises from the length gap of the $\SUO$-Casimir in the continuous series. If $m_{\mathrm{int}} < \frac{1}{2}$ or $m_{\mathrm{max}} < \frac{1}{2}$, then the integration in Sector II, or in Sectors I and II, respectively, are empty and do not contribute to the partition function.} In contrast, the unbounded sum in the third sector poses new challenges because of the particular form of the Regge action for spatially flat cosmology. Considering fixed boundary data $(l_0,l_1)$, the Regge action goes as
\begin{equation}
S_{\mathrm{R}}(l_0,l_1,m)\underset{m\gg 1}{\longrightarrow} \frac{1}{m}\,.
\end{equation}
Consequently, the exponential $\e^{iS_{\mathrm{R}}}$ goes to one for $m\rightarrow \infty$, a depiction of which is given in Fig.~\ref{fig:freezing}. The asymptotic freezing of oscillations obstructs the sum in III to converge. Notice, that this is not an effect of merely considering a single strut length as bulk variable. We have numerically checked that the same issue occurs for a space-like bulk slice. Furthermore, the same behavior is expected in the continuum as we stress below.

A similar behavior is observed if a massless scalar field is coupled to the system, that is, for the partition function
\begin{equation}
Z_{\textsc{iii}}(l_0,l_1,\phi_0,\phi_1) = \sum_{m\in\mathbb{N}/2}\hat{A}^{\textsc{iii}}(\mu = 0)\,.
\end{equation}
A visualization of the freezing oscillations is given in Fig.~\ref{fig:freezing}. This is an immediate consequence of the kinetic term of the scalar field action which, for fixed boundary data $(l_0, l_1, \phi_0,\phi_1)$, goes as
\begin{equation}
S_\phi(\mu = 0) \underset{m\gg 1}{\longrightarrow} \frac{(l_0+l_1)^2}{8m}(\phi_0-\phi_1)^2\,.
\end{equation}
Even if the spin-foam integrand exhibits saddle points, as it is the case for the exemplary plot, expectation values of geometric observables such as the edge length diverge. That is because the action gets effectively stationary in the limit $m\rightarrow \infty$. Notice in particular that while an upper cutoff $m_{\mathrm{max}}$ regularizes $Z$, the result is cutoff dependent and $m_{\max}$ cannot be removed. We have checked numerically that
\begin{equation}
\lim\limits_{m_{\mathrm{max}}\rightarrow \infty}\frac{\sum_{m}m\hat{A}^{\textsc{iii}}(\mu=0)}{\sum_{m'}\hat{A}^{\textsc{iii}}(\mu=0)} = \infty\,.
\end{equation}
It is a minimal consistency check that expectation values of the semi-classical spin-foam partition function relate to the solutions of classical Regge calculus. Thus, we conclude that the amplitudes $\hat{A}^{\textsc{iii}}_3$ and $\hat{A}^{\textsc{iii}}(\mu = 0)$, defined in Eqs.~\eqref{eq:a3} and~\eqref{eq:final amplitude}, respectively, are badly behaved as amplitudes for discrete gravity path integrals.

In the next section, we show that a non-zero mass acts as a regulator of the partition function rendering finite expectation values. Before that, let us comment on the issue of freezing oscillations in preceding works in the context of continuum quantum cosmology~\cite{Feldbrugge:2017kzv} and effective spin-foam cosmology~\cite{Dittrich:2023rcr}. In continuum quantum cosmology for spatially flat spacetimes coupled to a massless scalar field, we expect that a similar issue of freezing oscillations occurs. Following~\cite{Feldbrugge:2017kzv}, the relevant variable being integrated over in the continuum is the lapse function $N(t)$ (the lapse integration does not matter for the argument here). The action of the coupled system is given by
\begin{equation}
\int\dd{t}\frac{a^3}{N}\left[\frac{1}{8\pi G_\mathrm{N}}\left(-3\left(\frac{\dot{a}}{a}\right)^2+N^2\frac{k}{a^2}-N^2\Lambda\right)+\frac{1}{2}\left(\dot{\phi}^2-N^2\mu^2\phi^2\right)\right]\,.
\end{equation}
Clearly, in the case of $\mu = k = \Lambda = 0$, the action goes as $\sim 1/N$, which is in correspondence with the discrete setting where the action goes as $1/m$ with $m$ the strut length. Furthermore, the lapse can be removed completely from the action by the time re-parametrization $\dd{\tau} = N\dd{t}$. Thus, the Lorentzian cosmological path integral,
\begin{equation}
Z = \int\mathcal{D}N\e^{iS}\,,
\end{equation}
yields divergent expectation values. Importantly, the physical system considered in~\cite{Feldbrugge:2017kzv} is given by $\mu = 0$ but $k = 1$ and $\Lambda > 0$. As a result, parts of the action are linear in the lapse function, and we expect this behavior to be crucial for the convergence of $Z$. Similarly, in the context of effective spin-foams~\cite{Dittrich:2023rcr}, the discrete spacetime is considered to be spatially spherical with a cosmological constant $\Lambda > 0$. As a result, the corresponding effective spin-foam partition function is convergent and expectation values of the strut length are finite.

\subsection{Mass term regularization: strut length expectation values in Sector III}\label{sec:mass term regularization}

\begin{figure}
    \centering
    \begin{subfigure}{0.5\textwidth}
    \includegraphics[width=\linewidth]{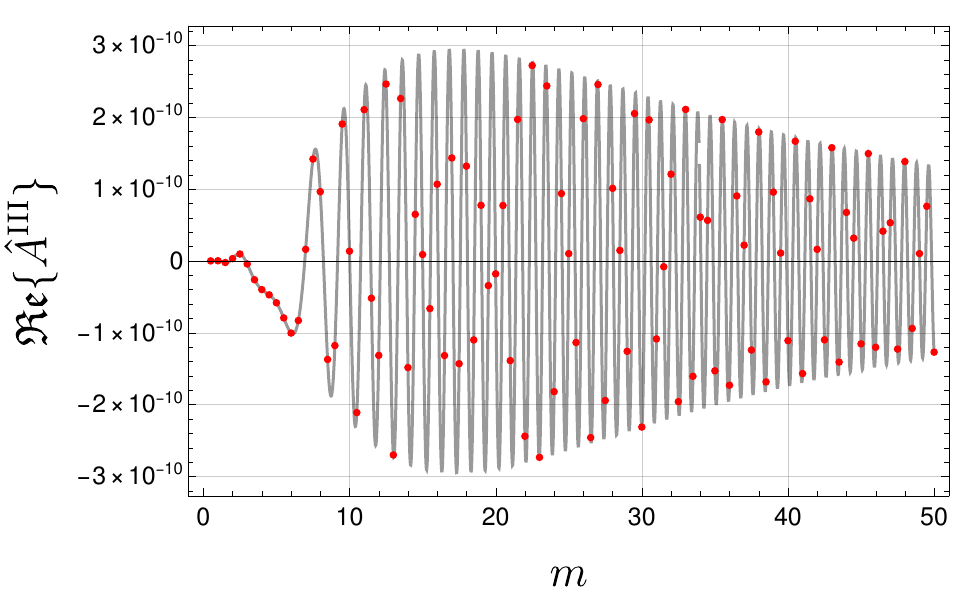}
    \end{subfigure}%
    \begin{subfigure}{0.5\textwidth}
    \includegraphics[width=\linewidth]{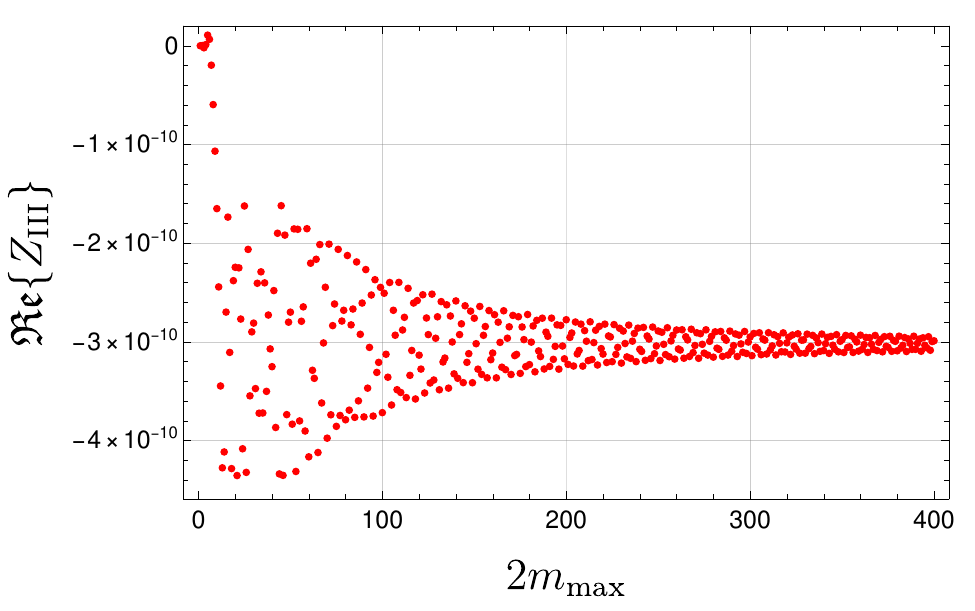}
    \end{subfigure}
    \caption{Left: real part of the amplitude $\hat{A}^{\textsc{iii}}$ as a function of the strut length $m$. A saddle point occurs at $m\approx 5$. Oscillations do not freeze due to the non-vanishing scalar field mass. Right: real part of partial sums of $\hat{A}^{\textsc{iii}}$ with upper summation cut-off $m_{\mathrm{max}}$. The imaginary part of the amplitude and the corresponding partial sums are of a similar form. For visualization, boundary values $l_0 = 10$, $l_1 = 30$, $\phi_0 = 2$, $\phi_1 = 4$ have been chosen as well as a mass $\mu = 0.06$.}
   \label{fig:ZIII}
\end{figure}

The freezing oscillations of the effective spin-foam amplitude necessitate to consider either spatial curvature, a non-vanishing cosmological constant or a different matter content. For the remainder, we consider the latter option and show that a non-vanishing scalar field mass yields indeed a convergent partition function. A depiction of the amplitude and the partition function for non-zero mass is given in Fig.~\ref{fig:ZIII}. 

\begin{figure}
    \centering
    \includegraphics[width=0.6\linewidth]{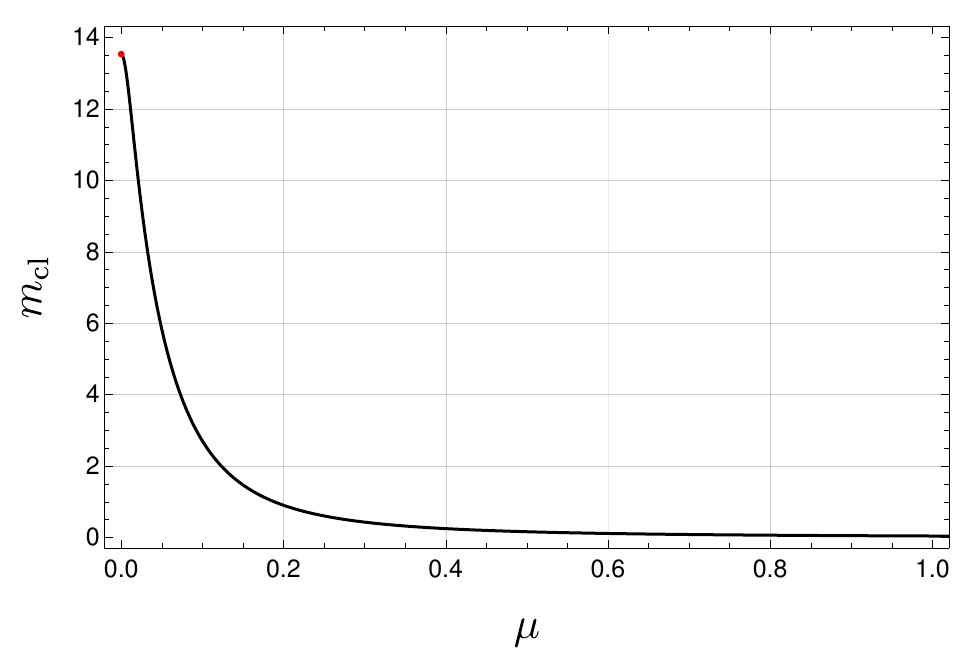}
    \caption{Classical solutions of the Regge equation for the strut length, $m_{\mathrm{cl}}$, as a function of the scalar field mass $\mu$. The boundary for this graph is given by $l_0 = 10$, $l_1 = 30$, $\phi_0 = 2$ and $\phi_1 = 4$ which have been chosen for demonstrational purposes.}
    \label{fig:strutmass}
\end{figure}

In the following, we compute the expectation value of the strut length $m$ in Sector III as a function of varying boundary data as well as for varying mass. To compare these expectation values with the classical solutions, we study the equation of motion for a single 3-frustum. Given boundary data $(l_0,l_1,\phi_0,\phi_1)$, classical strut length correspond to solutions of the Regge equation\footnote{Notice that in Sectors I and II, no classical solutions exist. See~\cite{Jercher:2023csk} for a detailed discussion.},
\begin{equation}\label{eq:Regge equation strut}
\begin{aligned}
 0 =&\, \pdv{(S_{\textsc{iii}}+S_\phi)}{m} \\[7pt]
=& \,4\left[\frac{\pi}{2}-\cos^{-1}\left(\frac{(l_0-l_1)^2}{4m^2+(l_0-l_1)^2}\right)\right] + {\pdv{w_0}{m}} (\phi_0-\phi_1)^2-{\pdv{M_0}{m}} (\phi_0^2+\phi_1^2)\,,
\end{aligned}
\end{equation}
where $w_0,M_0$ are defined in Eq.~\eqref{eq:w and M}. Using \textsc{FindRoot} of \textsc{Mathematica}, this transcendental equation can be solved for $m$, constituting the classical solution $m_{\mathrm{cl}}$. For fixed boundary data $(l_0,\phi_0,l_1,\phi_1)$, $m_{\mathrm{cl}}$ is depicted in Fig.~\ref{fig:strutmass} as a function of $\mu$.

Besides the real and imaginary parts of $\langle m\rangle_{\textsc{iii}}$, we compute the relative deviation $\delta$ between $\mathfrak{Re}\{\langle m\rangle_{\textsc{iii}}\}$ and $m_{\mathrm{cl}}$, defined as
\begin{equation}\label{eq:relerr}
\delta = \frac{\abs{\mathfrak{Re}\{\langle m\rangle_{\textsc{iii}}\}-m_{\mathrm{cl}}}}{m_{\mathrm{cl}}}\,.
\end{equation}
Also, we compute the relative variance $\mathrm{rVar}$, defined as
\begin{equation}\label{eq:rVar}
\mathrm{rVar}(m)_{\textsc{iii}} = \frac{\sqrt{\abs{\mathfrak{Re}\{\langle m^2\rangle_{\textsc{iii}}\}-\mathfrak{Re}\{\langle m\rangle_{\textsc{iii}}\}^2}}}{\mathfrak{Re}\{\langle m\rangle_{\textsc{iii}}\}}\,,
\end{equation}
providing a measure of the standard deviation relative to the expectation value.

\paragraph{Varying edge length $\vb*{l}\mathbf{_1}$.} 

\begin{figure}
    \centering
    \begin{subfigure}{0.45\textwidth}
    \includegraphics[width=\linewidth]{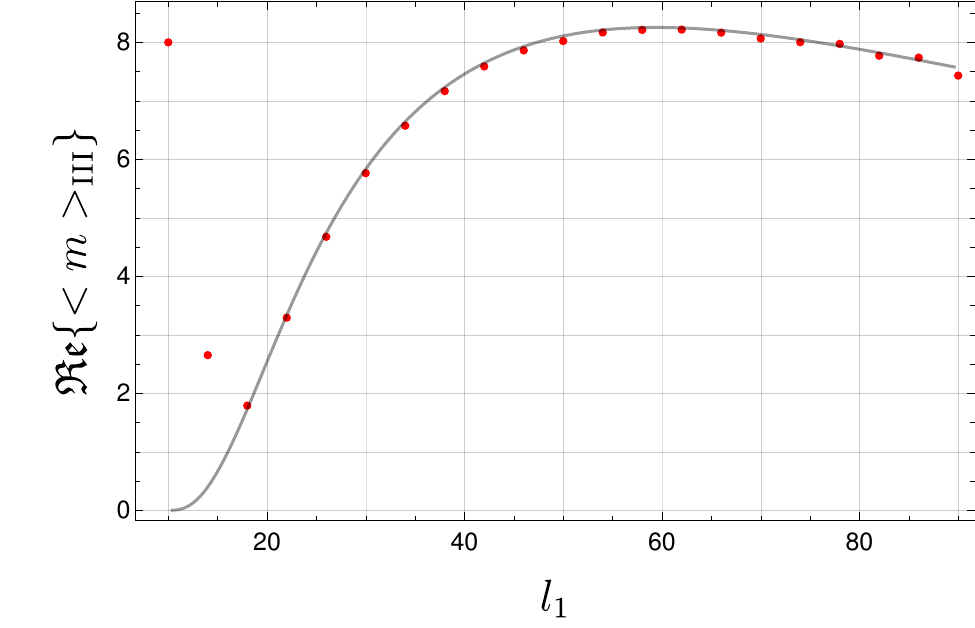}
    \vspace{-7mm}
    \caption{Real part of $\langle m\rangle_{\textsc{iii}}$ and classical solutions $m_{\mathrm{cl}}$.}
    \label{fig:b expval a Re}
    \end{subfigure}\hspace{0.05\textwidth}
    \begin{subfigure}{0.45\textwidth}
    \includegraphics[width=\linewidth]{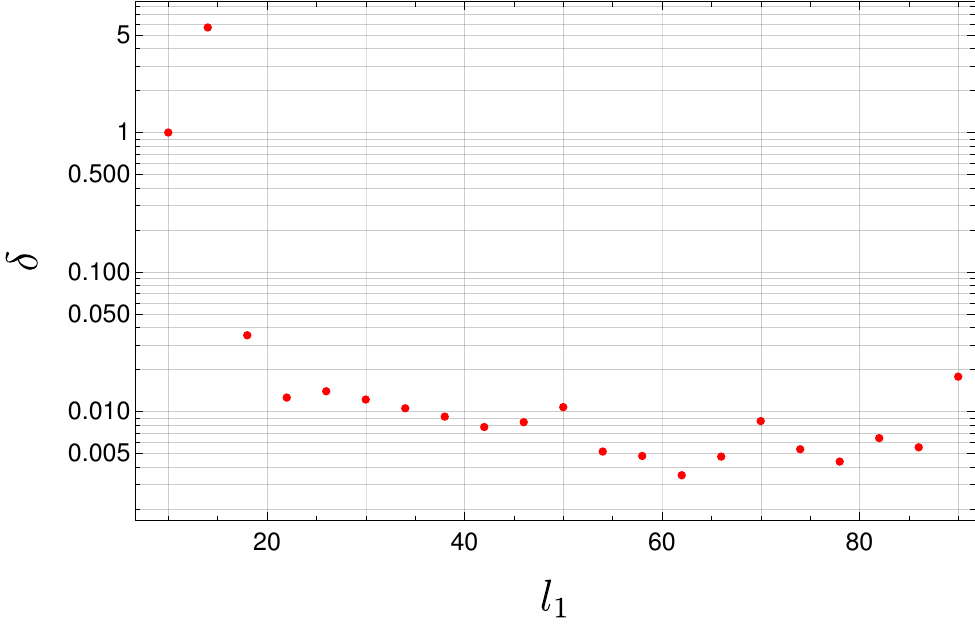}
    \vspace{-7mm}
    \caption{Relative deviation between expectation value and classical solution.}
    \label{fig:b expval a relerr}
    \end{subfigure}\\[5mm]
    \begin{subfigure}{0.45\textwidth}
    \includegraphics[width=\linewidth]{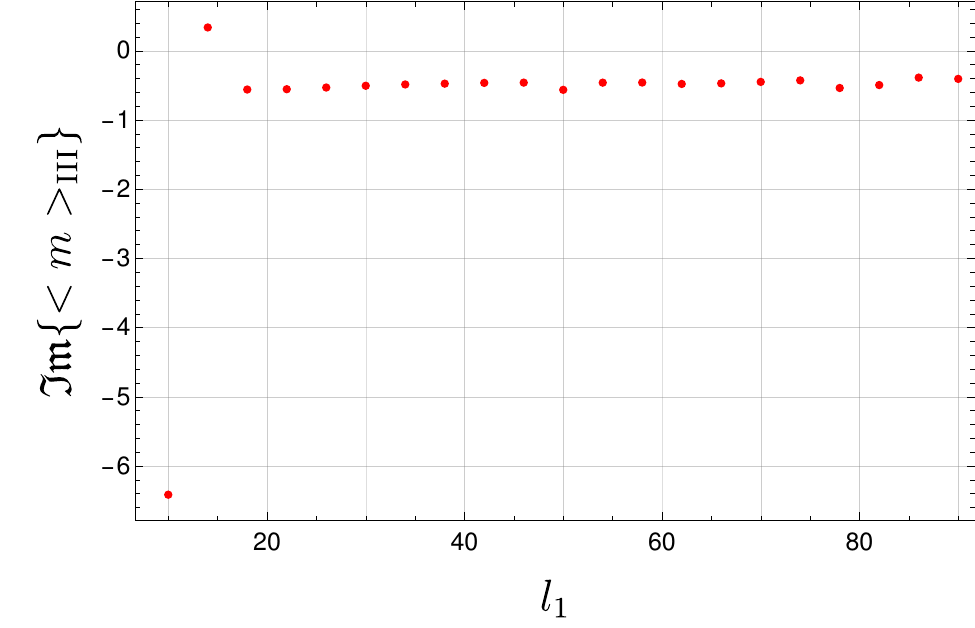}
    \vspace{-7mm}
    \caption{Imaginary part of $\langle m\rangle_{\textsc{iii}}$.}
    \label{fig:b expval a Im}
    \end{subfigure}\hspace{0.05\textwidth}
    \begin{subfigure}{0.45\textwidth}
    \includegraphics[width=\linewidth]{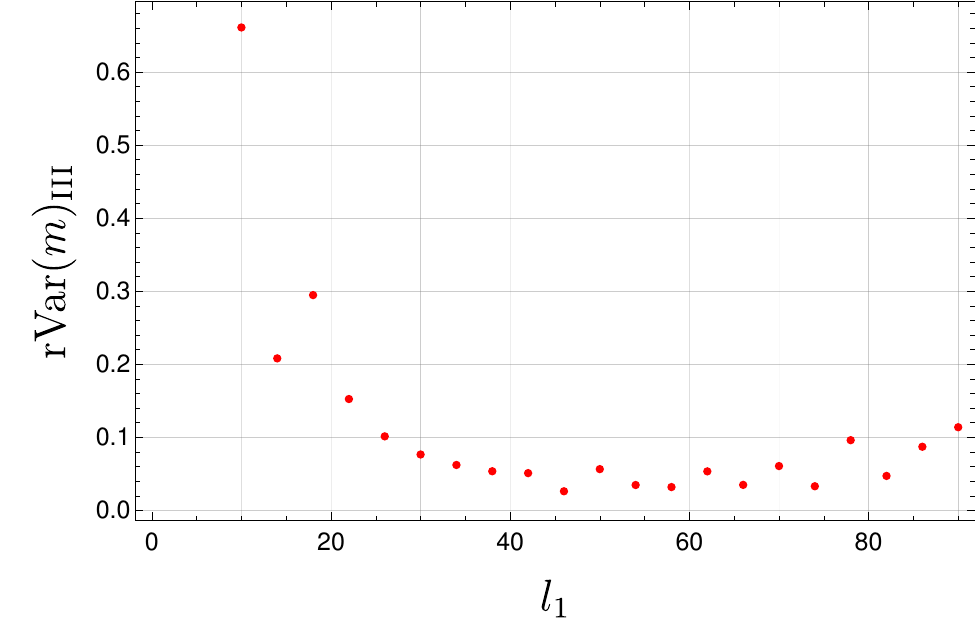}
    \vspace{-7mm}
    \caption{Relative variance of $m$ in Sector III.}
    \label{fig:b expval a rVar}
    \end{subfigure}
    \caption{(a): real part of the strut length expectation value computed in Sector III (red dots) and the classical solutions $m_{\mathrm{cl}}$ (gray graph). (b): the corresponding relative deviation between $\langle m\rangle_{\textsc{iii}}$ and the classical solution as defined in Eq.~\eqref{eq:relerr}. (c): the imaginary part of the strut length expectation value. (d): relative variance of the strut length, defined as in Eq.~\eqref{eq:rVar}. All the quantities above are depicted as a function of the edge length $l_1 = 10+4n$ with $n\in\{0,\dots,20\}$. We have fixed the initial square length to $l_0= 10$ as well as $\phi_0=2$, $\phi_1 =4$ and $\mu = 0.05$.}
    \label{fig:b expval a}
\end{figure}

For fixed edge length $l_0$ as well as fixed scalar field values $\phi_0,\phi_1$ on the boundary, the expectation value of the strut length in Sector III can be computed as a function of the spatial edge length $l_1$. An exemplary plot of the functional dependence of the real part of $\langle m\rangle_{\textsc{iii}}$ on $l_1$ is given in Fig.~\ref{fig:b expval a Re}. As for the classical solutions, the expectation value of $m$ is not monotonic in $l_1$ but reaches a maximum and then slowly decreases.

The relative deviation $\delta$ of (the real part of) the expectation value and the classical solution decreases for larger $l_1$ as shown in Fig.~\ref{fig:b expval a relerr}. For small $l_1$ the expectation value of $m$ is small and the saddle point of the amplitude cannot be resolved with the spectrum $m\in\mathbb{N}/2$. In particular, the classical solutions $m_{\mathrm{cl}}$ for $l_1 = 10$ and $l_1 = 14$ lie below the minimum value of $m = 0.5$, i.e. the gap of the spectrum and can therefore not be probed. As discussed in~\cite{Dittrich:2023rcr}, a refinement of the spectrum is necessary to resolve these deviations, which holds true also for the system considered here. For values of $l_1$ that are much larger than the ones depicted in Fig.~\ref{fig:b expval a relerr} numerical errors are expected to increase. That is because the frequency of oscillations of $A_{\textsc{iii}}$ increases while the amplitude of these oscillations decreases. Extending Wynn's epsilon algorithm to this regime therefore requires arbitrary precision arithmetic.

The imaginary part of the expectation value is approximately given by $\mathfrak{Im}\{\langle m\rangle_{\textsc{iii}}\} = -\frac{1}{2}$ for those configurations that can be sufficiently captured by the spectrum $m\in\mathbb{N}/2$. A plot thereof is provided in Fig.~\ref{fig:b expval a Im}. 

The relative variance rVar, defined in Eq.~\eqref{eq:rVar} and plotted in Fig.~\ref{fig:b expval a rVar}, attains large values for small $l_1$. These deviations decrease for larger values of $l_1$ meaning that the expectation value is more sharply peaked. From the set of data considered here, the relative variance does not go to zero  asymptotically but rather to a constant value for further increasing values of $l_1$.  

\paragraph{Varying scalar field $\vb*{\phi}\mathbf{_1}$.}

\begin{figure}
    \centering
    \begin{subfigure}{0.45\textwidth}
    \includegraphics[width=\linewidth]{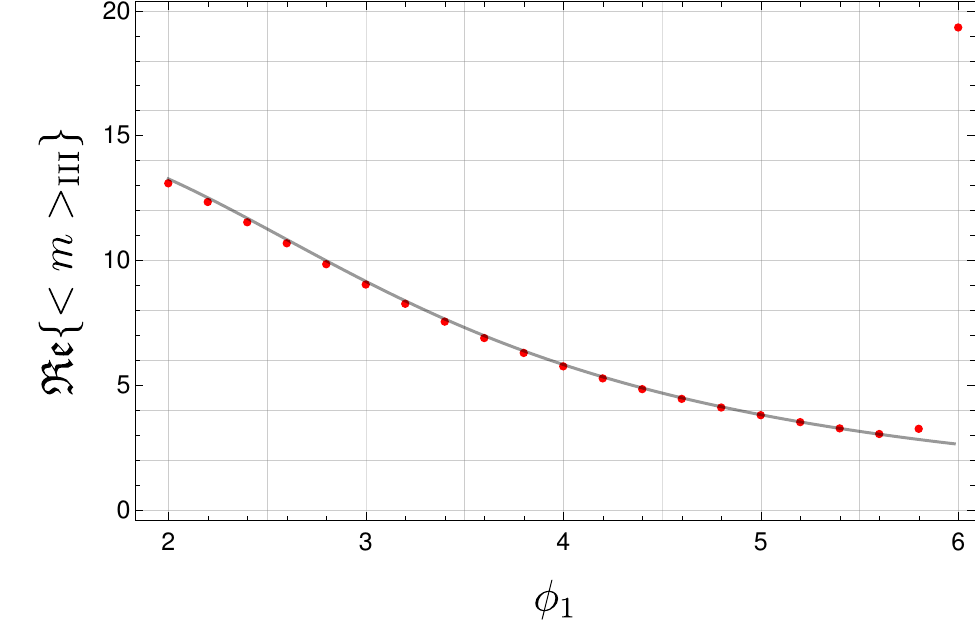}
    \vspace{-7mm}
    \caption{Real part of $\langle m\rangle_{\textsc{iii}}$ and classical solutions $m_{\mathrm{cl}}$.}
    \label{fig:b expval phi Re}
    \end{subfigure}\hspace{0.05\textwidth}
    \begin{subfigure}{0.45\textwidth}
    \includegraphics[width=\linewidth]{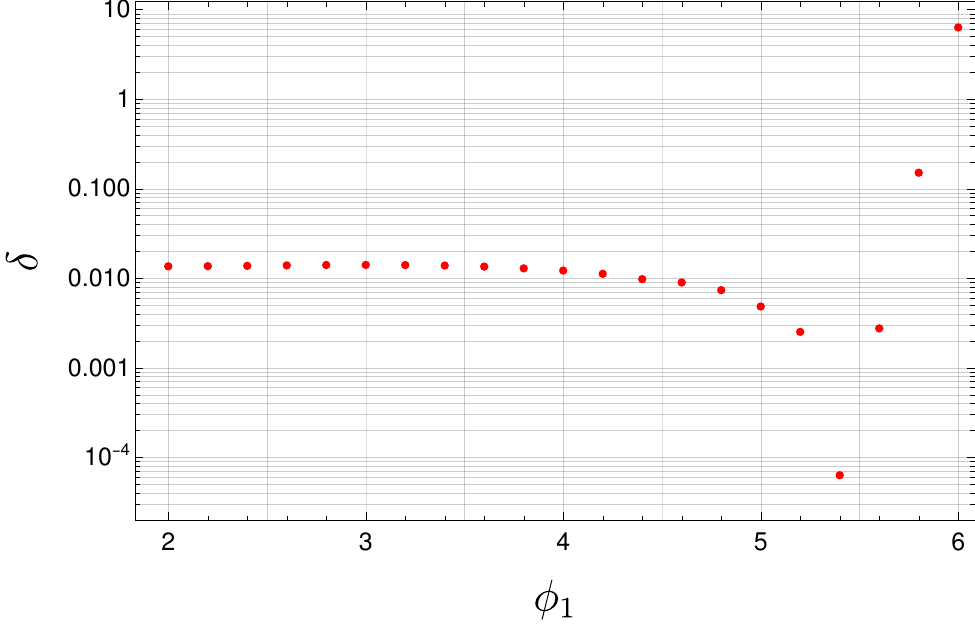}
    \vspace{-7mm}
    \caption{Relative deviation between expectation value and classical solution.}
    \label{fig:b expval phi relerr}
    \end{subfigure}\\[5mm]
    \begin{subfigure}{0.45\textwidth}
    \includegraphics[width=\linewidth]{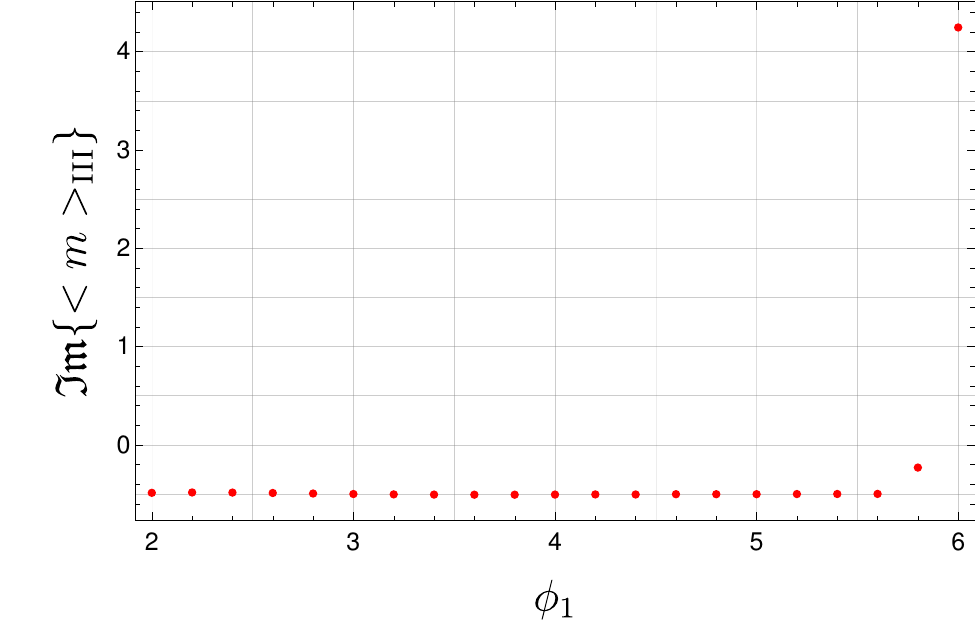}
    \vspace{-7mm}
    \caption{Imaginary part of $\langle m\rangle_{\textsc{iii}}$.}
    \label{fig:b expval phi Im}
    \end{subfigure}\hspace{0.05\textwidth}
    \begin{subfigure}{0.45\textwidth}
    \includegraphics[width=\linewidth]{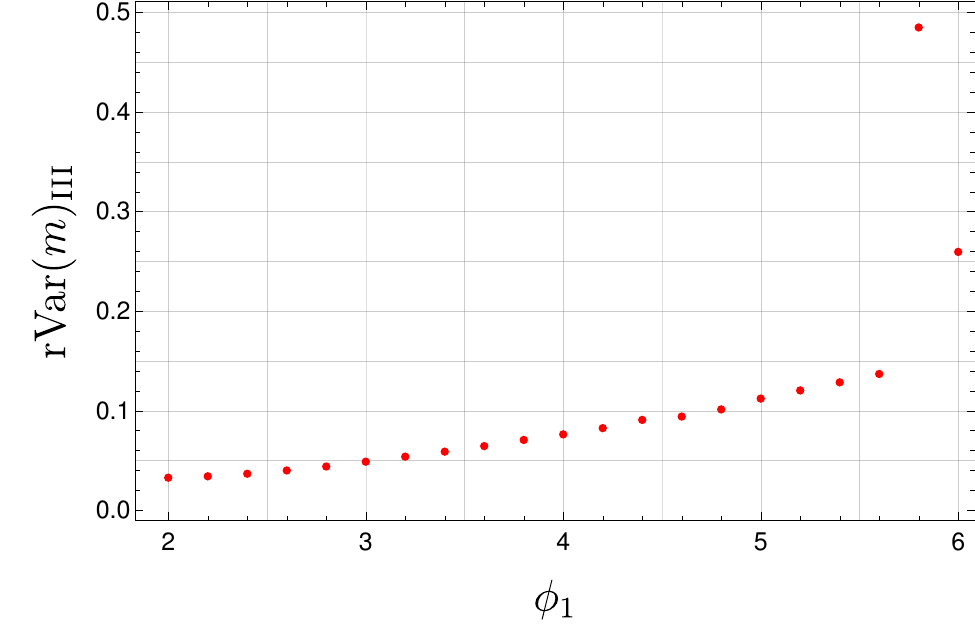}
    \vspace{-7mm}
    \caption{Relative variance of $m$ in Sector III.}
    \label{fig:b expval phi rVar}
    \end{subfigure}
    \caption{(a): real part of the strut length expectation value computed in Sector III (red dots) and the classical solutions $m_{\mathrm{cl}}$ (gray graph). (b): the corresponding relative deviation $\delta$ between $\langle m\rangle_{\textsc{iii}}$ and the classical solution, as defined in Eq.~\eqref{eq:relerr}. (c): imaginary part of the strut length expectation value. (d): relative variance of the strut length according to Eq.~\eqref{eq:rVar}. The quantities above are computed as a a function of varying scalar field value $\phi_1 = 2+0.02n$ with $n\in\{0,\dots,20\}$. We have fixed the geometric data to $l_0= 10$ and $l_1 = 30$ as well as an initial scalar field value $\phi_0=2$. Here, the mass is given by $\mu = 0.05$.}
    \label{fig:b expval phi}
\end{figure}

For fixed geometric data $l_0,l_1$ as well as a fixed initial scalar field value $\phi_0$, one can compute $\langle m\rangle_{\textsc{iii}}$ as a function of the scalar field value $\phi_1$. A plot of the real part of this expectation value is provided in Fig.~\ref{fig:b expval phi Re}. We find a monotonically decreasing classical solutions $m_{\mathrm{cl}}$ and expectation values of $m$ for increasing scalar field value and thus also for an increasing difference in scalar field values $\phi_0-\phi_1$. 

The relative deviation between $\mathfrak{Re}\{\langle m\rangle_{\textsc{iii}}\}$ and $m_{\mathrm{cl}}$, depicted in Fig.~\ref{fig:b expval phi relerr}, remains constant and small for small values of $\phi_1$ and slightly decreases for larger $\phi_1$. For further growing scalar field values, we observe that $\delta$ grows large. That is because of the rapid oscillations at increasing $\phi_1$ as well as the discrete spectrum, $m\in\mathbb{N}/2$,  which is not sufficient to resolve the saddle point . The situation is thus similar to the $l_1$-dependence discussed above, requiring a refinement of the spectrum for expectation values close to classicality.

The imaginary part of the expectation value of $m$, given in Fig.~\ref{fig:b expval phi Im} is approximately given by $\mathfrak{Im}\{\langle m\rangle_{\textsc{iii}}\} = -\frac{1}{2}$ and appears to be independent of the scalar field values $\phi_1$ as long as the discrete spectrum is sufficient to resolve the saddle point. 

A plot of the relative variance is given in Fig.~\ref{fig:b expval phi rVar}. Interestingly, it increases for larger values of $\phi_1$ while the relative deviation $\delta$ slightly decreases. That is, the expectation value tends to be closer to the corresponding classical value with increasing fluctuations as measured by rVar. 

\paragraph{Mass-dependence.} 

\begin{figure}
    \centering
    \begin{subfigure}{0.45\textwidth}
    \includegraphics[width=\linewidth]{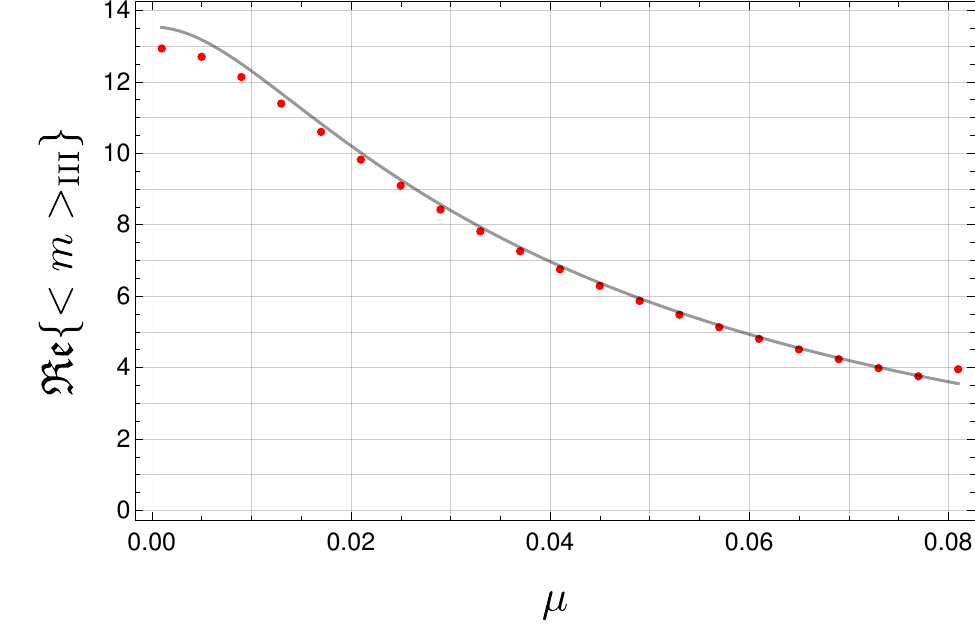}
    \vspace{-7mm}
    \caption{Real part of $\langle m\rangle_{\textsc{iii}}$ and classical solutions $m_{\mathrm{cl}}$.}
    \label{fig:b expval m Re}
    \end{subfigure}\hspace{0.05\textwidth}
    \begin{subfigure}{0.45\textwidth}
    \includegraphics[width=\linewidth]{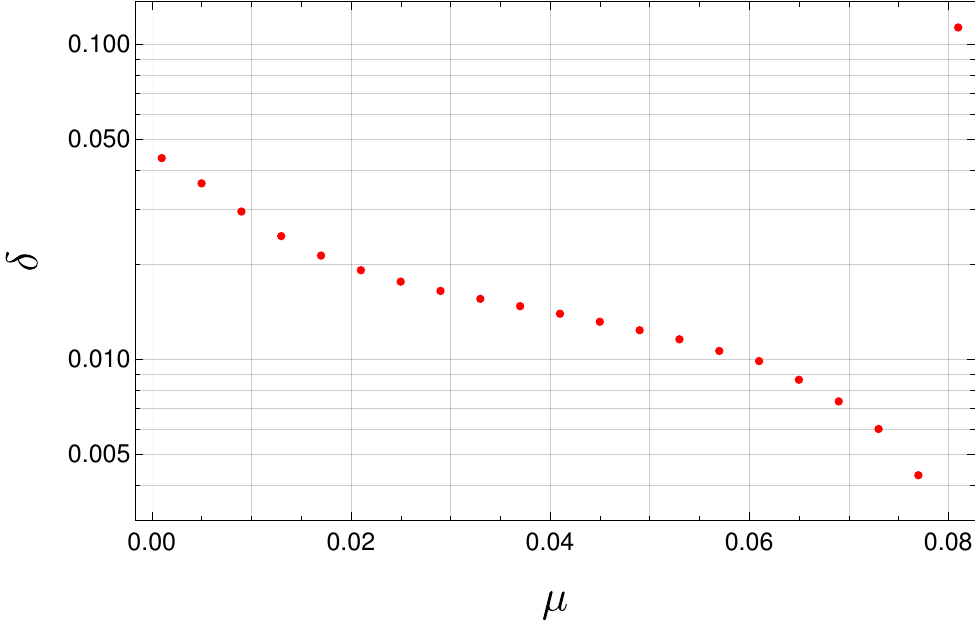}
    \vspace{-7mm}
    \caption{Relative deviation between expectation value and classical solution.}
    \label{fig:b expval m relerr}
    \end{subfigure}\\[5mm]
    \begin{subfigure}{0.45\textwidth}
    \includegraphics[width=\linewidth]{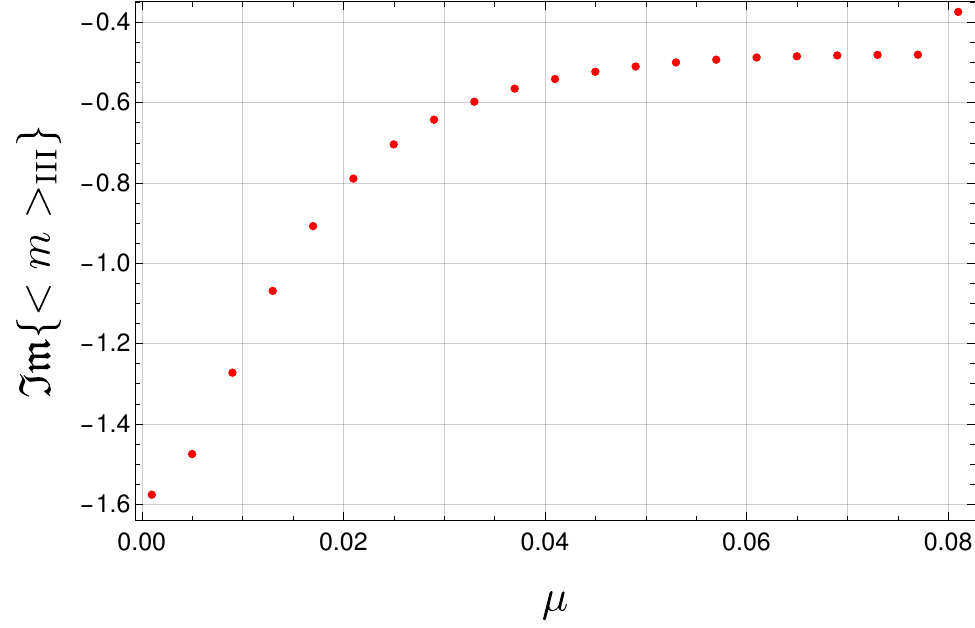}
    \vspace{-7mm}
    \caption{Imaginary part of $\langle m\rangle_{\textsc{iii}}$.}
    \label{fig:b expval m Im}
    \end{subfigure}\hspace{0.05\textwidth}
    \begin{subfigure}{0.45\textwidth}
    \includegraphics[width=\linewidth]{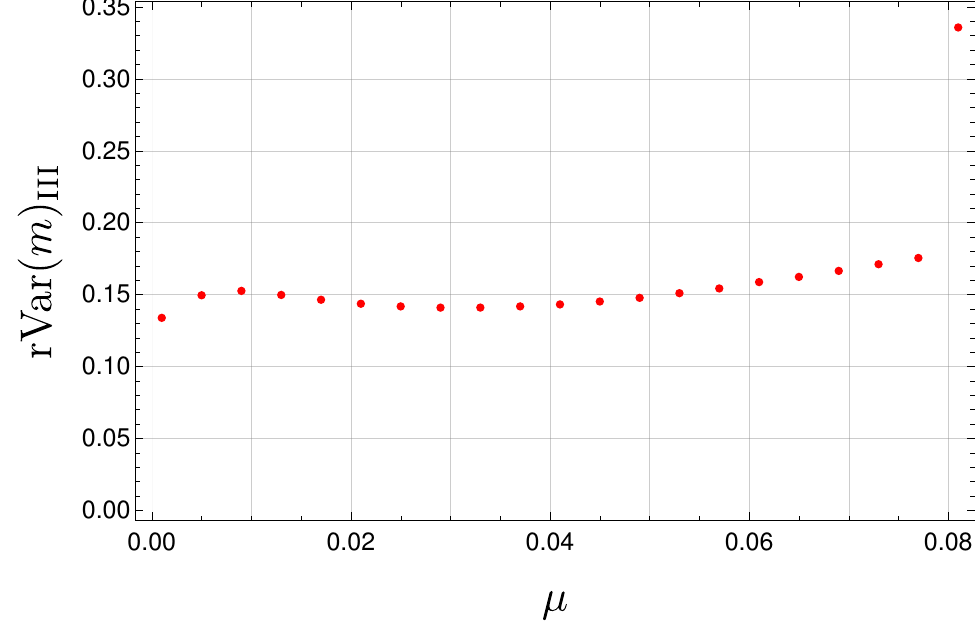}
    \vspace{-7mm}
    \caption{Relative variance of $m$ in Sector III.}
    \label{fig:b expval m rVar}
    \end{subfigure}
    \caption{(a): real part of the strut length expectation value computed in Sector III (red dots) and the classical solutions $m_{\mathrm{cl}}$ (gray graph). (b): the corresponding relative deviation between $\langle m\rangle_{\textsc{iii}}$ and the classical solution as defined in Eq.~\eqref{eq:relerr}. (c): the imaginary part of the strut length expectation value. (d): relative variance of the strut length, defined as in Eq.~\eqref{eq:rVar}. All the quantities are computed for varying scalar field mass $\mu = 10^{-3}(1+4n)$ with $n\in\{0,\dots,20\}$. We have fixed the boundary data to $l_0= 10, l_1 = 30$, $\phi_0=2$ and $\phi_1=4$.}
    \label{fig:b expval m}
\end{figure}

Given fixed boundary data $(l_0,\phi_0,l_1,\phi_1)$, the classical theory exhibits the solution $m_{\mathrm{cl}}(\mu)$ which is a continuous function of the scalar field mass in particular at the point $\mu=0$, as depicted Fig.~\ref{fig:strutmass}. That is,
\begin{equation}
    \lim\limits_{\mu\rightarrow 0}m_{\mathrm{cl}}(\mu) = m_{\mathrm{cl}}(0)\,.
\end{equation}
Also in the quantum theory, the strut length expectation value is given as a continuous function of the mass parameter in the regime $0<\mu\ll 1$ as plotted in Fig.~\ref{fig:b expval m Re}. There, the real part of $\langle m\rangle_{\textsc{iii}}$ shows good agreement with the classical solutions for small mass values. For non-zero masses, the dependence of $\mathfrak{Re}\{\expval{m}_{\textsc{iii}}\}$ on $\mu$ follows an inverse square law, similar to what has been obtained in~\cite{Ali:2022vhn}. Remarkably, given the results of Sec.~\ref{sec:freezing oscillations}, the strut length expectation is \emph{discontinuous} in the mass at the point $\mu=0$, i.e.
\begin{equation}
    \infty > \lim\limits_{\mu\rightarrow 0^+}\langle m\rangle_{\textsc{iii}} \neq \eval{\langle m\rangle_{\textsc{iii}}}_{\mu = 0} = \infty\,.
\end{equation}
Following the discussion of Sec.~\ref{sec:freezing oscillations}, this is to be expected since the introduction of a non-zero mass $\mu$ guarantees oscillations linear in the summation variable. In the continuum, the mass term explicitly breaks the lapse-independence of the action, similar to the breaking of gauge symmetry in Proca theory and massive gravity~\cite{Veltman1970}. The results observed here serve as a discrete analogon of such phenomenons.\footnote{Notice that this effect is to be distinguished from the breaking of translation symmetry in the scalar field. As noted in~\cite{Ali:2022vhn}, for periodic boundary conditions on the scalar field, the path integral diverges in the massless case and is finite in the massive case due to the breaking translation symmetry $\phi\rightarrow \phi+a$ for some $a$.}

Starting from mass values close to zero, the relative deviation $\delta$ of $\mathfrak{Re}\{\expval{m}_{\textsc{iii}}\}$ and the classical solutions $m_{\mathrm{cl}}$ decreases for increasing $\mu$. That is because the mass term is the only term in the gravity and matter action that is linear in the summation variable $m$, leading to constant frequency oscillations at large values of $m$. As discussed in Sec.~\ref{sec:freezing oscillations}, these oscillations ensure the convergence of the partition sum. However, if the oscillations become very rapid, then the discrete points $m\in\mathbb{N}/2$ in the length spectrum do not suffice to resolve the saddle point of the amplitude. This explains the large value of $\delta$ at the largest mass value depicted in Fig.~\ref{fig:b expval m relerr}. As for the cases above, a refinement of the length spectrum would lead to results closer to classicality. 

The imaginary part of $\langle m\rangle_{\textsc{iii}}$ converges to $-\frac{1}{2}$ for increasing mass values until the discreteness of the spectrum leads to deviations. A plot is given in Fig.~\ref{fig:b expval m Im}.

Finally, the relative variance is plotted in Fig.~\ref{fig:b expval m rVar}, showing an almost constant behavior for intermediate masses and increasing values for larger masses. Interestingly, this is in contrast to the behavior of $\delta$. Thus, $\mathfrak{Re}\{\expval{m}_{\textsc{iii}}\}$ agrees better with the classical solutions for large $\mu$ accompanied however with larger fluctuations as measured by rVar. 

\paragraph{Summary.} We investigated in this section the expectation value of the strut length $m$ computed via the effective cosmological partition function in Sector III for a variety of boundary data and mass parameters. The real part of the expectation value generically agrees with the classical solutions $m_{\mathrm{cl}}$ obtained from the Regge equation in Eq.~\eqref{eq:Regge equation strut}. We have identified two numerical sources for large deviations $\delta$ between this real part and $m_{\mathrm{cl}}$. Most importantly, the discreteness of the spectrum of time-like length $m\in\mathbb{N}/2$ can lead to an insufficient resolution of the saddle point of the amplitude $\hat{A}_{\textsc{iii}}$. This issue has been discussed in~\cite{Dittrich:2023rcr} where a refinement of the spectrum has been suggested to resolve this issue. Another numerical challenge that affected in particular the behavior for varying edge length $l_1$ is that of  highly oscillatory summands combined with small amplitudes. This issue can be potentially resolved by utilizing arbitrary precision arithmetic. Given these sources of numerical errors, it is challenging to decouple these from actual quantum fluctuations. To that end, we computed the relative variance, defined in Eq.~\eqref{eq:rVar}, which for varying scalar field value and mass appears to be decoupled from $\delta$. Therefore, the relative variance potentially provides a measure of quantum fluctuations in these cases. Lastly, we have demonstrated in Fig.~\ref{fig:b expval a Im},~\ref{fig:b expval phi Im} and~\ref{fig:b expval m Im} that the imaginary part of the expectation value is approximated by $-\frac{1}{2}$ for a range of different boundary data. The existence of a non-zero imaginary part has been deemed a quantum effect in~\cite{Dittrich:2023rcr} and is to be expected from the perspective of the general boundary formulation, discussed in Sec.~\ref{sec:Effective cosmological partition function}. Whether real expectation values can be obtained by considering different boundary states or different observables is unclear. We leave the physical interpretation of this imaginary part as an intriguing open question to future research. 

\subsection{Strut length expectation value in Sectors I and II}\label{sec:including I and II}

For the preceding computations, we restricted to time-like struts in the bulk. However, the full effective partition function contains not only a sum over the strut length but also over its causal character. Therefore, a complete evaluation of the partition function and expectation values requires the inclusion of Sectors I and II, considered in the following.

As discussed in Sec.~\ref{sec:Asymptotic vertex amplitude and measure factors}, we have derived an analytical formula for the Hessian determinant in Sector II. Following Eq.~\eqref{eq:Z single frustum} the contribution to $Z$ from Sector II consists of a bounded integral
\begin{equation}
Z_{\textsc{ii}}(l_0,l_1,\phi_0,\phi_1) = \int\limits_{1/2}^{m_{\mathrm{int}}}\dd{m}\hat{A}^{\textsc{ii}}(l_0,l_1,m,\phi_0,\phi_1)\,,
\end{equation}
which can be computed straightforwardly using numerical integration techniques.\footnote{Here and in the remainder, we utilize either the \textsc{Cuba}-integration package in \textsc{Julia}: \href{https://github.com/giordano/Cuba.jl}{https://github.com/giordano/Cuba.jl} for Monte-Carlo integration (see also Ref.~\cite{Hahn:2005pf}), the \textsc{QuadGK} package in \textsc{Julia}: \href{https://juliamath.github.io/QuadGK.jl/stable/}{https://juliamath.github.io/QuadGK.jl/stable/} using quadrature methods or the \textsc{NIntegrate}-function in \textsc{Mathematica}.} We remind the reader that with our choice of identifying geometrical edge length with the underlying spin-foam variables, space-like edges exhibit a minimal length of $1/2$. The evaluation of the partition function in Sector I,
\begin{equation}
Z_\textsc{i}(l_0,l_1,\phi_0,\phi_1) = \int\limits_{m_{\mathrm{int}}}^{m_{\mathrm{max}}}\dd{m}\hat{A}^{\textsc{i}}(l_0,l_1,m,\phi_0,\phi_1)\,,
\end{equation}
is more challenging since an analytical formula of the Hessian determinant at $\vartheta\neq 1$ is not at hand. In principle, the numerical integration algorithms can also be applied to $\hat{A}^{\textsc{i}}$ without an explicit formula of $\det H_\vartheta$. In this case the determinant $\det H_\vartheta$ needs to be computed for every sampling point of the integrand. Unfortunately, many samples and integrand evaluations are required for convergence due to rapid oscillations. As a result, convergent numerical integration requires an unfeasible amount of computation time.

To surpass these obstacles and explore the effects of Sector I on the total partition function and the strut expectation value, we interpolate the Hessian determinant numerically between a large number of discrete points and use this function then for the numerical integration. 

\begin{figure}
    \centering
    \begin{subfigure}{0.45\textwidth}
    \includegraphics[width=\linewidth]{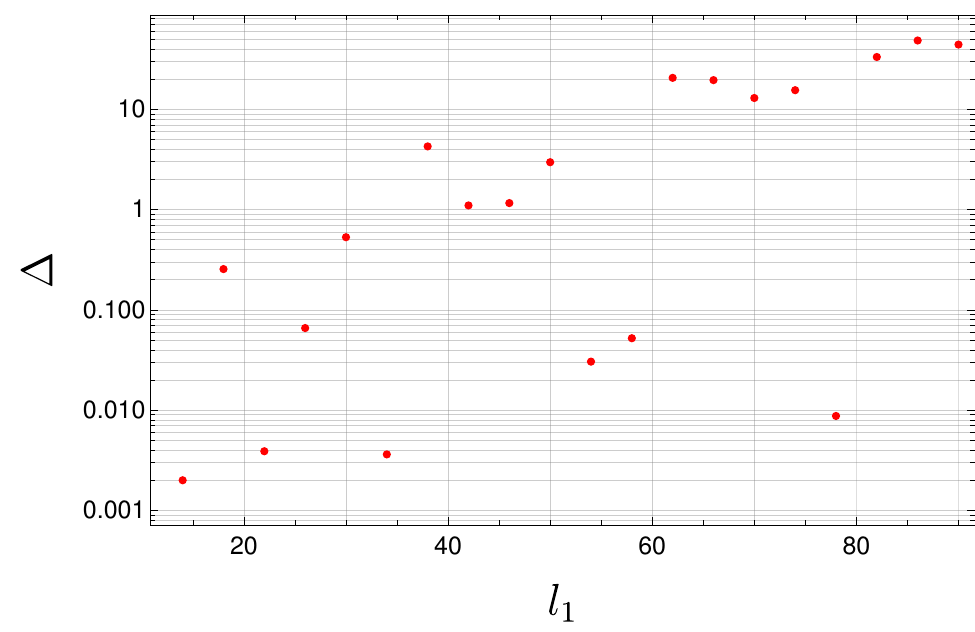}
    \end{subfigure}\hspace{0.05\textwidth}
    \begin{subfigure}{0.45\textwidth}
    \includegraphics[width=\linewidth]{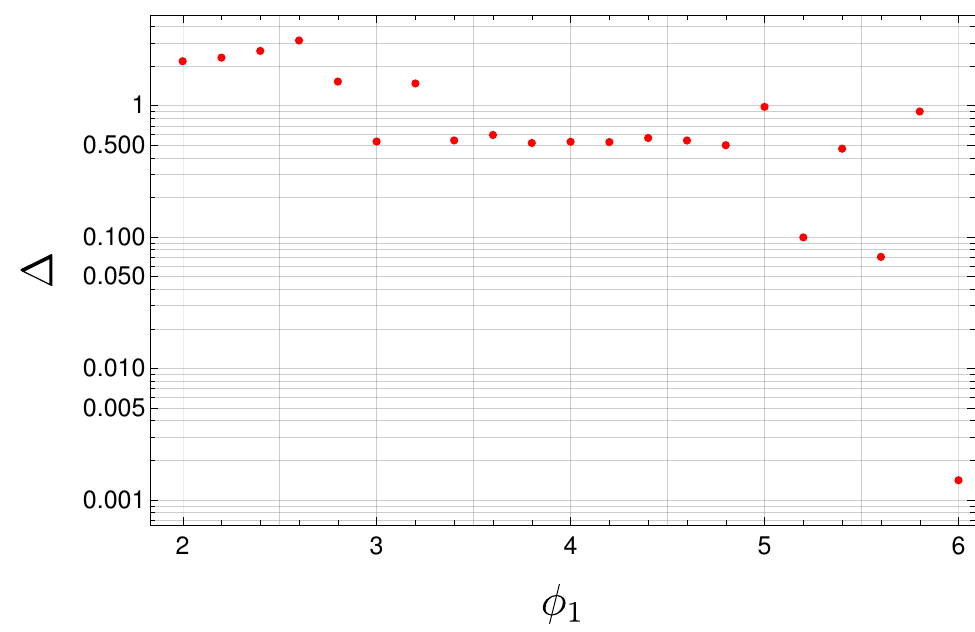}
    \end{subfigure}\\[5mm]
    \begin{subfigure}{0.45\textwidth}
    \includegraphics[width=\linewidth]{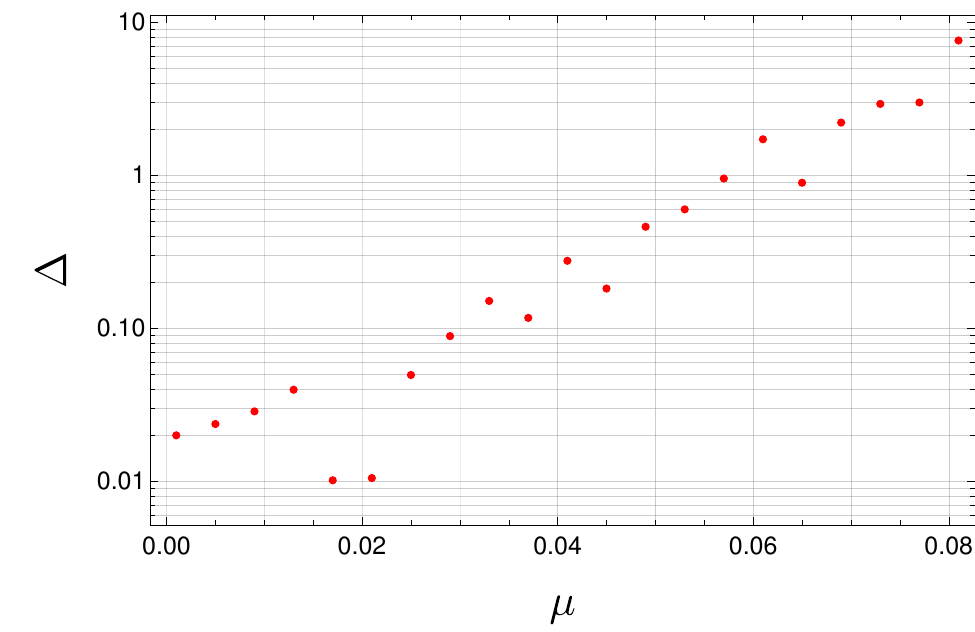}
    \end{subfigure}
\caption{Relative deviation $\Delta$ of the squared strut length expectation values $\langle m^2\rangle$ and $\langle m^2\rangle_{\textsc{iii}}$ as defined in Eq.~\eqref{eq:Delta}. Top left: $\Delta$ as a function of varying edge length $l_1 = 10+4n$ with $n\in\{1,\dots,20\}$ and $l_0=10$, $\phi_0=2$, $\phi_1=4$ and $\mu=0.05$ fixed. Top right: $\Delta$ as a function of varying scalar field value $\phi_1=2+0.02n$ with $n\in\{0,\dots,20\}$ and $l_0 = 10$, $l_1=30,\phi_0=2$ and $\mu=0.05$ fixed. Bottom: $\Delta$ as a function of varying scalar field mass $\mu = 10^{-3}(1+4n)$ with $n\in\{0,\dots,20\}$ and $l_0=10$, $l_1 =30$, $\phi_0=2$ and $\phi_1 = 4$ fixed.}
    \label{fig:IIIvstot}
\end{figure}

The influence of Sectors I and II is quantified by first computing the squared strut length expectation value, defined as
\begin{equation}
\langle m^2\rangle = -\langle m^2\rangle_{\textsc{i}}-\langle m^2\rangle_{\textsc{ii}}+\langle m^2\rangle_{\textsc{iii}}\,,
\end{equation}
where we explicitly take into account the causal character of the strut length. The deviation between $\langle m^2\rangle$ and $\langle m^2\rangle_{\textsc{iii}}$ is then measured by\footnote{The information on the sign of $\langle m^2\rangle_{\textsc{iii}}-\langle m^2\rangle$ is not captured by $\Delta$ due to the absolute value entering its definition. We notice that for the data points in Fig.~\ref{fig:IIIvstot}, there is no definite sign, i.e. some differences are positive while others are negative.}
\begin{equation}\label{eq:Delta}
\Delta = \frac{\abs{\mathfrak{Re}\{\langle m^2\rangle_{\textsc{iii}}\}-\mathfrak{Re}\{\langle m^2\rangle\}}}{\mathfrak{Re}\{\langle m^2\rangle_{\textsc{iii}}\}}\,.
\end{equation}
Our results for the deviation $\Delta$ are summarized in Fig.~\ref{fig:IIIvstot}.

In the top left panel of Fig.~\ref{fig:IIIvstot}, $\Delta$ is plotted for varying final edge length $l_1$. In this case, we observe that the deviations tend to increase for larger values $l_1$ with more than half of them being larger than $0.1$. Therefore, we consider the deviations between $\langle\cdot\rangle_{\textsc{iii}}$ and $\langle\cdot\rangle$ as substantial. 

In the top right panel of Fig.~\ref{fig:IIIvstot}, the deviation $\Delta$ is depicted as a function of varying scalar field value. We observe that almost all the data points evaluate to $\Delta > 0.1$, therefore constituting substantial deviations between $\langle\cdot\rangle_{\textsc{iii}}$ and $\langle\cdot\rangle$. 

In the bottom panel of Fig.~\ref{fig:IIIvstot}, the deviation $\Delta$ is plotted as a function of the scalar field mass $\mu$. We observe a linear dependence between $\Delta$ and $\mu$ with larger deviations for larger mass values. For mass values $\mu < 0.03$, deviations are below $0.1$ while the rise to $\Delta > 1.0$ for $\mu > 0.06$. 

Our results show that only in a regime of small scalar field mass $\mu$, the full partition function with space-like and time-like struts can be satisfactorily approximated with the partition function in Sector III. Outside this regime, the deviations $\Delta$ are substantial. This behavior is sourced by a lack of an exponential suppression which can be traced back to the failure of the asymptotics to reproduce complex Lorentzian deficit angles, see the discussion in Sec.~\ref{sec:causality violations}. Although a suppression mechanism for causality violations is missing for the effective cosmological model constructed here, we emphasize that in the present setting one can consistently exclude these configurations as done in Secs.~\ref{sec:mass term regularization} and ~\ref{sec:1slice}. Thus, the effective partition function restricted to Sector III still provides a viable model for quantum cosmology.   

In the following, we compare our results to those obtained from effective spin-foam models, where causality violating configurations are exponentially suppressed.  

\subsection{Comparison to effective spin-foam models}\label{sec:comparison to ESF}

Inspired from quantum cosmology in a continuum path integral~\cite{Feldbrugge:2017kzv}, Lorentzian quantum Regge calculus and effective spin-foams have been applied to spatially spherical cosmologies with a positive cosmological constant~\cite{Asante:2021phx,Dittrich:2023rcr}. In this section, we set up the effective spin-foam partition function for 3-dimensional spatially flat cosmology with the ingredients of~\cite{Dittrich:2023rcr} to compare the results with the model developed in this work. 

The effective spin-foam amplitudes with a coupled massive scalar field are given by
\begin{equation}
A^{\mathrm{sl}}_{\textsc{esf}} = \upmu_{\mathrm{sl}}\e^{i(S_{\mathrm{I,II}}+S_\phi)},\qquad A^{\mathrm{tl}}_{\textsc{esf}} = \upmu_{\mathrm{tl}}\e^{i(S_{\textsc{iii}}+S_\phi)}\,,
\end{equation}
where $S_\textsc{i},S_{\textsc{ii}}$ and $S_{\textsc{iii}}$ are the Lorentzian Regge actions given in Eqs.~\eqref{eq:S_I}--\eqref{eq:S_III}. Importantly, the Regge actions $S_{\textsc{i}}$ and $S_{\textsc{ii}}$ entering $A_{\textsc{esf}}^{\mathrm{sl}}$ are complex valued in contrast to the amplitudes obtained in Sec.~\ref{sec:Asymptotic vertex amplitude and measure factors} from a semi-classical limit of spin-foams. Notice that the signs of the imaginary parts of $S_{\textsc{i}}$ and $S_{\textsc{ii}}$ have been chosen such that $\mathfrak{Im}\{S\}>0$ for which $\e^{iS}$ gives exponential suppression. A detailed discussion of this choice is given in~\cite{Asante:2021phx,Dittrich:2023rcr}. 

The functions $\upmu_{\mathrm{tl}}$ and $\upmu_{\mathrm{sl}}$ are the measure factors for time-like and space-like struts, respectively. Following the discussion at the end of Sec.~\ref{sec:Asymptotic vertex amplitude and measure factors} different choices of the measure factors are conceivable. For comparison with~\cite{Dittrich:2023rcr,Asante:2021phx}, we define 
\begin{equation}\label{eq:ESF measure}
\upmu(m)_\epsilon = \frac{\epsilon}{2}\sqrt{3i(l_0+l_1)}\frac{m}{\left(\frac{(l_0-l_1)^2}{2}+\epsilon m^2\right)^{\frac{3}{4}}}\,,
\end{equation}
with $\epsilon=+$ ($\epsilon = -$) for time-like (space-like) struts. These factors can be derived by adapting the continuum measure used in~\cite{Feldbrugge:2017kzv} to the context of discrete Regge calculus.

The domain of length variables in effective spin-foams is imported from the spectrum of boundary spins in the full spin-foam model. As previously discussed and noted in~\cite{Borissova:2024txs,Borissova:2024pfq}, this amounts to space-like struts having a continuum spectrum $[\frac{1}{2},\infty
)$ and time-like struts having a discrete spectrum $\mathbb{N}/2$. Thus, the effective spin-foam partition function for a single 3-frustum with boundary data $(l_0,\phi_0,l_1,\phi_1)$ is given by
\begin{equation}\label{eq:Z_ESF}
Z_{\textsc{esf}} = \sum_{m\in\mathbb{N}/2}A_{\textsc{esf}}^{\mathrm{tl}}(l_0,l_1,m,\phi_0,\phi_1) + \int\limits_{1/2}^{m_{\mathrm{max}}}\dd{m}A_{\textsc{esf}}^{\mathrm{sl}}(l_0,l_1,m,\phi_0,\phi_1)\,.
\end{equation}
Expectation values of the strut length or functions thereof are then computed accordingly. Along the lines of the previous sections, we compute expectation values first only from the partition function in Sector III and then study the influence of the causality violating configurations in Sectors I and II.

\begin{figure}
    \centering
    \begin{subfigure}{0.45\textwidth}
    \includegraphics[width=\linewidth]{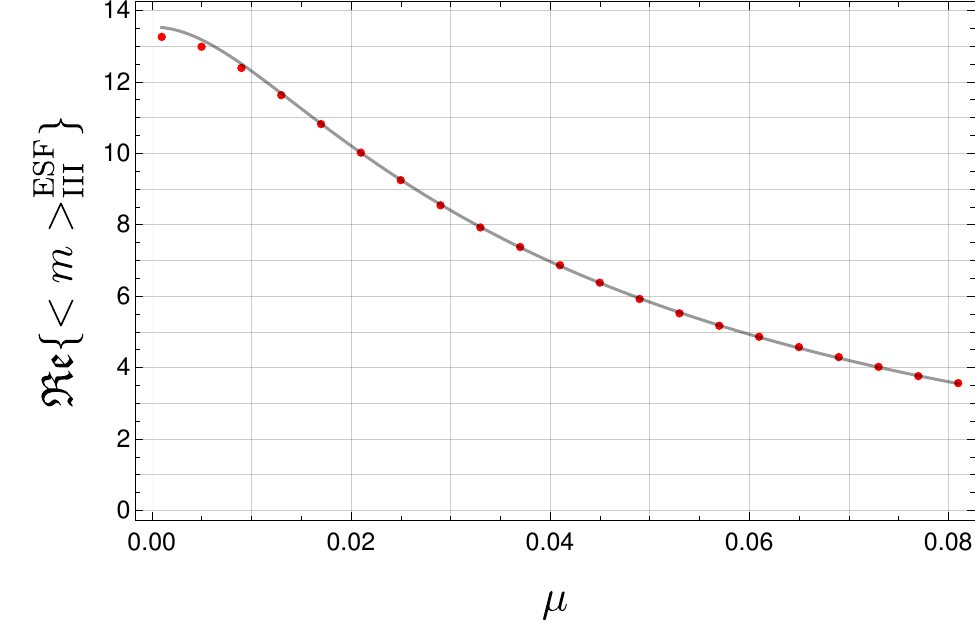}
    \vspace{-7mm}
    \caption{Real part of $\langle m\rangle_{\textsc{iii}}^{\textsc{esf}}$ and classical solutions $m_{\mathrm{cl}}$.}
    \label{fig:b expval m ESF Re}
    \end{subfigure}\hspace{0.05\textwidth}
    \begin{subfigure}{0.45\textwidth}
    \includegraphics[width=\linewidth]{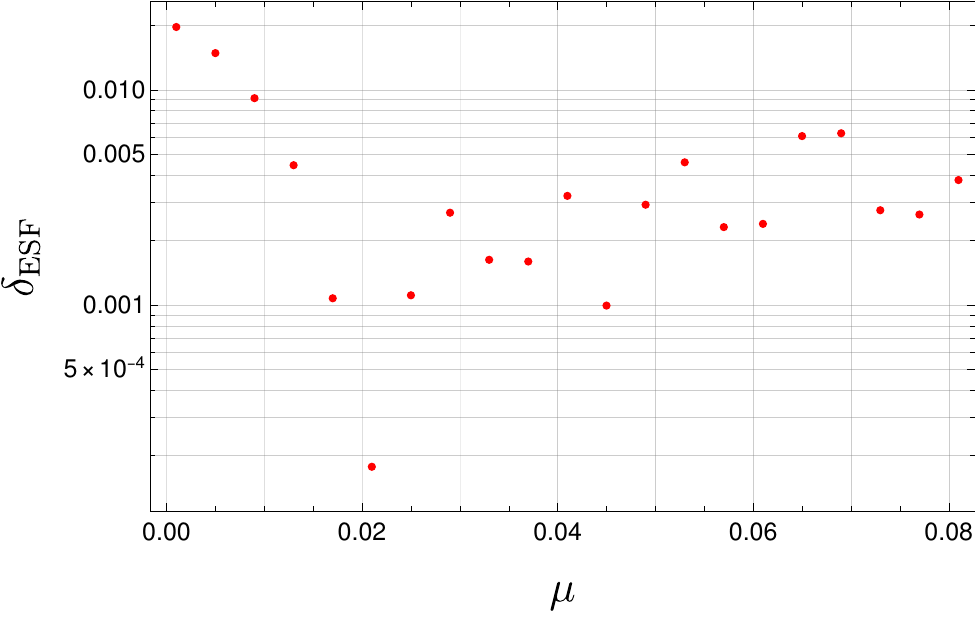}
    \vspace{-7mm}
    \caption{Relative deviation between expectation value and classical solution.}
    \label{fig:b expval m ESF relerr}
    \end{subfigure}\\[5mm]
    \begin{subfigure}{0.45\textwidth}
    \includegraphics[width=\linewidth]{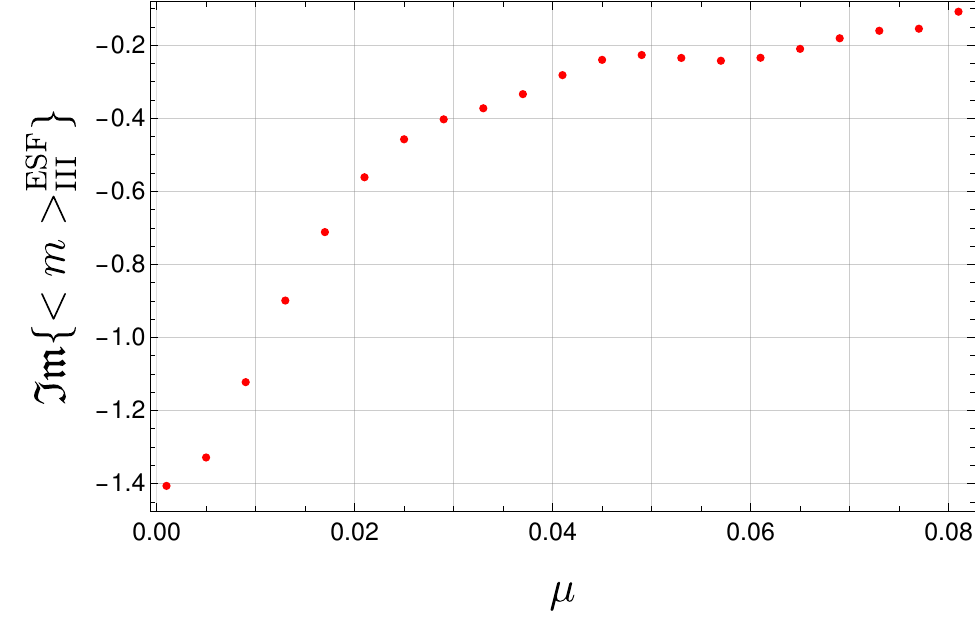}
    \vspace{-7mm}
    \caption{Imaginary part of $\langle m\rangle_{\textsc{iii}}$.}
    \label{fig:b expval m ESF Im}
    \end{subfigure}\hspace{0.05\textwidth}
    \begin{subfigure}{0.45\textwidth}
    \includegraphics[width=\linewidth]{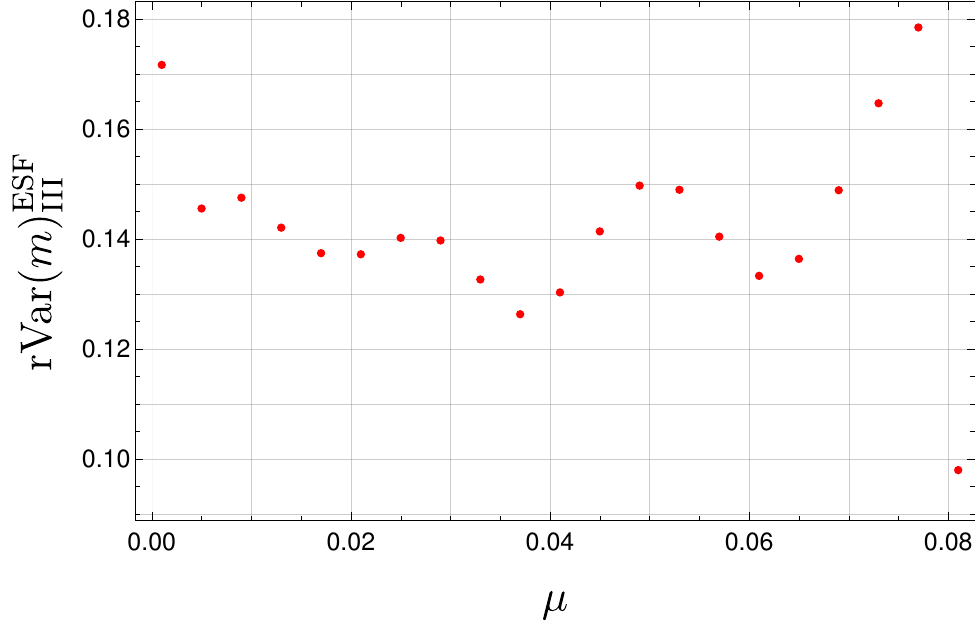}
    \vspace{-7mm}
    \caption{Relative variance of $m$ in Sector III.}
    \label{fig:b expval m ESF rVar}
    \end{subfigure}
    \caption{(a): real part of the strut length expectation value computed from effective spin-foams in Sector III (red dots) and the classical solutions $m_{\mathrm{cl}}$ (gray graph). (b): the corresponding relative deviation between $\langle m\rangle_{\textsc{iii}}^{\textsc{esf}}$ and the classical solution as defined in Eq.~\eqref{eq:relerr}. (c): the imaginary part of the strut length expectation value. (d): relative variance of the strut length, defined as in Eq.~\eqref{eq:rVar}. All the quantities are computed for varying scalar field mass $\mu = 10^{-3}(1+4n)$ with $n\in\{0,\dots,20\}$. We have fixed the boundary data to $l_0= 10, l_1 = 30$, $\phi_0=2$ and $\phi_1=4$.}
    \label{fig:b expval m ESF}
\end{figure}

The real part of the strut length expectation value, depicted in Fig.~\ref{fig:b expval m ESF Re} as a function of the scalar field mass, generically agrees 
with the classical solutions. The expectation value is finite arbitrarily close to $\mu\rightarrow 0$ while it diverges at $\mu = 0$ due to asymptotically freezing oscillations of the amplitudes, equivalent to the discussion in Sec.~\ref{sec:mass term regularization}.  

The relative deviation $\delta_{\textsc{esf}}$ between $\mathfrak{Re}\{\langle m\rangle_{\textsc{iii}}^{\textsc{esf}}\}$ and the classical solution is generically smaller than the one obtained from the effective cosmological partition function utilized in the previous sections. To that end compare Figs.~\ref{fig:b expval m relerr} and~\ref{fig:b expval m ESF relerr}. Since the exponentiated Regge actions are the same for both models, the deviations necessarily arise from the different measure terms. The measure $\upmu_{\mathds{1}}$ discussed in Sec.~\ref{sec:Asymptotic vertex amplitude and measure factors} is an involved function of the strut length $m$ compared to the simple form of $\upmu_{\mathrm{tl}}$ defined in Eq.~\eqref{eq:ESF measure}. Tentatively, the stronger deviations observed in Sec.~\ref{sec:mass term regularization} could therefore constitute quantum corrections arising from the measure factor. 

Visualized in Fig.~\ref{fig:b expval m ESF Im}, the imaginary part of the strut length expectation value is non-constant and shows oscillatory behavior. In particular, it does not show the curious tendency of $\ev{m}_{\textsc{iii}} = 1/2$. A similar form of the imaginary part has been observed for the cosmological effective spin-foam models considered in~\cite{Dittrich:2023rcr}.

Also the relative variance, depicted in Fig.~\ref{fig:b expval m ESF rVar}, is non-constant and shows stronger oscillatory behavior than for the results of Sec.~\ref{sec:mass term regularization}.

While we considered here a varying scalar field mass, one can repeat the numerical evaluation of $\langle m\rangle_{\textsc{iii}}^{\textsc{esf}}$ for different boundary data such as a varying scalar field value $\phi_1$ or a varying edge length $l_1$. We have checked that in these cases, a similar behavior can be observed: expectation values which are closer to the classical value than the ones computed in Sec.~\ref{sec:mass term regularization} and an imaginary part and a relative variance which oscillate. 

\begin{figure}
    \centering
    \includegraphics[width=0.45\textwidth]{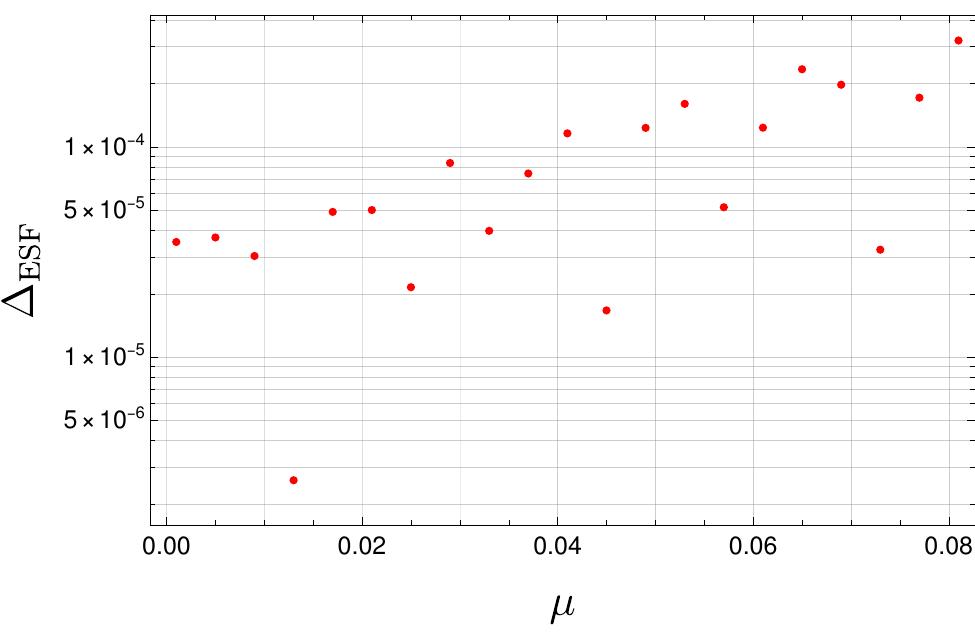}
    \caption{Relative deviation $\Delta_{\textsc{esf}}$ of the squared strut length expectation values $\langle m^2\rangle$ and $\langle m^2\rangle_{\textsc{iii}}$ as defined in Eq.~\eqref{eq:Delta}. Here, $\Delta_{\textsc{esf}}$ is plotted as a function of varying scalar field mass $\mu = 10^{-3}(1+4n)$ with $n\in\{0,\dots,20\}$ and $l_0=10$, $l_1 =30$, $\phi_0=2$ and $\phi_1 = 4$ fixed.}
    \label{fig:IIIvstot ESF}
\end{figure}

The results presented in Fig.~\ref{fig:b expval m ESF} arise from utilizing effective spin-foam amplitudes in Sector III where the strut is time-like. Similar to Sec.~\ref{sec:including I and II}, we consider in the following the effect of including also Sector I and II on the expectation value of the squared strut length. In particular, we measure the deviation between $\langle m^2\rangle_{\textsc{iii}}^{\textsc{esf}}$ and $\langle m^2\rangle^{\textsc{esf}}$ via the relative deviation $\Delta$ defined in Eq.~\eqref{eq:Delta}. Our results  are depicted in Fig.~\ref{fig:IIIvstot ESF} for varying scalar field mass $\mu$. 

We observe the same tendency as in Sec.~\ref{sec:including I and II}, namely that the numerical values of $\Delta_{\textsc{esf}}$ increase with larger values of the mass $\mu$. However, all the values of $\Delta_{\textsc{esf}}$ depicted here lie below $4\cdot 10^{-4}$. The negligible influence of causality violating configurations is rooted in the exponential suppression, arising from the imaginary parts of the deficit angles. Therefore, also in the 3-dimensional spatially flat setting, effective spin-foam models exhibit a suppression mechanism of causality violations~\cite{Dittrich:2023rcr}. At the same time, the small values of $\Delta_{\textsc{esf}}$ justify to only utilize the amplitudes of Sector III for computing expectation values. In this regime, the results of the model developed in Sec.~\ref{sec:A proposal for an effective cosmological amplitude} are compatible with those obtained in effective spin-foams. 


\section{Outlook: space-like bulk slices}\label{sec:1slice}

In this section we give an outlook on the evaluation of the effective partition function for a discretization consisting of $\mathcal{V} = 2$ cubes containing a space-like slice in the bulk. The dynamical variables are given by two strut lengths, one spatial edge length and one scalar field value. Studying this setting lies the foundation for future investigations of physically interesting scenarios such as a quantum bounce.

The results of the previous section have shown that the contributions of Sector I and II to the strut length expectation value are generically non-negligible. However, in the symmetry reduced setting here, the path integral can be consistently restricted to causally regular configurations by hand without numerical expenses. Furthermore, the computations of Sec.~\ref{sec:including I and II} were numerically costly due to the highly oscillatory nature of the amplitudes in these sectors. For these reasons, we restrict in this section to the amplitudes of Sector III and compute the partition function and expectation values only with respect to these amplitudes.

\subsection{Classical equations of motion}

The classical dynamics of spatially flat (3+1)-dimensional cosmologies has been studied in~\cite{Jercher:2023csk} for a minimally coupled massless scalar field. These results can in part be transferred to the present setting, with the difference being that the volume of hinges entering the Regge action are lengths rather than areas, and the scalar field is considered to be massive. 

\begin{figure}
    \centering
    \includegraphics[width=0.6\linewidth]{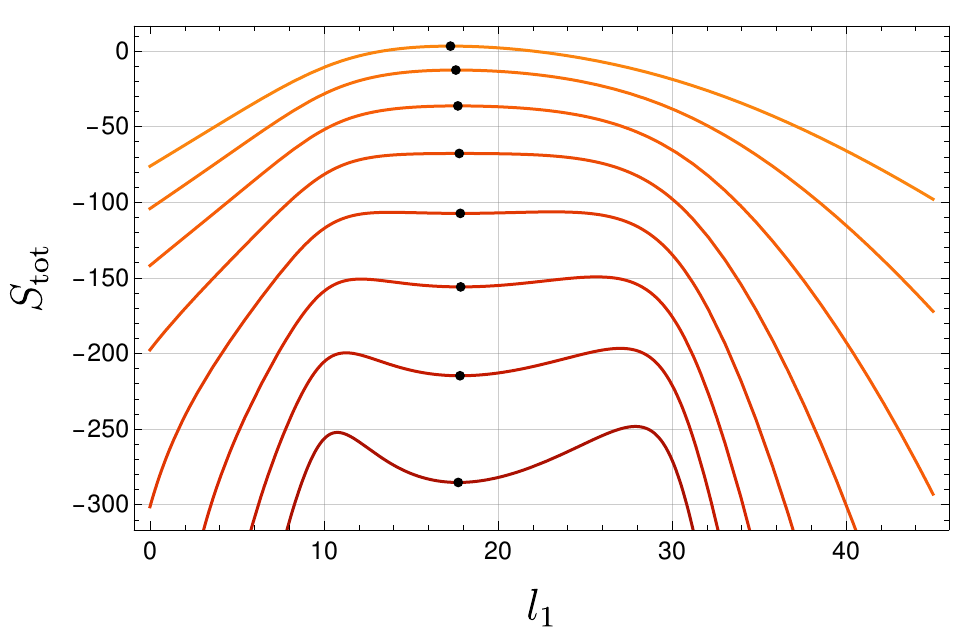}
    \caption{Total Regge action for the discretization $\mathcal{X}_2$ as a function of the spatial edge length $l_1$ in the bulk with varying scalar field mass $\mu\in\{0.01,\dots,0.08\}$ where darker red corresponds to larger $\mu$. The bulk strut length $m_0$ and $m_1$ as well as the bulk scalar field $\phi_1$ are evaluated on solutions of the equations of motion. Values of $l_1$ that solve the equations of motion are marked with a black dot. Notice that for $\mu \geq 0.05$, multiple extrema are visible, all of which satisfy $\partial S_{\mathrm{tot}}/\partial l_1 = 0$. However, we explicitly checked that for the non-marked extrema, the functions $\partial S_{\mathrm{tot}}/\partial m_0 $ and $\partial S_{\mathrm{tot}}/\partial m_1$ do not vanish and thus, these points do not solve the whole set of dynamical equations. Boundary values have been fixed to $l_0 = 10$, $l_2 = 30$, $\phi_0 = 2$ and $\phi_2=4$ for demonstrational purposes.}
    \label{fig:S tot m}
\end{figure}

The total action on $\mathcal{X}_2$ governing the dynamics of the causally regular sector is given by
\begin{equation}
    S_{\mathrm{tot}} = \sum_{n=0}^1\left[S_\textsc{iii}(l_n,l_{n+1},m_n)+S_\phi(l_n,l_{n+1},m_n,\phi_n,\phi_{n+1})\right]\,,
\end{equation} 
with $S_{\textsc{iii}}$ and $S_\phi$ defined in Eqs.~\eqref{eq:S_III} and~\eqref{eq:S_phi}, respectively. The variables $\partial\psi = (l_0,\phi_0,l_2,\phi_2)$ lie on the boundary while $(m_0,m_1,l_1,\phi_1)$ are bulk variables and thus dynamical. As shown in Sec.~\ref{sec:Minimally coupled massive scalar field}, the scalar field equations can be explicitly solved in terms of the bulk and boundary variables with $\phi_1^{\mathrm{sol}}$ explicitly given in Eq.~\eqref{eq:phi sol}. The remaining equations of motion are 
\begin{equation}
\pdv{S_{\mathrm{tot}}}{m_0} = 0\,,\qquad \pdv{S_{\mathrm{tot}}}{l_1} = 0\,,\qquad \pdv{S_{\mathrm{tot}}}{m_1} = 0,
\end{equation}
which form a set of coupled transcendental equations.  Guessing initial values close to the actual solution (e.g. visually by plotting the Regge action), these equations can be explicitly solved using the \textsc{FindRoot} method of \textsc{Mathematica}. A plot of the Regge action as a function of the bulk variable $l_1$ is given in Fig.~\ref{fig:S tot m}. 

For the later purpose of comparison to expectation values, the classical solutions of the bulk variables are summarized in Tab.~\ref{tab:exp vals} for boundary data $l_0 = 10$, $l_2 = 30$, $\phi_0 = 2$, $\phi_2 = 4$ and scalar field mass $\mu = 0.05$.

\subsection{Evaluation of the partition function}\label{sec:Evaluation of the partition function}

The key object for computing expectation values is the partition function which, restricted to time-like strut length, is explicitly given by
\begin{multline}\label{eq:Z1slice}
Z_{\mathcal{X}_2}(\partial\psi) = \sum_{m_0,m_1\in\mathbb{N}/2}\int_{1/2}^{\infty}\dd{l_1}\int_{\R}\dd{\phi_1} \\ A_f(l_1)\hat{A}^{\textsc{iii}}(l_0,l_1,m_0,\phi_0,\phi_1)\hat{A}^{\textsc{iii}}(l_1,l_2,m_1,\phi_1,\phi_2)\,,
\end{multline}
for boundary data $\partial\psi = (l_0,\phi_0,l_2,\phi_2)$.

Explicit evaluation of the partition function is challenging because of several reasons. First of all, the integrand is highly oscillatory in all the variables, requiring costly numerical integration for the variable $l_1$ and an extension of sequence acceleration methods to multiple variables for the $m_0$ and $m_1$-summations. Only the $\phi_1$-integration is straightforward to compute as it is given by a Fresnel integral which is the extension of a Gaussian integral to an oscillating exponent quadratic in the integration variable. 

The strategy we employ to compute $Z_{\mathcal{X}_2}$ is the following. First, we analytically perform the $\phi_1$-integration. Then, we compute the $l_1$-integral numerically for a large set of bulk strut length $2m_0,2m_1\in\{1,\dots,2 m_{\mathrm{max}}\}$. Finally, we apply a series transformation to the remaining sum over $m_0$ and $m_1$ to accelerate its convergence. We detail these steps next.

\subsubsection{Integration of bulk scalar field}\label{sec:Integration of bulk scalar field}

The relevant part of the $\phi_1$-integration entering Eq.~\eqref{eq:Z1slice} is given by $I_{\phi_1}[1]$,
having defined the functional 
\begin{equation}
    I_{\phi_1}[\phi_1] = \int_\R\dd{\phi_1}\e^{i\left(S_\phi(l_0,l_1,m_0,\phi_0,\phi_1)+S_\phi(l_1,l_2,m_1,\phi_1,\phi_2)\right)}\phi_1 \,,
\end{equation}    
where $S_\phi$ is given in Eq.~\eqref{eq:S_phi}. Notice that we factored out the exponentiated Regge action as well as the measure factors, both of which are independent of $\phi_1$. Exploiting that the scalar field action is quadratic in $\phi_1$, the integration can be performed analytically, yielding
\begin{equation}
I_{\phi_1}[1] = \upmu_{\phi_1}\e^{iS_\phi^{\mathrm{eff}}}
\end{equation}
with
\begin{equation}
\begin{aligned}
    S_\phi^{\mathrm{eff}} &= \left((\phi_0-\phi_2)^2\frac{w_0w_1}{w_0 + w_1 -M_0 - M_1}\right)\\[7pt]
    &-\left(\phi_0^2\left(M_0+\frac{w_0(M_0 + M_1)}{w_0+w_1-M_0-M_1}\right)+\phi_2^2\left(M_1+\frac{w_1(M_0+M_1)}{w_0+w_1-M_0-M_1}\right)\right)\,,
\end{aligned}
\end{equation}
and
\begin{equation}\label{eq:mu phi}
    \upmu_{\phi_1} = \sqrt{\frac{i\pi}{w_0+w_1-M_0-M_1}}\,.
\end{equation} 
Recall that $w_n$ and $M_n$ are defined in Eq.~\eqref{eq:w and M}. Notice that by integrating out the bulk scalar field, the remaining amplitude does not factorize into amplitudes for each spacetime slab and can therefore be considered as non-local.

For the purpose of computing the expectation value of $\phi_1$, the integral $I_{\phi_1}[\phi_1]$
can be computed similarly, yielding
\begin{equation}
I_{\phi_1}[\phi_1] = I_{\phi_1}[1]\frac{\phi_0 w_0 + \phi_2w_1}{w_0+w_1-M_0-M_1}\,.
\end{equation}
Notice that the additional factor is precisely given by the solution of the scalar field equations of motion provided in Eq.~\eqref{eq:phi sol}.

For non-vanishing scalar field mass $\mu\neq 0$, the measure term $\upmu_{\phi_1}$ as well as the effective scalar field action $S_\phi^{\mathrm{eff}}$ contain inverse factors of $w_0+w_1-M_0-M_1$. Thus, for fixed boundary data and bulk struts, the remaining amplitude diverges as a function of $l_1$ when $w_0+w_1-M_0-M_1 = 0$. The point of divergence, $l_1 = l_1^{\mathrm{div}}$, is inversely proportional to the scalar field mass, $l_1^{\mathrm{div}}\sim \mu^{-\alpha}$ for some $\alpha > 0$. Thus, in the massless limit, the divergence is pushed to infinity and thus effectively absent. We checked explicitly that the divergence of $\upmu_{\phi_1}$ does not impede the $l_1$-integration detailed down below. However, when computing the scalar field expectation value or higher powers of it, the degree of divergence increases which can lead to infinities. We discuss this case in Sec.~\ref{sec:Summation of bulk strut lengths}.

\subsubsection{Integration of bulk spatial edge length}

The integration of the spatial edge length $l_1$ can be performed numerically for a large set of bulk strut length $2m_0, 2m_1\in\{1,\dots,2 m_{\mathrm{max}}\}$ and fixed boundary data $\partial\psi$. As a result, one obtains an \emph{effective} amplitude\footnote{Notice that with the restriction to time-like struts, the integration in $l_1$ is unbounded from above, irrespective of the boundary lengths $l_0$ or $l_2$.}
\begin{equation}\label{eq:l1-integration}
A_{\mathrm{eff}}(m_0, m_1;\partial\psi) = \int_{1/2}^{\infty}\dd{l_1}A_v^{\mathrm{asy}}(l_0,l_1,m_0)A_v^{\mathrm{asy}}(l_1,l_2,m_1)I_{\phi_1}[1]\,,
\end{equation}
which can be stored as a matrix of size $2m_{\max}\times 2m_{\max}$. For computing the expectation value of the spatial edge length $l_1$, the same integral is being evaluated but with an insertion of a factor of $l_1$. In the following, we refer to the integrand of the equation above as $A_{l_1}$.

Due to its oscillatory nature, numerically evaluating the integral is costly. However, we stick to numerical integration techniques as they yield reliable results which can be cross-checked using different methods like Monte Carlo integration and quadrature. Discretizing the variable $l_1$ and performing the resulting sum up to some upper cutoff converges only very slowly in the limit of removing the cutoff. Following the discussion of~\cite{Dittrich:2023rcr}, the integrand is also not suited for an application of Wynn's algorithm. That is because schematically, $A_{l_1}\sim \e^{i(l_1^3+\dots)}$, i.e. its oscillations are non-constant which yields pseudo-saddle points that obstruct the convergence of the Shanks transform.

\begin{figure}
    \centering
        \begin{subfigure}{0.45\textwidth}
        \includegraphics[width=\linewidth]{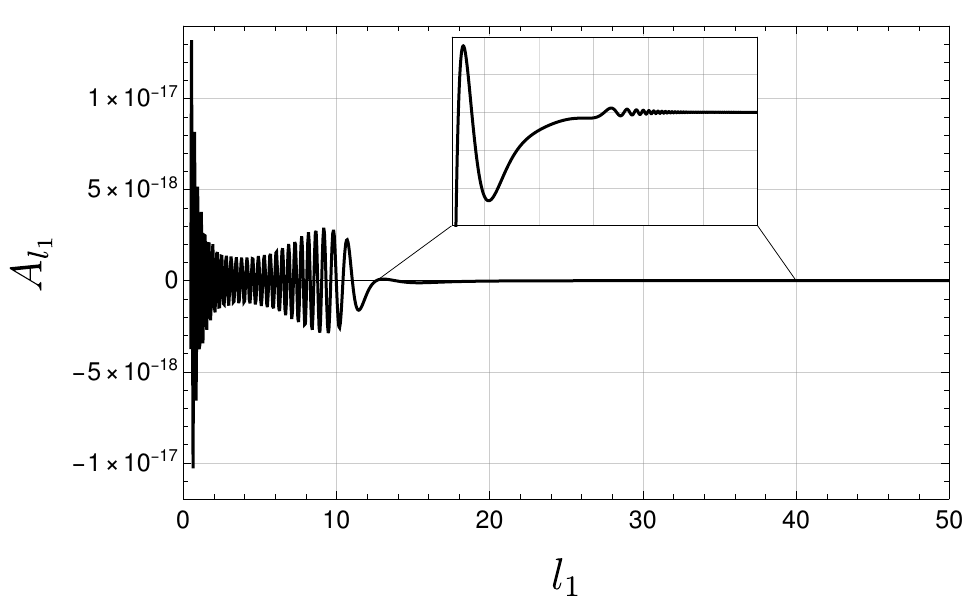}
        \end{subfigure}\hspace{0.05\textwidth}
        \begin{subfigure}{0.45\textwidth}
        \includegraphics[width=\linewidth]{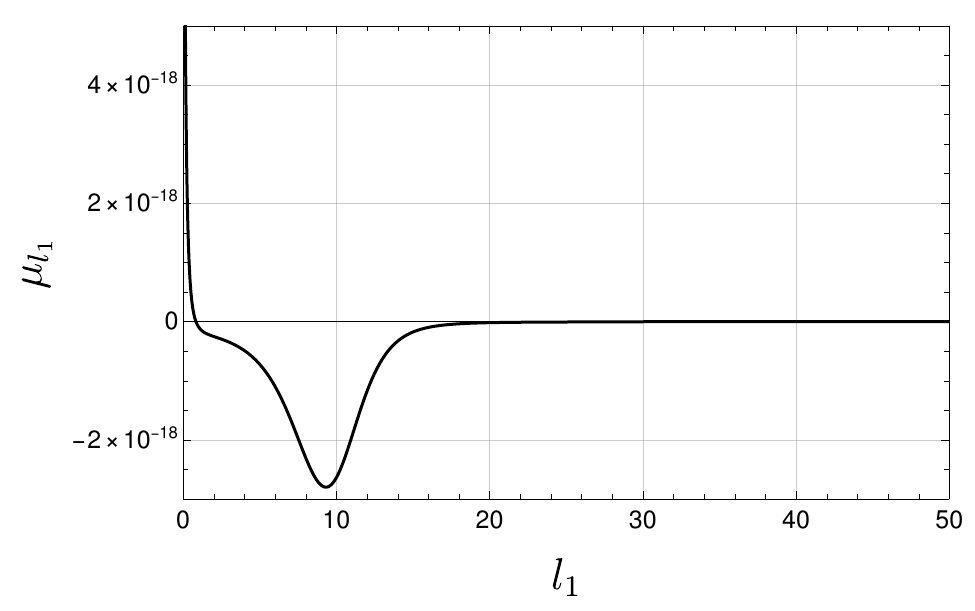} 
        \end{subfigure}
    \caption{Left: integrand of the $l_1$-integration in Eq.~\eqref{eq:l1-integration}. Right: the measure factor of $A_{l_1}$ in Eq.~\eqref{eq:l1-integration}. Both plots are given for bulk struts fixed to $m_0 = 2$, $m_1 = 3$, boundary data $l_0 = 10$, $l_2=30$, $\phi_0 = 2$, $\phi_2 = 4$ and scalar field mass $\mu = 0.05$. In the interval $l_1\in [12,40]$, the curve is enlarged to present the behavior close to the saddle points.}
    \label{fig:l1-integrand}
    \end{figure}

The amplitude $A_{l_1}$ and the associated measure are depicted in Fig.~\ref{fig:l1-integrand} for an exemplary set of boundary data and bulk strut length close to the classical solution given in Tab.~\ref{tab:exp vals}. We observe the following properties: 1) The amplitude diverges for $l_1\rightarrow 0$ as $l_1^{-1}$ which follows from the measure $\upmu_{\mathds{1}}$. However, due to our choice of importing the length gap as  $l\in[\frac{1}{2},\infty)$, the amplitude is finite. 2) $A_{l_1}$ is decaying as $l_1^{-33/2}$ for large values of $l_1$ which is advantageous for convergence of the integration, and yields a suppression of the aforementioned divergence of the factor $\upmu_{\phi_1}$. However, the suppression is already significant in the region of the classical solution $l_1^{\mathrm{cl}} = 17.83$ which spoils the resolution of this saddle point. This is a crucial point for pursuing computations. 3) The saddle point extends over a large range of $l_1$ values. This is an immediate consequence of the action which exhibits a plateau around the true solution of the equations of motion, as Fig.~\ref{fig:S tot m} demonstrates. 

\begin{figure}
    \centering
    \includegraphics[width=0.6\linewidth]{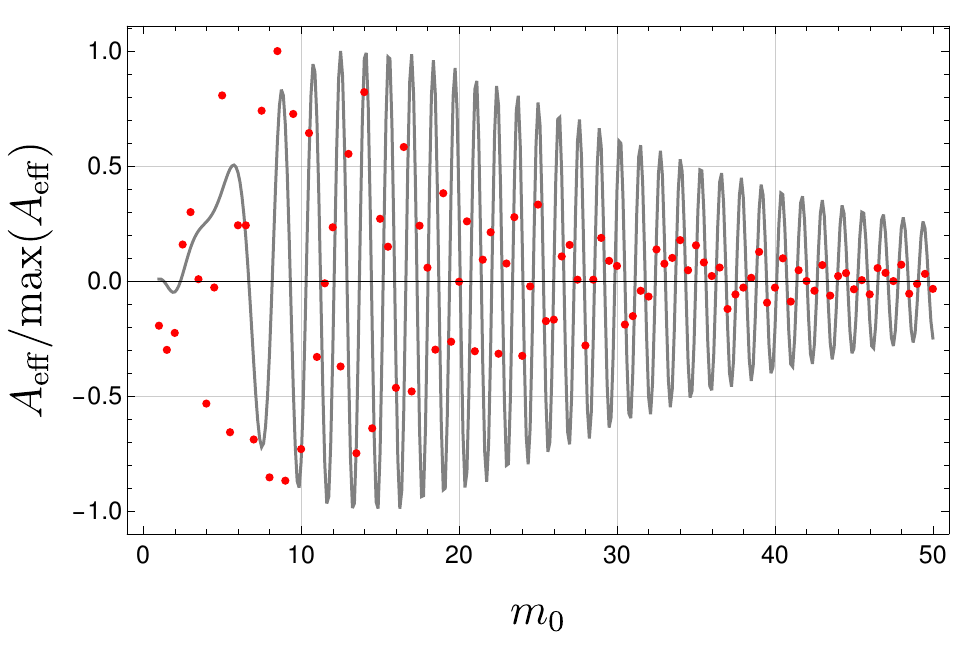}
    \caption{The effective amplitude $A_{\mathrm{eff}}$ normalized by $\mathrm{max}(A_{\mathrm{eff}})$ (red dots), and the amplitude $A_{l_1}$ with solutions $l_1^{\mathrm{cl}}(m_0,m_1)$ inserted and similarly normalized (gray plot). The bulk strut length is fixed close to the classical solution, $m_1 = 3$, the boundary data is given by $l_0 = 10$, $l_2=30$, $\phi_0=2$, $\phi_2=4$, and $\mu=0.05$. The saddle point displayed by the gray plot is not visible for the effective amplitude obtained from the $l_1$-integration.}
    \label{fig:A_eff}
\end{figure}

To investigate the semi-classical properties of the effective amplitude $A_{\mathrm{eff}}$ before computing the sum over the bulk strut length $m_0$ and $m_1$, it can be compared to $A_{l_1}$ with the classical solutions inserted. More precisely, in this setting the Regge equation $\partial S_{\mathrm{tot}}/\partial l_1$ can be solved for $l_1$ as a function of the bulk strut length, yielding $l^{\mathrm{cl}}_1(m_0,m_1)$. Then, $A_{\mathrm{eff}}(m_0,m_1)$ can be compared to $A_{l_1}(l_1^{\mathrm{cl}}(m_0,m_1),m_0,m_1)$ in particular in regard to its saddle points. For fixed boundary data and fixed bulk strut length $m_1 = 3$ close to the classical solution, a comparison of these two amplitudes as a function of $m_0$ is presented in Fig.~\ref{fig:A_eff}. 

The plot in Fig.~\ref{fig:A_eff} shows that $A_{\mathrm{eff}}$ does not display a saddle point at the classical solution $m_0^{\mathrm{cl}}$. Therefore, one can already expect at this point significant deviations from semi-classics. Indeed, this is what we find when computing expectation values of $m_0, m_1$ and $l_1$. The reason for these deviations is point 2) given above: the amplitude $A_{l_1}$ is strongly suppressed in the region of the saddle point. This explanation is supported by the toy model example in Sec.~\ref{sec:A toy model}, where the saddle point region is not suppressed and expectation values comparable to classical solutions are obtained. Thus, we observe here an intricate relation between measure factors and semi-classical physics from discrete gravity path integrals.   

\subsubsection{Summation of bulk strut lengths}\label{sec:Summation of bulk strut lengths}

The remaining task is to compute the sum over the two time-like strut length in the bulk, $m_0, m_1\in\mathbb{N}/2$,
\begin{equation}
Z_{\mathcal{X}_2}(\partial\psi) = \sum_{m_0 \in\mathbb{N}/2}\sum_{m_1 \in\mathbb{N}/2}A_{\mathrm{eff}}(m_0, m_1;\partial\psi)\,.
\end{equation}
To apply the series acceleration techniques introduced in Sec.~\ref{sec:Wynn} to a sum over multiple variables, we introduce a common cutoff $N \leq m_{\max}$ of the strut length variables. There are different schemes to implement this cutoff, such as a triangular one with $m_0+m_1\leq N$ or a rectangular one with $m_0\leq N$ and $m_1\leq N$, characterizing a subset $J_N\subset \mathbb{N}/2\times\mathbb{N}/2$. While the choice of cutoff scheme does not matter in the limit of $N\rightarrow\infty$, it is expected to be relevant for finite $N$. Choosing a subset $J_N$, the residual partition function defines partial sums
\begin{equation}
\mathfrak{S}_N = \sum_{(m_0,m_1)\in J_N}A_{\mathrm{eff}}(m_0,m_1,\partial\psi)\,,
\end{equation}
to which Wynn's algorithm can be applied. Notice that strut length expectation values can be computed by inserting corresponding factors of $m_0$ and $m_1$ into the expression above.

The measure of $A_{\mathrm{eff}}$ falls off as $m^{-7/2}$ for large bulk strut length $m$ and thus exhibits good convergence. In fact, Wynn's $\epsilon$-algorithm shows fast convergence already for values $N = 40$ independent of the cutoff scheme $J_N$. Thus, only a small subset of configurations $(m_0,m_1)$ needs to be probed for $\mathfrak{S}_N$ to converge.

\paragraph{Results.} With the measure of the effective cosmological amplitude, the bulk strut length summation converges and Wynn's $\epsilon$-algorithm yields results that do not depend strongly on the cutoff scheme $J_N$. A summary of the expectation values of bulk strut lengths, the bulk spatial edge length and the bulk scalar field is given in Tab.~\ref{tab:exp vals}. For the numerical simulations, we have chosen $N = 200$, but smaller values of $N$ are sufficient to find convergence.

\begin{table}[]
    \centering
    \begin{tabular}{| c | c | c | c | c | c |}
    \hline
    \multicolumn{2}{| c |}{} & $m_0^{\mathrm{cl}}$ & $m_1^{\mathrm{cl}} $ & $l_1^{\mathrm{cl}} $ & $\phi_1^{\mathrm{cl}} $\\
    \hline
    \multicolumn{2}{| c |}{cl. solutions} & $2.20$ & $3.04$ & $17.83$ & $3.15$\\
    \hline
    \hline
    $\quad \text{amplitude}\quad$   &   $\quad J_N\quad $   &   $\quad \ev{m_0}\quad$   &  $\quad\ev{m_1}\quad$ & $\quad \ev{l_1}\quad$ & $\quad\ev{\phi_1}\quad$ \\
    \hline
    \multirow{2}{*}{$Z_{\mathcal{X}_2}$}  &   triang.  &   $22.05+ 5.06i$  &   $12.76-1.00 i$ & $2.56-1.83i$ & $1.43-2.13i$   \\
      &   rect.  &   $22.00+5.45i$  &   $13.13-0.84i$ & $2.56 -1.81i$ & $0.32-1.01i$ \\
    \multirow{2}{*}{$Z_{\mathcal{X}_2}^{\mathrm{toy}}$}  &  triang.  &   $3.82+0.27i$  &   $3.11 -0.64i$  & $21.26+1.32i$ & $3.27+0.14i$ \\
      &   rect.  &   $3.82+0.31i$  &  $3.12-0.58i$  & $21.26+1.32i$ & $3.27+0.14i$ \\
    \hline
    \end{tabular}
    \caption{Upper part: solutions of the classical equations of motion for the bulk variables $m_0$, $m_1$, $l_1$ and $\phi_1$. Lower part: expectation values of the same bulk variables for the effective partition function $Z_{\mathcal{X}_2}$ and the toy model $Z_{\mathcal{X}_2}^{\mathrm{toy}}$ of Sec.~\ref{sec:A toy model}. $J_N$ denotes the subset of $\{m_0,m_1\}$ used to define the partial sums $\mathfrak{S}_N$ to which Wynn's algorithm is being applied. All the data is computed for boundary values $l_0=10$, $l_2=30$, $\phi_0=2$, $\phi_2=4$ and a scalar field mass $\mu = 0.05$.}
    \label{tab:exp vals}
\end{table}

Comparing the real part of expectation values to the classical solutions we find strong deviations for all observables. More precisely, the values of $\mathfrak{Re}\{\ev{m_0}\}$ and $\mathfrak{Re}\{\ev{m_1}\}$ are significantly larger than the respective classical solutions, and $\mathfrak{Re}\{\ev{l_1}\}$ is much smaller than $l_1^{\mathrm{cl}}$. For all three observables, the triangular and rectangular cutoff schemes of the bulk strut length yield approximately the same result. Also the scalar field value $\mathfrak{Re}\{\ev{\phi_1}\}$ does not agree with the classical solution and its value depends furthermore on the chosen scheme $J_N$. Notice that for $\ev{\phi_1}$ we encountered divergences at large $(m_0,m_1)$ requiring cutting off the sequence at $N' < N$. 

By explicitly computing the expectation values of geometric and matter observables we have found what was already indicated above: the expectation values computed with the effective amplitude $Z_{\mathcal{X}_2}$ do not reproduce classical results despite the presence of saddle points. This is a result of the effective measure which suppresses the region of the $l_1$-integration where the saddle point is located, see Fig.~\ref{fig:l1-integrand}. In particular, one obtains similar results for different mass parameters and boundary data. This demonstrates an intricate interplay between the path integral measure and semi-classical behavior. We remind the reader that the measure utilized here is given by a product of measures obtained from a stationary phase approximation of the single vertex amplitude. Whether the measure obtained from a stationary phase approximation of the full amplitude is such that semi-classical behavior can be obtained from $Z_{\mathcal{X}_2}$ remains as an intriguing open question.

To substantiate the relation between measure and semi-classics we construct in the following section a toy model equipped with a measure that allows to capture the saddle points of the amplitude.

\subsection{A toy model}\label{sec:A toy model}

To verify that the strategy outlined above is \textit{in principle} suited to reproduce semi-classical expectation values, we construct in this section a toy measure with the following properties: 1) finiteness at the minimum values $l_1 = m_0 = m_1 = 0.5$, 2) a sufficiently strong decay at large $l_1, m_0$ and $m_1$, and 3) the saddle point of the $l_1$-integrand can be sufficiently resolved in contrast to what has been observed in Fig.~\ref{fig:l1-integrand}.

\begin{figure}
    \centering
    \begin{subfigure}{0.45\textwidth}
    \includegraphics[width=\linewidth]{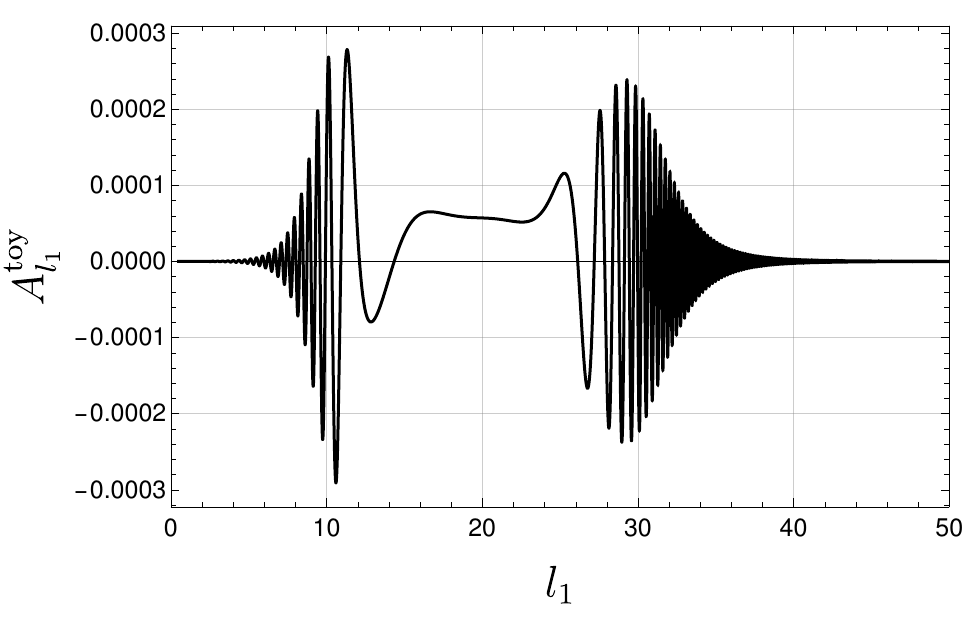} 
    \end{subfigure}\hspace{0.05\textwidth}
    \begin{subfigure}{0.45\textwidth}
    \includegraphics[width=\linewidth]{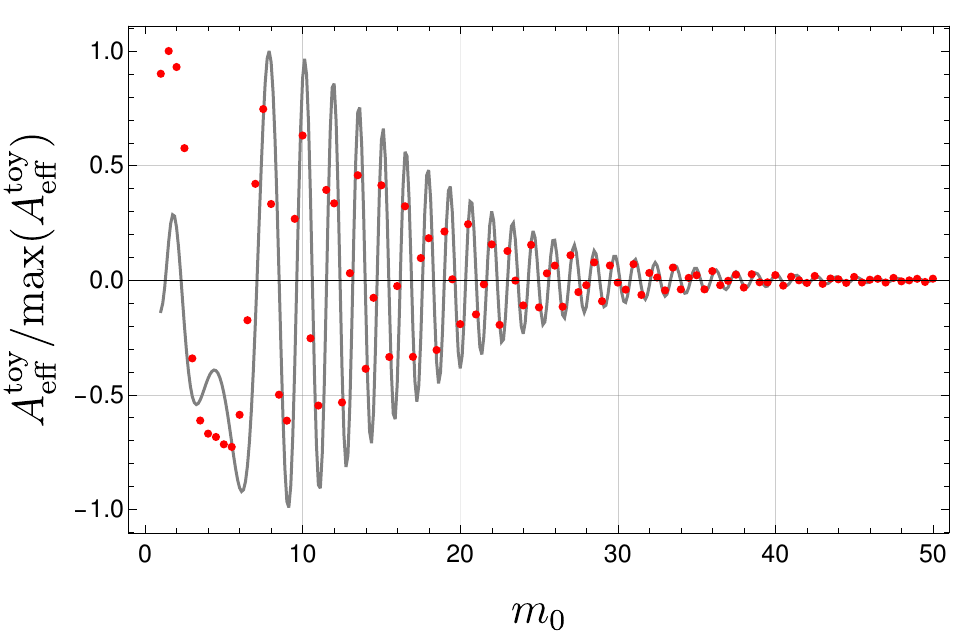}
    \end{subfigure}
    \caption{Left: $l_1$-integrand of the toy model for fixed bulk strut length $m_0 = 2$ and $m_1 = 3$. Right: toy amplitude $A^{\mathrm{toy}}_{\mathrm{eff}}$ obtained from the $l_1$-integration normalized by $\mathrm{max}(A^{\mathrm{toy}}_{\mathrm{eff}})$ (red dots), and the amplitude $A^{\mathrm{toy}}_{l_1}$ with solutions $l_1^{\mathrm{cl}}(m_0,m_1)$ inserted and similarly normalized (gray plot). The other bulk strut length is fixed to $m_1 = 3$. Both graphics are generated for boundary data $l_0=10$, $l_2=30$, $\phi_0=2$, $\phi_2=4$ and a scalar field mass $\mu = 0.05$.}
    \label{fig:toy}
\end{figure}

A measure that satisfies all the above properties is given by
\begin{equation}
\upmu_{\mathrm{toy}} = \frac{m_0l_1}{\left(\frac{(l_0-l_1)^2}{2}+m_0^2\right)^2} \frac{m_1l_1}{\left(\frac{(l_1-l_2)^2}{2}+m_1^2\right)^2}\,,
\end{equation}
factorizing into measure factors associated to the first and second $3$-frustum, respectively. The resulting toy partition function is defined as
\begin{equation}
Z_{\mathcal{X}_2}^{\mathrm{toy}}(\partial\psi) = \sum_{m_0,m_1}\int_{1/2}^{\infty}\dd{l_1}\int_\R\dd{\phi_1}\upmu_{\mathrm{toy}}\e^{iS_{\mathrm{tot}}}\,.
\end{equation}
The scalar field integration is performed as outlined in Sec.~\ref{sec:Integration of bulk scalar field} and the resulting divergent measure factor $\upmu_{\phi_1}$, defined in Eq.~\eqref{eq:mu phi}, does not interfere with the subsequent computation steps. In the left panel of Fig.~\ref{fig:toy}, the $l_1$-integrand is depicted for bulk strut length fixed close to the classical solution. For the toy measure $\upmu_{\mathrm{toy}}$, the saddle point in $l_1$ is clearly visible and not suppressed, in contrast to the case before. The right panel of Fig.~\ref{fig:toy} shows the normalized amplitude after integrating out the spatial bulk edge $l_1$. The saddle point in $m_0$ is resolved and approximately at the same position as the one of the amplitude $A_{l_1}^{\mathrm{toy}}$ with the classical solutions $l_1^{\mathrm{cl}}(m_0,m_1)$ inserted.  

Finally, the summation over the bulk strut length $m_0,m_1$ can be performed using a subset $J_N$ of configurations and applying Wynn's $\epsilon$-algorithm to the resulting partial sums $\mathfrak{S}_N$. We used $N = 200$ for the numerical simulations, but already values of $N=40$ suffice for convergence of the series acceleration. Furthermore, the results of the summation are approximately independent of the cutoff scheme $J_N$.  

\paragraph{Results.} Expectation values of geometric and matter observables are presented in Tab.~\ref{tab:exp vals} for boundary values $l_0 = 10$, $l_2=30$, $\phi_0=2$, $\phi_2=4$, and a scalar field mass $\mu=0.05$. We find that $\mathfrak{Re}\{\ev{m_1}_{\mathrm{toy}}\}$ and $\mathfrak{Re}\{\ev{\phi_1}_{\mathrm{toy}}\}$ are close to the classical solution with a relative deviation $<5\%$. The value of $\mathfrak{Re}\{\ev{m_0}_{\mathrm{toy}}\}$ is substantially larger $(>43\%)$ than the respective classical solution. The spatial edge length expectation value is slightly larger than $l_1^{\mathrm{cl}}$ by approximately $16\%$. The imaginary parts of the expectation values are small compared to the real parts.

In Sec.~\ref{sec:Numerical evaluation: Strut in the bulk}, expectation values of the bulk strut length have shown to deviate from the classical solution either if the spectrum is too coarse to resolve the saddle point or if the classical solution is too close to the length gap. The same argument applies for the deviation of $\mathfrak{Re}\{\ev{m_0}_{\mathrm{toy}}\}$ from the classical solution computed here. 
Given the form of the $l_1$-integrand in the left panel of Fig.~\ref{fig:toy}, an expectation value of $\mathfrak{Re}\{\ev{l_1}_{\mathrm{toy}}\}\approx 20$ is to be expected. Indeed, the observed deviation from $l_1^{\mathrm{cl}}$ arises from a plateau of the total action that is being displayed in Fig.~\ref{fig:S tot m}. Therefore, expectation values of $l_1$ are expected to be closer to the classical solution if the total action exhibits a clear extremum at this point which in turn depends on the boundary data $\partial\psi$ and the mass parameter $\mu$.

The toy example shows that semi-classical physics can be obtained from an effective partition function on an extended cellular complex by following the recipe outlined in Sec.~\ref{sec:Evaluation of the partition function}. We identified the measure of the $l_1$-integration as the most important ingredient for the success of this strategy. While the effective amplitude derived in Sec.~\ref{sec:A proposal for an effective cosmological amplitude} is sufficiently decaying in $m_0,m_1$ and $l_1$ and finite at the respective lower bounds, the saddle points in $l_1$ cannot be sufficiently resolved due to the measure factors. In contrast, using the toy measure $\upmu_{\mathrm{toy}}$ which does not suppress the region of saddle points, expectation values close to classicality are being obtained.

\section{Discussion and conclusion}\label{sec:Discussion}

Building on  the (2+1) coherent spin-foam model introduced in~\cite{Simao:2024don}, we constructed an effective model for Lorentzian quantum cosmology coupled to a massive scalar field. In contrast to previously introduced effective spin-foam models, our model comes equipped with a measure derived from the stationary phase approximation of a single frustum. The asymptotic amplitude does not exhibit exponential suppression of causality violations as it only contains the real part of the Lorentzian Regge action. We have shown that for a single 3-frustum with time-like struts the real part of the strut length expectation value agrees well with the classical solution and is a discontinuous function of the scalar field mass at $\mu = 0$. Despite not exhibiting saddle points, causality violating configurations are generically non-negligible due to the lack of exponential suppression in contrast to effective spin-foam computations. For two glued 3-frusta with time-like struts and a spatial slice in the bulk, the classicality of expectation values is intimately connected to the measure factor; we further proposed a toy measure to address this point, yielding geometric and matter expectation values close to the classical solutions.

Before proceeding with a more extensive discussion of our results, and for the reader's convenience, we first summarize the main points of this article. In short, we have concluded:
\begin{itemize}
    \itemsep0em 
    \item An \textit{effective cosmological model} can be derived from the semi-classical limit of the full quantum (2+1) spin-foam model, accounting for \textit{different causal character configurations}; see Sec.~\ref{sec:model}.
    \item A \textit{scalar field} can be coupled to the model. A \textit{non-vanishing mass $\mu$ is required for the partition function to converge}. Indeed, expectation values are discontinuous at $\mu = 0$. This gives credence to the usefulness of oscillating clock fields, as discussed in \cite{Bojowald:2021uqo}; see Secs.~\ref{sec:Minimally coupled massive scalar field},~\ref{sec:freezing oscillations} and~\ref{sec:mass term regularization}.
    \item \textit{Expectation values} computed from the model on a single frustum agree with \textit{classical solutions when restricted to causally regular configurations}. Including the causally violating sector introduces deviations, conceivably due to the \textit{lack of an exponential suppression mechanism} as in \cite{Dittrich:2023rcr}; see Secs.~\ref{sec:mass term regularization}--\ref{sec:comparison to ESF}.
    \item Crucially, \textit{time-like cells are required for reproducing classical solutions} from expectation values, strengthening the claim that spin-foam models ought to account for different assignments of causal character; see Sec.~\ref{sec:mass term regularization}.
    \item In pushing the model to a \textit{bulk space-like slice, classical solutions are not recovered}. The computed expectation values deviate from classical results, regardless of which causal sectors are considered; see Sec.~\ref{sec:1slice}.
    \item The \textit{measure} of the partition function has proven to be an \textit{essential contributor} to the computed expectation values. This indicates that it is not sufficient to merely consider the behavior of the action exponential in studying the semi-classical limit of spin-foams; see Sec.~\ref{sec:1slice}.
\end{itemize}

Let us delve in more deeply. The starting point of our construction was the (2+1) coherent model~\cite{Simao:2024don}, which we introduced in Sec.~\ref{sec:The vertex, the amplitude and the partition function} for cuboidal combinatorics and arbitrary causal characters of sub-cells. A semi-classical limit of the vertex was discussed in Sec.~\ref{sec:The semi-classical limit of the vertex} forming a crucial ingredient of the effective amplitude constructed thereafter. The results of~\cite{Jercher:2024kig} already hinted at the particular role of causality violating configurations in the effective model: only real-valued deficit angles enter the semi-classical vertex amplitude as otherwise, the critical point equations of the stationary phase approximation are not satisfied. 

A sequence of simplifications of the full spin-foam amplitude was conducted in Sec.~\ref{sec:Spin-foam amplitude simplifications} to construct a numerically feasible effective model. After inserting a resolution of the identity at every face the first modification was performed. This step consisted of fixing the integration domain of the group integrations to effectively performing a symmetry reduction. To ensure semi-classical gluing conditions, the coherent state pairing corresponding to space-like edges were additionally supplemented with the Gaussian constraints introduced in~\cite{Simao:2024don}. As a result, we obtained a product of vertex amplitudes with fixed coherent states peaked on 3-frusta geometries. The last step modification consisted of replacing the quantum amplitude by its semi-classical form. 

Thereafter, we analyzed the measure factor of the amplitude arising as the inverse square root of the Hessian determinant from the stationary phase approximation. Subsequently, in Secs.~\ref{sec:Minimally coupled massive scalar field} and~\ref{sec:Effective cosmological partition function}, we minimally coupled a massive scalar field and finalized the setup of the effective cosmological partition function. We concluded Sec.~\ref{sec:A proposal for an effective cosmological amplitude} with a discussion of causality violations, demonstrating that these are present in the effective model and that they are not accompanied by a complex Regge action but rather its real part. We investigated the consequences thereof in Secs.~\ref{sec:including I and II} and~\ref{sec:comparison to ESF}.

In Sec.~\ref{sec:Numerical evaluation: Strut in the bulk}, we numerically evaluated the effective cosmological partition function for a single 3-frustum, using numerical integration and the acceleration techniques summarized in Sec.~\ref{sec:Wynn}, which have been previously applied in~\cite{Dittrich:2023rcr}. For vanishing scalar field mass $\mu = 0$, vanishing cosmological constant and spatial flatness, the causally regular partition function does not converge due to an asymptotically stationary action, as to be expected also in the continuum. Working instead with a non-vanishing mass $\mu$ which acts as a regulator of the effective path integral, we studied the strut length expectation value $\expval{m}_{\textsc{iii}}$ for varying boundary data and scalar field mass in the causally regular sector. A good agreement of the real part of $\expval{m}_{\textsc{iii}}$ with the classical solutions is obtained in the regime where the saddle point of the amplitude can be sufficiently resolved with the discrete spectrum $m\in\mathbb{N}/2$. Outside this regime, i.e. when the classical solution is below the length gap $m = 1/2$ or insufficiently many points are probing the saddle point, better agreement with classicality would require a refinement of the length spectrum or a smaller gap, as already pointed out in~\cite{Dittrich:2023rcr}. The imaginary part of the expectation value shows a tendency towards the value $-\frac{1}{2}$ and is in particular non-vanishing. We will elaborate on this point again below. We explicitly demonstrated that $\expval{m}_{\textsc{iii}}$ is a discontinuous function of the mass $\mu$ since it diverges at $\mu = 0$ but remains finite in the limit $\mu\rightarrow 0^+$. 

The results of Sec.~\ref{sec:including I and II} demonstrate that due to the lack of a suppression mechanism, causality violating configurations can spoil the expectation value of the strut length, driving it away from the classical solution. This result is corroborated by the effective spin-foam computations of Sec.~\ref{sec:comparison to ESF}, where an exponential suppression of these configurations exists, rendering their contributions to $\expval{m}_{\textsc{iii}}$ negligible.

The last section of this article was concerned with evaluating the partition function for two glued 3-frusta with time-like struts containing a spatial slice in the bulk. We outlined a strategy to compute $Z_{\mathcal{X}_2}$, starting with an analytical integration of the scalar field variable $\phi_1$. The next step consisted of the spatial edge length integration yielding an effective amplitude that depends on the two bulk strut length $(m_0,m_1)$. The remaining infinite summations were performed by adapting series acceleration techniques to the case of multiple variables. We found that with the amplitude constructed in Sec.~\ref{sec:A proposal for an effective cosmological amplitude}, the outlined strategy converged. However, strong deviations between expectation values and classical solutions emerge. As the main reason for these deviations, we identified the measure factor entering the $l_1$-integration which leads to a strong suppression of the saddle points in $l_1$. The resulting effective amplitude does not exhibit a saddle point in the bulk strut lengths $m_0$ and $m_1$ and thus, an agreement of expectation values and classical solutions is not being reproduced. This result suggests that semi-classicality in this sense sensitively depends on the path integral measure. We strengthened this interpretation by constructing a toy measure that does not suppress the saddle points in $l_1$, and the resulting expectation values showed indeed agreement with classical solutions.  
\\
\\
Effective spin-foam models~\cite{Asante:2021zzh,Asante:2020iwm,Asante:2020qpa} have proven numerically feasible and suitable to study physically interesting scenarios such as cosmology~\cite{Dittrich:2023rcr,Asante:2021phx}, the de Sitter horizon entropy~\cite{Dittrich:2024awu} or the role of spike configurations in Lorentzian spin-foams~\cite{Borissova:2024pfq,Borissova:2024txs}. These models are however proposed ad hoc, leaving it as an open question if and how they can be obtained from fundamental spin-foam models. In the cosmological setting, this question motivated the construction and investigation of the present model. Tentatively, the results of this work suggest that while causality violating configurations generically appear in the semi-classical limit, their exponential suppression (or enhancement) does not emerge. In the (2+1) model considered here, that is because only the real part of deficit angles is obtained. Of course, our argument is limited to the (2+1)-dimensional symmetry reduced setting. Furthermore, the semi-classical limit was taken at each spin-foam vertex locally. Thus, spin-foams may still possess a mechanism for suppressing causality violations which may be found for 2-complexes consisting of multiple vertices and bulk edges. This constitutes an intriguing new research direction, which can be explored in the asymptotic regime via complex critical points~\cite{Han:2021kll,Han:2023cen} and for the full calculation via numerical methods like \textsc{sl2cfoam-next}~\cite{Dona:2019dkf,Gozzini:2021kbt,Dona:2022dxs}; for the latter recent advances in numerical methods~\cite{Dona:2023myv,Steinhaus:2024qov,Asante:2024eft} may prove to be vital.

The expectation values computed in Secs.~\ref{sec:Numerical evaluation: Strut in the bulk} and~\ref{sec:1slice} generically carry a non-vanishing imaginary part. This is a consequence of utilizing complex amplitudes and has been deemed a quantum effect in~\cite{Dittrich:2023rcr}. However, a physical interpretation of these imaginary parts remains as an open question. In particular, it would be interesting to relate the expectation values of the path integral to expectation values obtained in an operator-state picture. In the latter, the length operator is given by the $\SUO$-Casimir, acting diagonally on spin network states and carrying real eigenvalues. The definition of expectation values in Sec.~\ref{sec:Effective cosmological partition function} fits into the general boundary framework by Oeckl~\cite{Oeckl:2003vu,Oeckl:2005bv,Oeckl:2006rs,Oeckl:2011qd}. It is noted in~\cite{Oeckl:2011qd} that real expectation values may be obtained by a suitable choice of boundary states, leaving their definition for the present setting as an interesting future research avenue.  

Time-like building blocks proved in this article to be a crucial ingredient in identifying a causally regular sector which exhibits saddle points at classical solutions. Their importance has been noticed in the context of Lorentzian Regge calculus~\cite{Dittrich:2021gww,Jercher:2023csk} and complex critical point computations~\cite{Han:2024ydv}. Also in the context of GFT condensate cosmology, time-like polyhedra were required to obtain GR-like cosmological perturbations and to define a relational Lorentzian reference frame~\cite{Jercher:2023nxa,Jercher:2023kfr}. We hope that these results strengthen the arguments for studying models that incorporate a causally complete set of discrete Lorentzian geometries. 

The results of Sec.~\ref{sec:1slice} present a promising outlook for future investigations on cosmological partition functions with spatial slices in the bulk. Despite the restriction to causal regularity and the simplification of the measure, the partition function 
$Z_{\mathcal{X}_2}^{\mathrm{toy}}$ can be utilized to study physically and conceptually interesting questions. These include a bouncing scenario, where initial and final spatial edges on the boundary are equal but boundary scalar field values evolve. Although there are no classical solutions for this boundary data, the partition function as well as expectation values can be computed. Such computations could potentially reveal a quantum bounce, connecting a contracting and expanding branch of the universe and yielding a non-zero transition probability in the sense of Oeckl's general boundary formulation. Conceptually interesting is also the role of the scalar field as a relational clock. In particular, the partition function 
$Z_{\mathcal{X}_2}^{\mathrm{toy}}$ offers the possibility to study under which conditions clock fluctuations dominate. Also, the influence of the scalar field mass on its properties as a clock can be studied. In this case, results from a path integral formulation of quantum cosmology can be compared to the results of~\cite{Bojowald:2021uqo,Martinez:2023fsd} in a Hamiltonian framework.

\paragraph{Acknowledgements}  The authors thank Seth Asante, Bianca Dittrich and Steffen Gielen for helpful discussions and Tim St\"{o}tzel for support with the numerical implementation. AFJ, JDS and SSt gratefully acknowledge support by the Deutsche Forschungsgemeinschaft (DFG, German Research Foundation) Grant No 422809950. AFJ and JDS acknowledge support by the DFG under Grant No 406116891 within the Research Training Group RTG 2522/1. AFJ is grateful for the generous financial support by the MCQST via the seed funding Aost 862981-8 granted to Jibril Ben Achour by the DFG under Germany’s Excellence Strategy – EXC-2111 – 390814868.

\appendix
\section{Facts and Conventions on SU(1,1)} 
\label{app:su11}

\subsection{Unitary irreducible representations}

\noindent The unitary irreducible representations of $\mathrm{SU}(1,1)$ were classified for the first time by Bargmann in \cite{Bargmann:1946me}, and the analysis of generalized eigenstates was carried out by Lindblad in \cite{Lindblad:1969zz}. The following collection of facts is lifted directly from the two authors; we follow the conventions of the latter. 

The $\mathfrak{su}(1,1)$ algebra is spanned by the generators $F^i=(L^3, K^1, K^2)$, defined in the fundamental representation with the standard Pauli matrices as $\varsigma^i/2=(\sigma_3/2, i\sigma_2/2, -i\sigma_1/2)$. The respective subgroups read
\begin{equation}
  \begin{gathered}
  e^{i \alpha L^3}=\begin{pmatrix}
    e^{i \frac{\alpha}{2}} & 0 \\
    0 & e^{-i \frac{\alpha}{2}}
  \end{pmatrix}\,, \quad   e^{i t K^1}=\begin{pmatrix}
    \cosh \frac{t}{2} & i\sinh\frac{t}{2} \\
    -i\sinh\frac{t}{2} & \cosh \frac{t}{2}
  \end{pmatrix}\,, \\
    e^{i u K^2}=\begin{pmatrix}
     \cosh \frac{u}{2} & \sinh\frac{u}{2} \\
    \sinh\frac{u}{2} & \cosh \frac{u}{2}
  \end{pmatrix}\,,
\end{gathered}
\end{equation}
and the Casimir element is given by $Q=(L^3)^2-(K^1)^2-(K^2)^2$. There are two families of unitary irreducible representations characterized by the eigenvalues of $Q$ and $L^3$, called the \textit{discrete} and \textit{continuous series}. Regarding the first, the Hilbert space $\mathcal{D}^q_k$ is spanned by the orthonormal states
\begin{equation}
  \begin{gathered}
  Q\ket{k,m}=k(k+1)\ket{k,m}\,, \quad  k \in -\frac{\mathbb{N}}{2}\,,\\
  L^3 \ket{k,m}=m\ket{k,m}\,, \quad m\in q(- k  + \mathbb{N}^0)\,, \quad q=\pm\,.
  \end{gathered}
\end{equation}
For the continuous series, an orthonormal basis for the Hilbert space $\mathcal{C}^\delta_j$ is given by 
\begin{equation}
  \begin{gathered}
  Q\ket{j,m}=j(j+1)\ket{j,m}\,, \quad j=-\frac{1}{2}+is\,,\quad s\in\mathbb{R}^+\,,\\
  L^3 \ket{j,m}=m\ket{j,m}\,, \quad m\in \delta + \mathbb{Z}\,, \quad \delta\in\{0,\frac{1}{2}\}\,. 
  \end{gathered}
\end{equation}
An alternative orthonormal basis of $\mathcal{C}^\delta_j$ can be obtained from generalized eigenstates of the non-compact operator $K^2$. The eigenstates satisfy
\begin{equation}
  \begin{gathered}
  Q\ket{j,\lambda,\sigma}=j(j+1)\ket{j,\lambda,\sigma}\,, \quad j=-\frac{1}{2}+is\,,\quad s\in\mathbb{R}^+\,,\\
  K^2 \ket{j,\lambda,\sigma}=\lambda \ket{j,m}\,, \quad \lambda \in \mathbb{C}\,, \\
  P \ket{j,\lambda,\sigma}=(-1)^\sigma \ket{j,\lambda,\sigma}\,, \quad \sigma\in\{0,1\}\,,
  \end{gathered}
\end{equation} 
where $P$ is an outer automorphism of the Lie algebra taking $(L^3, K^1, K^2)\mapsto (-L^3, -K^1, K^2)$. They are complete and orthonormal in the sense that
\begin{equation}
  \begin{gathered}
  \sum_\sigma \int_{\mathbb{R}+i\alpha} \diff\lambda \,  \braket{j,m|j, \lambda, \sigma} \braket{j, \overline{\lambda}, \sigma|j,n}=\delta_{m,n}\,, \quad \alpha \in \mathbb{R}\,, \\
  \sum_m  \braket{j, \overline{\lambda'}, \sigma'|j,m} \braket{j,m|j, \lambda, \sigma} = \delta(\lambda-\lambda')\,, \quad \Im \lambda =\Im \lambda'\,,
  \end{gathered}
\end{equation}
and indeed there is a family of bases $\left\{\, \ket{j, \lambda+i \alpha, \sigma} \, | \, \lambda \in \mathbb{R}\,, \sigma \in \{0,1\}\right\}_{\alpha}$ indexed by $\alpha \in \mathbb{R}$. 

\subsection{Parametrization and Haar measure} 
\label{sec:parsu11}

Among the possible parametrizations of the group the following (coherently normalized) are useful for this work~\cite{Lindblad:1970tv,Ruehl1970}
\begin{enumerate}[label=\arabic*), leftmargin=*, itemsep=0.4ex, before={\everymath{\displaystyle}}]
\item
$
  g=e^{i \alpha L^3} e^{i t K^1} e^{i u K^2}\,, \quad \diff g= (4\pi)^{-2} \cosh t \, \diff \alpha\, \diff t \, \diff u\,,
$

with $0 \leq \alpha < 4\pi\,,\; -\infty<t,u<\infty$;
\item
$
  g=e^{i \alpha L^3} e^{i u K^2} e^{i \beta L^3}\,, \quad \diff g= (4\pi)^{-2} \sinh u \, \diff \alpha\, \diff t \, \diff \beta \,,
$

with $ 0 \leq \alpha < 4\pi\,, \;  0\leq\beta<2\pi\,, \;  u\geq 0$;
\item \label{param3}
$
g=\begin{pmatrix}
  \alpha & \beta \\ \overline{\beta} & \overline{\alpha}
\end{pmatrix}\,, \quad   \diff g= \pi^{-2} \delta(|\alpha|^2-|\beta|^2-1) D\alpha D\beta\,, \quad \\ D\alpha=\frac{i}{2}\diff \alpha \wedge \diff \overline{\alpha}\,,\quad \mathrm{\textit{m.m.}} \, D\beta\,;
$
\item
$
g=\begin{pmatrix}
  \alpha & \beta \\ \overline{\beta} & \overline{\alpha}
\end{pmatrix}\,, \quad \diff g=(2\pi)^{-2} |c|^{-1}\diff a\,\diff b\,\diff c\,,
$

with $a=\frac{(\alpha-\overline{\alpha})-(\beta-\overline{\beta})}{2i}\,, \; b=\frac{(\overline{\alpha}-\alpha)+(\overline{\beta}-\beta)}{2i}\,, \; c=\frac{(\alpha+\overline{\alpha})-(\beta+\overline{\beta})}{2}\,.
$
\end{enumerate}

\subsection{Harmonic analysis}

There exists a Plancherel formula for $L^2(\mathrm{SU}(1,1))$. According to \cite{Pukanszky:1963,Takahashi:1961,Ruehl1970}, given any function $f\in \mathcal{C}_0^\infty$,
\begin{equation}
\label{idexp}
\begin{gathered}
  f(\mathds{1})=\sum_\delta \int_{-\infty}^\infty \diff s\, s\tanh(\pi s)^{1-4\delta} \mathrm{Tr}_{\mathcal{C}^\delta_j}\left[\int_{\mathrm{SU}(1,1)}\diff g\, f(g) D^{j}(g)  \right] \\
  +\sum_{q} \sum_{2k=-1}^{-\infty} (-2k-1) \mathrm{Tr}_{\mathcal{D}^q_k}\left[\int_{\mathrm{SU}(1,1)}\diff g\, f(g) D^{k}(g)  \right]\,,
  \end{gathered}
\end{equation}
where the Haar measure is as above. Observe that the lowest $k=-\frac{1}{2}$ discrete representation is absent from the decomposition of $f$. Setting $f(g):=\int\diff h \overline{f}_1(h) f_2(gh)$ it follows from \eqref{idexp} that
\begin{equation}
\begin{gathered}
  \int \diff g \, \overline{f}_1(g) f_2(g)=\sum_\delta \int_{-\infty}^\infty \diff s\, s\tanh(\pi s)^{1-4\delta} (\overline{c}_1)^{j(\delta)}_{\lambda\sigma,\lambda'\sigma'} (c_2)^{j(\delta)}_{\lambda\sigma,\lambda'\sigma'}  \\
   +\sum_{q} \sum_{2k=-1}^{-\infty} (-2k-1) (\overline{c}_1)^{k(q)}_{m m'} (c_2)^{k(q)}_{m m'}\,, 
\end{gathered}
\end{equation}
where all lower repeated indices are appropriately contracted, and the Fourier coefficients read
\begin{equation}
\begin{gathered}
  (c_i)^{k(q)}_{m m'}= \int\diff g \, f_i(g) D^{k(q)}_{mm'}(g)\,, \\
  (c_i)^{j(\delta)}_{\lambda\sigma,\lambda'\sigma'}= \int\diff g \, f_i(g) D^{j(\delta)}_{\lambda\sigma, \lambda' \sigma'}(g)\,.
\end{gathered}
\end{equation}
The coefficients for the continuous series could of course also be written with respect to the $L^3$ eigenbasis. 
 Yet another consequence of equation \eqref{idexp} are the orthogonality relations
\begin{align}
  &\int \diff g \, \overline{D}^{k(q)}_{m n}(g) D^{k'(q')}_{m' n'}(g) = \frac{\delta_{q,q'} \delta_{k,k'} }{-2k-1}\;   \delta_{m,m'} \delta_{n,n'}\,, \\
    &\int \diff g \, \overline{D}^{j(\delta)}_{m,n}(g) D^{j'(\delta')}_{m' n'}(g)= \frac{\delta_{\delta,\delta'} \delta(j-j')}{s\tanh(\pi s)^{1-4\delta}}\;  \delta_{m,m'} \delta_{n,n'}\,,\\
  &\int \diff g \, \overline{D}^{j(\delta)}_{\lambda\sigma,\mu \epsilon}(g) D^{j'(\delta')}_{\lambda' \sigma', \mu' \epsilon '}(g) = \frac{ \delta_{\delta,\delta'} \delta(j-j')}{s\tanh(\pi s)^{1-4\delta}}\; \delta(\lambda-\lambda') \delta(\mu-\mu') \,,
  \label{coefforto}
\end{align}
which hold for the matrix coefficients of the discrete and continuous series in the $L^3$ eigenbasis, and the coefficients of the continuous series in the $K^2$ eigenbasis, respectively.

\bibliographystyle{JHEP}
\bibliography{references}

\providecommand{\href}[2]{#2}\begingroup\raggedright\begin{thebibliography}{100}

\bibitem{Ashtekar:2021kfp}
A.~Ashtekar and E.~Bianchi, \emph{{A short review of loop quantum gravity}}, \href{https://doi.org/10.1088/1361-6633/abed91}{\emph{Rept. Prog. Phys.} {\bfseries 84} (2021) 042001} [\href{https://arxiv.org/abs/2104.04394}{{\ttfamily 2104.04394}}].

\bibitem{Calcagni:2019kzo}
G.~Calcagni, S.~Kuroyanagi, S.~Marsat, M.~Sakellariadou, N.~Tamanini and G.~Tasinato, \emph{{Gravitational-wave luminosity distance in quantum gravity}}, \href{https://doi.org/10.1016/j.physletb.2019.135000}{\emph{Phys. Lett. B} {\bfseries 798} (2019) 135000} [\href{https://arxiv.org/abs/1904.00384}{{\ttfamily 1904.00384}}].

\bibitem{Perez:2013uz}
A.~P{\'e}rez, \emph{{The Spin-Foam Approach to Quantum Gravity}}, {\emph{Living Rev. Relativity} {\bfseries 16} (2013) 3} [\href{https://arxiv.org/abs/1205.2019}{{\ttfamily 1205.2019}}].

\bibitem{Oriti:2006ts}
D.~Oriti, \emph{{The group field theory approach to quantum gravity}},  in \emph{Approaches to Quantum Gravity: Toward a New Understanding of Space, Time and Matter}, D.~Oriti, ed., (Cambridge, UK), Cambridge University Press (2007) [\href{https://arxiv.org/abs/gr-qc/0607032}{{\ttfamily gr-qc/0607032}}].

\bibitem{Freidel:2005qe}
L.~Freidel, \emph{{Group field theory: An Overview}}, \href{https://doi.org/10.1007/s10773-005-8894-1}{\emph{Int. J. Theor. Phys.} {\bfseries 44} (2005) 1769} [\href{https://arxiv.org/abs/hep-th/0505016}{{\ttfamily hep-th/0505016}}].

\bibitem{Marchetti:2020umh}
L.~Marchetti and D.~Oriti, \emph{{Effective relational cosmological dynamics from Quantum Gravity}}, \href{https://doi.org/10.1007/JHEP05(2021)025}{\emph{JHEP} {\bfseries 05} (2021) 025} [\href{https://arxiv.org/abs/2008.02774}{{\ttfamily 2008.02774}}].

\bibitem{Oriti:2016qtz}
D.~Oriti, L.~Sindoni and E.~Wilson-Ewing, \emph{{Emergent Friedmann dynamics with a quantum bounce from quantum gravity condensates}}, \href{https://doi.org/10.1088/0264-9381/33/22/224001}{\emph{Class. Quant. Grav.} {\bfseries 33} (2016) 224001} [\href{https://arxiv.org/abs/1602.05881}{{\ttfamily 1602.05881}}].

\bibitem{Gielen:2016dss}
S.~Gielen and L.~Sindoni, \emph{{Quantum Cosmology from Group Field Theory Condensates: a Review}}, \href{https://doi.org/10.3842/SIGMA.2016.082}{\emph{SIGMA} {\bfseries 12} (2016) 082} [\href{https://arxiv.org/abs/1602.08104}{{\ttfamily 1602.08104}}].

\bibitem{Jercher:2021bie}
A.F.~Jercher, D.~Oriti and A.G.A.~Pithis, \emph{{Emergent cosmology from quantum gravity in the Lorentzian Barrett-Crane tensorial group field theory model}}, \href{https://doi.org/10.1088/1475-7516/2022/01/050}{\emph{JCAP} {\bfseries 01} (2022) 050} [\href{https://arxiv.org/abs/2112.00091}{{\ttfamily 2112.00091}}].

\bibitem{Agullo:2016tjh}
I.~Agullo and P.~Singh, \emph{{Loop Quantum Cosmology}},  in \emph{{Loop Quantum Gravity}: {The First 30 Years}}, A.~Ashtekar and J.~Pullin, eds., pp.~183--240, WSP (2017), \href{https://doi.org/10.1142/9789813220003_0007}{DOI} [\href{https://arxiv.org/abs/1612.01236}{{\ttfamily 1612.01236}}].

\bibitem{Bahr:2017bn}
B.~Bahr, S.~Kl{\"o}ser and G.~Rabuffo, \emph{{Towards a Cosmological subsector of Spin Foam Quantum Gravity}}, {\emph{Phys. Rev. D} {\bfseries 96} (2017) 086009} [\href{https://arxiv.org/abs/1704.03691}{{\ttfamily 1704.03691}}].

\bibitem{Bianchi:2010ej}
E.~Bianchi, C.~Rovelli and F.~Vidotto, \emph{{Towards spinfoam cosmology}}, {\emph{Phys. Rev. D} {\bfseries 82} (2010) 084035} [\href{https://arxiv.org/abs/1003.3483}{{\ttfamily 1003.3483}}].

\bibitem{Dittrich:2023rcr}
B.~Dittrich and J.~Padua-Arg\"uelles, \emph{{Lorentzian Quantum Cosmology from Effective Spin Foams}}, \href{https://doi.org/10.3390/universe10070296}{\emph{Universe} {\bfseries 10} (2024) 296} [\href{https://arxiv.org/abs/2306.06012}{{\ttfamily 2306.06012}}].

\bibitem{Han:2024ydv}
M.~Han, H.~Liu, D.~Qu, F.~Vidotto and C.~Zhang, \emph{{Cosmological Dynamics from Covariant Loop Quantum Gravity with Scalar Matter}},  \href{https://arxiv.org/abs/2402.07984}{{\ttfamily 2402.07984}}.

\bibitem{Engle:2007em}
J.~Engle, R.~Pereira and C.~Rovelli, \emph{{Loop-Quantum-Gravity Vertex Amplitude}}, {\emph{Phys. Rev. Lett.} {\bfseries 99} (2007) 161301} [\href{https://arxiv.org/abs/0705.2388}{{\ttfamily 0705.2388}}].

\bibitem{Engle:2008fj}
J.~Engle, E.R.~Livine, R.~Pereira and C.~Rovelli, \emph{{LQG vertex with finite Immirzi parameter}}, {\emph{Nucl. Phys. B} {\bfseries 799} (2008) 136} [\href{https://arxiv.org/abs/0711.0146}{{\ttfamily 0711.0146}}].

\bibitem{Bianchi:2011ym}
E.~Bianchi, T.~Krajewski, C.~Rovelli and F.~Vidotto, \emph{{Cosmological constant in spinfoam cosmology}}, \href{https://doi.org/10.1103/PhysRevD.83.104015}{\emph{Phys. Rev. D} {\bfseries 83} (2011) 104015} [\href{https://arxiv.org/abs/1101.4049}{{\ttfamily 1101.4049}}].

\bibitem{Rennert:2013pfa}
J.~Rennert and D.~Sloan, \emph{{A Homogeneous Model of Spinfoam Cosmology}}, \href{https://doi.org/10.1088/0264-9381/30/23/235019}{\emph{Class. Quant. Grav.} {\bfseries 30} (2013) 235019} [\href{https://arxiv.org/abs/1304.6688}{{\ttfamily 1304.6688}}].

\bibitem{Sarno:2018ses}
G.~Sarno, S.~Speziale and G.V.~Stagno, \emph{{2-vertex Lorentzian Spin Foam Amplitudes for Dipole Transitions}}, \href{https://doi.org/10.1007/s10714-018-2360-x}{\emph{Gen. Rel. Grav.} {\bfseries 50} (2018) 43} [\href{https://arxiv.org/abs/1801.03771}{{\ttfamily 1801.03771}}].

\bibitem{Gozzini:2019nbo}
F.~Gozzini and F.~Vidotto, \emph{{Primordial Fluctuations From Quantum Gravity}}, \href{https://doi.org/10.3389/fspas.2020.629466}{\emph{Front. Astron. Astrophys. Cosmol.} {\bfseries 7} (2021) 629466} [\href{https://arxiv.org/abs/1906.02211}{{\ttfamily 1906.02211}}].

\bibitem{Frisoni:2022urv}
P.~Frisoni, F.~Gozzini and F.~Vidotto, \emph{{Markov chain Monte Carlo methods for graph refinement in spinfoam cosmology}}, \href{https://doi.org/10.1088/1361-6382/acc5d6}{\emph{Class. Quant. Grav.} {\bfseries 40} (2023) 105001} [\href{https://arxiv.org/abs/2207.02881}{{\ttfamily 2207.02881}}].

\bibitem{Frisoni:2023lvb}
P.~Frisoni, F.~Gozzini and F.~Vidotto, \emph{{Primordial fluctuations from quantum gravity: 16-cell topological model}},  \href{https://arxiv.org/abs/2312.02399}{{\ttfamily 2312.02399}}.

\bibitem{Han:2021kll}
M.~Han, Z.~Huang, H.~Liu and D.~Qu, \emph{{Complex critical points and curved geometries in four-dimensional Lorentzian spinfoam quantum gravity}}, \href{https://doi.org/10.1103/PhysRevD.106.044005}{\emph{Phys. Rev. D} {\bfseries 106} (2022) 044005} [\href{https://arxiv.org/abs/2110.10670}{{\ttfamily 2110.10670}}].

\bibitem{Han:2023cen}
M.~Han, H.~Liu and D.~Qu, \emph{{Complex critical points in Lorentzian spinfoam quantum gravity: Four-simplex amplitude and effective dynamics on a double-\ensuremath{\Delta}3 complex}}, \href{https://doi.org/10.1103/PhysRevD.108.026010}{\emph{Phys. Rev. D} {\bfseries 108} (2023) 026010} [\href{https://arxiv.org/abs/2301.02930}{{\ttfamily 2301.02930}}].

\bibitem{Han:2024lti}
M.~Han, H.~Liu and D.~Qu, \emph{{A Mathematica program for numerically computing real and complex critical points in 4-dimensional Lorentzian spinfoam amplitude}},  \href{https://arxiv.org/abs/2404.10563}{{\ttfamily 2404.10563}}.

\bibitem{Conrady:2010kc}
F.~Conrady and J.~Hnybida, \emph{{A spin foam model for general Lorentzian 4-geometries}}, \href{https://doi.org/10.1088/0264-9381/27/18/185011}{\emph{Class. Quant. Grav.} {\bfseries 27} (2010) 185011} [\href{https://arxiv.org/abs/1002.1959}{{\ttfamily 1002.1959}}].

\bibitem{Conrady:2010vx}
F.~Conrady, \emph{{Spin foams with timelike surfaces}}, \href{https://doi.org/10.1088/0264-9381/27/15/155014}{\emph{Class. Quant. Grav.} {\bfseries 27} (2010) 155014} [\href{https://arxiv.org/abs/1003.5652}{{\ttfamily 1003.5652}}].

\bibitem{Jercher:2023rno}
A.F.~Jercher, S.~Steinhaus and J.~Th\"urigen, \emph{{Curvature effects in the spectral dimension of spin foams}}, \href{https://doi.org/10.1103/PhysRevD.108.066011}{\emph{Phys. Rev. D} {\bfseries 108} (2023) 066011} [\href{https://arxiv.org/abs/2304.13058}{{\ttfamily 2304.13058}}].

\bibitem{Bahr:2015gxa}
B.~Bahr and S.~Steinhaus, \emph{{Investigation of the Spinfoam Path integral with Quantum Cuboid Intertwiners}}, \href{https://doi.org/10.1103/PhysRevD.93.104029}{\emph{Phys. Rev. D} {\bfseries 93} (2016) 104029} [\href{https://arxiv.org/abs/1508.07961}{{\ttfamily 1508.07961}}].

\bibitem{Bahr:2016dl}
B.~Bahr and S.~Steinhaus, \emph{{Numerical Evidence for a Phase Transition in 4D Spin-Foam Quantum Gravity}}, {\emph{Phys. Rev. Lett.} {\bfseries 117} (2016) } [\href{https://arxiv.org/abs/1605.07649}{{\ttfamily 1605.07649}}].

\bibitem{Bahr:2017klw}
B.~Bahr and S.~Steinhaus, \emph{{Hypercuboidal renormalization in spin foam quantum gravity}}, \href{https://doi.org/10.1103/PhysRevD.95.126006}{\emph{Phys. Rev. D} {\bfseries 95} (2017) 126006} [\href{https://arxiv.org/abs/1701.02311}{{\ttfamily 1701.02311}}].

\bibitem{Assanioussi:2020fml}
M.~Assanioussi and B.~Bahr, \emph{{Hopf link volume simplicity constraints in spin foam models}}, \href{https://doi.org/10.1088/1361-6382/abb117}{\emph{Class. Quant. Grav.} {\bfseries 37} (2020) 205003} [\href{https://arxiv.org/abs/2005.12004}{{\ttfamily 2005.12004}}].

\bibitem{Allen:2022unb}
C.~Allen, F.~Girelli and S.~Steinhaus, \emph{{Numerical evaluation of spin foam amplitudes beyond simplices}}, \href{https://doi.org/10.1103/PhysRevD.105.066003}{\emph{Phys. Rev. D} {\bfseries 105} (2022) 066003} [\href{https://arxiv.org/abs/2201.09902}{{\ttfamily 2201.09902}}].

\bibitem{Asante:2021zzh}
S.K.~Asante, B.~Dittrich and J.~Padua-Arguelles, \emph{{Effective spin foam models for Lorentzian quantum gravity}}, \href{https://doi.org/10.1088/1361-6382/ac1b44}{\emph{Class. Quant. Grav.} {\bfseries 38} (2021) 195002} [\href{https://arxiv.org/abs/2104.00485}{{\ttfamily 2104.00485}}].

\bibitem{Asante:2020iwm}
S.K.~Asante, B.~Dittrich and H.M.~Haggard, \emph{{Discrete gravity dynamics from effective spin foams}}, \href{https://doi.org/10.1088/1361-6382/ac011b}{\emph{Class. Quant. Grav.} {\bfseries 38} (2021) 145023} [\href{https://arxiv.org/abs/2011.14468}{{\ttfamily 2011.14468}}].

\bibitem{Asante:2020qpa}
S.K.~Asante, B.~Dittrich and H.M.~Haggard, \emph{{Effective Spin Foam Models for Four-Dimensional Quantum Gravity}}, \href{https://doi.org/10.1103/PhysRevLett.125.231301}{\emph{Phys. Rev. Lett.} {\bfseries 125} (2020) 231301} [\href{https://arxiv.org/abs/2004.07013}{{\ttfamily 2004.07013}}].

\bibitem{Barrett:1997tx}
J.W.~Barrett, M.~Rocek and R.M.~Williams, \emph{{A Note on area variables in Regge calculus}}, \href{https://doi.org/10.1088/0264-9381/16/4/025}{\emph{Class. Quant. Grav.} {\bfseries 16} (1999) 1373} [\href{https://arxiv.org/abs/gr-qc/9710056}{{\ttfamily gr-qc/9710056}}].

\bibitem{Asante:2018wqy}
S.K.~Asante, B.~Dittrich and H.M.~Haggard, \emph{{The Degrees of Freedom of Area Regge Calculus: Dynamics, Non-metricity, and Broken Diffeomorphisms}}, \href{https://doi.org/10.1088/1361-6382/aac588}{\emph{Class. Quant. Grav.} {\bfseries 35} (2018) 135009} [\href{https://arxiv.org/abs/1802.09551}{{\ttfamily 1802.09551}}].

\bibitem{Regge:1961ct}
T.E.~Regge, \emph{{General relativity without coordinates}}, {\emph{Nuovo Cimento} {\bfseries 19} (1961) 558}.

\bibitem{Dittrich:2021gww}
B.~Dittrich, S.~Gielen and S.~Schander, \emph{{Lorentzian quantum cosmology goes simplicial}}, \href{https://doi.org/10.1088/1361-6382/ac42ad}{\emph{Class. Quant. Grav.} {\bfseries 39} (2022) 035012} [\href{https://arxiv.org/abs/2109.00875}{{\ttfamily 2109.00875}}].

\bibitem{Asante:2021phx}
S.K.~Asante, B.~Dittrich and J.~Padua-Arg\"uelles, \emph{{Complex actions and causality violations: applications to Lorentzian quantum cosmology}}, \href{https://doi.org/10.1088/1361-6382/accc01}{\emph{Class. Quant. Grav.} {\bfseries 40} (2023) 105005} [\href{https://arxiv.org/abs/2112.15387}{{\ttfamily 2112.15387}}].

\bibitem{Jercher:2022mky}
A.F.~Jercher, D.~Oriti and A.G.A.~Pithis, \emph{{Complete Barrett-Crane model and its causal structure}}, \href{https://doi.org/10.1103/PhysRevD.106.066019}{\emph{Phys. Rev. D} {\bfseries 106} (2022) 066019} [\href{https://arxiv.org/abs/2206.15442}{{\ttfamily 2206.15442}}].

\bibitem{Liu:2018gfc}
H.~Liu and M.~Han, \emph{{Asymptotic analysis of spin foam amplitude with timelike triangles}}, \href{https://doi.org/10.1103/PhysRevD.99.084040}{\emph{Phys. Rev. D} {\bfseries 99} (2019) 084040} [\href{https://arxiv.org/abs/1810.09042}{{\ttfamily 1810.09042}}].

\bibitem{Simao:2021qno}
J.D.~Sim\~ao and S.~Steinhaus, \emph{{Asymptotic analysis of spin-foams with timelike faces in a new parametrization}}, \href{https://doi.org/10.1103/PhysRevD.104.126001}{\emph{Phys. Rev. D} {\bfseries 104} (2021) 126001} [\href{https://arxiv.org/abs/2106.15635}{{\ttfamily 2106.15635}}].

\bibitem{Simao:2024don}
J.D.~Sim\~ao, \emph{{A new 2+1 coherent spin-foam vertex for quantum gravity}}, \href{https://doi.org/10.1088/1361-6382/ad721e}{\emph{Class. Quant. Grav.} {\bfseries 41} (2024) 195015} [\href{https://arxiv.org/abs/2402.05993}{{\ttfamily 2402.05993}}].

\bibitem{Jercher:2023csk}
A.F.~Jercher and S.~Steinhaus, \emph{{Cosmology in Lorentzian Regge calculus: causality violations, massless scalar field and discrete dynamics}}, \href{https://doi.org/10.1088/1361-6382/ad37e9}{\emph{Class. Quant. Grav.} {\bfseries 41} (2024) 105008} [\href{https://arxiv.org/abs/2312.11639}{{\ttfamily 2312.11639}}].

\bibitem{Jercher:2024kig}
A.F.~Jercher, J.D.~Sim\~ao and S.~Steinhaus, \emph{{Partial absence of cosine problem in 3d Lorentzian spin foams}},  \href{https://arxiv.org/abs/2404.16943}{{\ttfamily 2404.16943}}.

\bibitem{Davids:1998bp}
S.~Davids, \emph{{Semiclassical limits of extended Racah coefficients}}, \href{https://doi.org/10.1063/1.533171}{\emph{J. Math. Phys.} {\bfseries 41} (2000) 924} [\href{https://arxiv.org/abs/gr-qc/9807061}{{\ttfamily gr-qc/9807061}}].

\bibitem{Garcia-Islas:2003ges}
J.M.~Garcia-Islas, \emph{{(2+1)-dimensional quantum gravity, spin networks and asymptotics}}, \href{https://doi.org/10.1088/0264-9381/21/2/009}{\emph{Class. Quant. Grav.} {\bfseries 21} (2004) 445} [\href{https://arxiv.org/abs/gr-qc/0307054}{{\ttfamily gr-qc/0307054}}].

\bibitem{Freidel:2000uq}
L.~Freidel, \emph{{A Ponzano-Regge model of Lorentzian 3-dimensional gravity}}, \href{https://doi.org/10.1016/S0920-5632(00)00775-1}{\emph{Nucl. Phys. B Proc. Suppl.} {\bfseries 88} (2000) 237} [\href{https://arxiv.org/abs/gr-qc/0102098}{{\ttfamily gr-qc/0102098}}].

\bibitem{Freidel:2005bb}
L.~Freidel and E.R.~Livine, \emph{{Ponzano-Regge model revisited III: Feynman diagrams and effective field theory}}, \href{https://doi.org/10.1088/0264-9381/23/6/012}{\emph{Class. Quant. Grav.} {\bfseries 23} (2006) 2021} [\href{https://arxiv.org/abs/hep-th/0502106}{{\ttfamily hep-th/0502106}}].

\bibitem{Asante:2022lnp}
S.K.~Asante, J.D.~Sim\~ao and S.~Steinhaus, \emph{{Spin-foams as semiclassical vertices: Gluing constraints and a hybrid algorithm}}, \href{https://doi.org/10.1103/PhysRevD.107.046002}{\emph{Phys. Rev. D} {\bfseries 107} (2023) 046002} [\href{https://arxiv.org/abs/2206.13540}{{\ttfamily 2206.13540}}].

\bibitem{Barrett:1993db}
J.W.~Barrett and T.J.~Foxon, \emph{{Semiclassical limits of simplicial quantum gravity}}, \href{https://doi.org/10.1088/0264-9381/11/3/009}{\emph{Class. Quant. Grav.} {\bfseries 11} (1994) 543} [\href{https://arxiv.org/abs/gr-qc/9310016}{{\ttfamily gr-qc/9310016}}].

\bibitem{Freidel:2002hx}
L.~Freidel, E.R.~Livine and C.~Rovelli, \emph{{Spectra of length and area in (2+1) Lorentzian loop quantum gravity}}, \href{https://doi.org/10.1088/0264-9381/20/8/304}{\emph{Class. Quant. Grav.} {\bfseries 20} (2003) 1463} [\href{https://arxiv.org/abs/gr-qc/0212077}{{\ttfamily gr-qc/0212077}}].

\bibitem{Perelomov:1986tf}
A.M.~Perelomov, \emph{{Generalized coherent states and their applications}} (1986).

\bibitem{Engle:2008ev}
J.~Engle and R.~Pereira, \emph{{Regularization and finiteness of the Lorentzian LQG vertices}}, \href{https://doi.org/10.1103/PhysRevD.79.084034}{\emph{Phys. Rev. D} {\bfseries 79} (2009) 084034} [\href{https://arxiv.org/abs/0805.4696}{{\ttfamily 0805.4696}}].

\bibitem{Ponzano:1968wi}
G.~Ponzano and T.E.~Regge, \emph{{Semiclassical limit of Racah coefficients}},  in \emph{Spectroscopic and group theoretical methods in physics}, F.~Bloch, ed., (Amsterdam), pp.~1--58, North-Holland (1968).

\bibitem{Hormander2003}
L.~H{\"o}rmander, \emph{The fourier transformation},  in \emph{The Analysis of Linear Partial Differential Operators I: Distribution Theory and Fourier Analysis}, (Berlin, Heidelberg), pp.~158--250, Springer Berlin Heidelberg (2003), \href{https://doi.org/10.1007/978-3-642-61497-2_8}{DOI}.

\bibitem{Dona:2017dvf}
P.~Don\`{a}, M.~Fanizza, G.~Sarno and S.~Speziale, \emph{{SU(2) graph invariants, Regge actions and polytopes}}, \href{https://doi.org/10.1088/1361-6382/aaa53a}{\emph{Class. Quant. Grav.} {\bfseries 35} (2018) 045011} [\href{https://arxiv.org/abs/1708.01727}{{\ttfamily 1708.01727}}].

\bibitem{Dona:2020yao}
P.~Dona and S.~Speziale, \emph{{Asymptotics of lowest unitary SL(2,C) invariants on graphs}}, \href{https://doi.org/10.1103/PhysRevD.102.086016}{\emph{Phys. Rev. D} {\bfseries 102} (2020) 086016} [\href{https://arxiv.org/abs/2007.09089}{{\ttfamily 2007.09089}}].

\bibitem{Barrett:2009gg}
J.W.~Barrett, R.J.~Dowdall, W.J.~Fairbairn, H.~Gomes and F.~Hellmann, \emph{{Asymptotic analysis of the EPRL four-simplex amplitude}}, \href{https://doi.org/10.1063/1.3244218}{\emph{J. Math. Phys.} {\bfseries 50} (2009) 112504} [\href{https://arxiv.org/abs/0902.1170}{{\ttfamily 0902.1170}}].

\bibitem{Barrett:2009mw}
J.W.~Barrett, R.J.~Dowdall, W.J.~Fairbairn, F.~Hellmann and R.~Pereira, \emph{{Lorentzian spin foam amplitudes: Graphical calculus and asymptotics}}, \href{https://doi.org/10.1088/0264-9381/27/16/165009}{\emph{Class. Quant. Grav.} {\bfseries 27} (2010) 165009} [\href{https://arxiv.org/abs/0907.2440}{{\ttfamily 0907.2440}}].

\bibitem{Kaminski:2017eew}
W.~Kaminski, M.~Kisielowski and H.~Sahlmann, \emph{{Asymptotic analysis of the EPRL model with timelike tetrahedra}}, \href{https://doi.org/10.1088/1361-6382/aac6a4}{\emph{Class. Quant. Grav.} {\bfseries 35} (2018) 135012} [\href{https://arxiv.org/abs/1705.02862}{{\ttfamily 1705.02862}}].

\bibitem{Asante:2024rrd}
S.K.~Asante and T.~Brysiewicz, \emph{{Solving the area-length systems in discrete gravity using homotopy continuation}}, \href{https://doi.org/10.1088/1361-6382/ad6dcc}{\emph{Class. Quant. Grav.} {\bfseries 41} (2024) 185006} [\href{https://arxiv.org/abs/2402.17080}{{\ttfamily 2402.17080}}].

\bibitem{Bahr:2016hwc}
B.~Bahr and S.~Steinhaus, \emph{{Numerical evidence for a phase transition in 4d spin foam quantum gravity}}, \href{https://doi.org/10.1103/PhysRevLett.117.141302}{\emph{Phys. Rev. Lett.} {\bfseries 117} (2016) 141302} [\href{https://arxiv.org/abs/1605.07649}{{\ttfamily 1605.07649}}].

\bibitem{Steinhaus:2018aav}
S.~Steinhaus and J.~Th{\"u}rigen, \emph{{Emergence of Spacetime in a restricted Spin-foam model}}, \href{https://doi.org/10.1103/PhysRevD.98.026013}{\emph{Phys. Rev. D} {\bfseries 98} (2018) 026013} [\href{https://arxiv.org/abs/1803.10289}{{\ttfamily 1803.10289}}].

\bibitem{Bahr:2018gwf}
B.~Bahr, G.~Rabuffo and S.~Steinhaus, \emph{{Renormalization of symmetry restricted spin foam models with curvature in the asymptotic regime}}, \href{https://doi.org/10.1103/PhysRevD.98.106026}{\emph{Phys. Rev. D} {\bfseries 98} (2018) 106026} [\href{https://arxiv.org/abs/1804.00023}{{\ttfamily 1804.00023}}].

\bibitem{Feldbrugge:2017kzv}
J.~Feldbrugge, J.-L.~Lehners and N.~Turok, \emph{{Lorentzian Quantum Cosmology}}, \href{https://doi.org/10.1103/PhysRevD.95.103508}{\emph{Phys. Rev. D} {\bfseries 95} (2017) 103508} [\href{https://arxiv.org/abs/1703.02076}{{\ttfamily 1703.02076}}].

\bibitem{Dittrich:2011vz}
B.~Dittrich and S.~Steinhaus, \emph{{Path integral measure and triangulation independence in discrete gravity}}, \href{https://doi.org/10.1103/PhysRevD.85.044032}{\emph{Phys. Rev.} {\bfseries D85} (2012) 044032} [\href{https://arxiv.org/abs/1110.6866}{{\ttfamily 1110.6866}}].

\bibitem{Borissova:2024pfq}
J.~Borissova, B.~Dittrich, D.~Qu and M.~Schiffer, \emph{{Spikes and spines in 3D Lorentzian simplicial quantum gravity}},  \href{https://arxiv.org/abs/2406.19169}{{\ttfamily 2406.19169}}.

\bibitem{Borissova:2024txs}
J.~Borissova, B.~Dittrich, D.~Qu and M.~Schiffer, \emph{{Spikes and spines in 4D Lorentzian simplicial quantum gravity}},  \href{https://arxiv.org/abs/2407.13601}{{\ttfamily 2407.13601}}.

\bibitem{Dittrich:2014rha}
B.~Dittrich, W.~Kami\'nski and S.~Steinhaus, \emph{{Discretization independence implies non-locality in 4D discrete quantum gravity}}, \href{https://doi.org/10.1088/0264-9381/31/24/245009}{\emph{Class. Quant. Grav.} {\bfseries 31} (2014) 245009} [\href{https://arxiv.org/abs/1404.5288}{{\ttfamily 1404.5288}}].

\bibitem{Fairbairn:2006dn}
W.J.~Fairbairn, \emph{{Fermions in three-dimensional spinfoam quantum gravity}}, \href{https://doi.org/10.1007/s10714-006-0395-x}{\emph{Gen. Rel. Grav.} {\bfseries 39} (2007) 427} [\href{https://arxiv.org/abs/gr-qc/0609040}{{\ttfamily gr-qc/0609040}}].

\bibitem{Speziale:2007mt}
S.~Speziale, \emph{{Coupling gauge theory to spinfoam 3d quantum gravity}}, \href{https://doi.org/10.1088/0264-9381/24/20/014}{\emph{Class. Quant. Grav.} {\bfseries 24} (2007) 5139} [\href{https://arxiv.org/abs/0706.1534}{{\ttfamily 0706.1534}}].

\bibitem{Mikovic:2001xi}
A.R.~Mikovic, \emph{{Spin foam models of matter coupled to gravity}}, \href{https://doi.org/10.1088/0264-9381/19/9/301}{\emph{Class. Quant. Grav.} {\bfseries 19} (2002) 2335} [\href{https://arxiv.org/abs/hep-th/0108099}{{\ttfamily hep-th/0108099}}].

\bibitem{Bianchi:2010bn}
E.~Bianchi, M.~Han, C.~Rovelli, W.~Wieland, E.~Magliaro and C.~Perini, \emph{{Spinfoam fermions}}, \href{https://doi.org/10.1088/0264-9381/30/23/235023}{\emph{Class. Quant. Grav.} {\bfseries 30} (2013) 235023} [\href{https://arxiv.org/abs/1012.4719}{{\ttfamily 1012.4719}}].

\bibitem{Han:2011as}
M.~Han and C.~Rovelli, \emph{{Spin-foam Fermions: PCT Symmetry, Dirac Determinant, and Correlation Functions}}, \href{https://doi.org/10.1088/0264-9381/30/7/075007}{\emph{Class. Quant. Grav.} {\bfseries 30} (2013) 075007} [\href{https://arxiv.org/abs/1101.3264}{{\ttfamily 1101.3264}}].

\bibitem{Freidel:2005me}
L.~Freidel and E.R.~Livine, \emph{{3D Quantum Gravity and Effective Noncommutative Quantum Field Theory}}, \href{https://doi.org/10.1103/PhysRevLett.96.221301}{\emph{Phys. Rev. Lett.} {\bfseries 96} (2006) 221301} [\href{https://arxiv.org/abs/hep-th/0512113}{{\ttfamily hep-th/0512113}}].

\bibitem{Livine:2024iyk}
E.R.~Livine and V.~Maris, \emph{{Matter coupled to 3d Quantum Gravity: One-loop Unitarity}},  \href{https://arxiv.org/abs/2406.03190}{{\ttfamily 2406.03190}}.

\bibitem{Kisielowski:2018oiv}
M.~Kisielowski and J.~Lewandowski, \emph{{Spin-foam model for gravity coupled to massless scalar field}}, \href{https://doi.org/10.1088/1361-6382/aafcc0}{\emph{Class. Quant. Grav.} {\bfseries 36} (2019) 075006} [\href{https://arxiv.org/abs/1807.06098}{{\ttfamily 1807.06098}}].

\bibitem{Oriti:2002bn}
D.~Oriti and H.~Pfeiffer, \emph{{A Spin foam model for pure gauge theory coupled to quantum gravity}}, \href{https://doi.org/10.1103/PhysRevD.66.124010}{\emph{Phys. Rev. D} {\bfseries 66} (2002) 124010} [\href{https://arxiv.org/abs/gr-qc/0207041}{{\ttfamily gr-qc/0207041}}].

\bibitem{Mikovic:2002uq}
A.R.~Mikovic, \emph{{Spin foam models of Yang-Mills theory coupled to gravity}}, \href{https://doi.org/10.1088/0264-9381/20/1/317}{\emph{Class. Quant. Grav.} {\bfseries 20} (2003) 239} [\href{https://arxiv.org/abs/gr-qc/0210051}{{\ttfamily gr-qc/0210051}}].

\bibitem{Ali:2022vhn}
M.~Ali and S.~Steinhaus, \emph{{Toward matter dynamics in spin foam quantum gravity}}, \href{https://doi.org/10.1103/PhysRevD.106.106016}{\emph{Phys. Rev. D} {\bfseries 106} (2022) 106016} [\href{https://arxiv.org/abs/2206.04076}{{\ttfamily 2206.04076}}].

\bibitem{Rovelli:1990ph}
C.~Rovelli, \emph{{What Is Observable in Classical and Quantum Gravity?}}, \href{https://doi.org/10.1088/0264-9381/8/2/011}{\emph{Class. Quant. Grav.} {\bfseries 8} (1991) 297}.

\bibitem{Hoehn:2019fsy}
P.A.~Hoehn, A.R.H.~Smith and M.P.E.~Lock, \emph{{Trinity of relational quantum dynamics}}, \href{https://doi.org/10.1103/PhysRevD.104.066001}{\emph{Phys. Rev. D} {\bfseries 104} (2021) 066001} [\href{https://arxiv.org/abs/1912.00033}{{\ttfamily 1912.00033}}].

\bibitem{Rovelli:2001bz}
C.~Rovelli, \emph{{Partial observables}}, \href{https://doi.org/10.1103/PhysRevD.65.124013}{\emph{Phys. Rev. D} {\bfseries 65} (2002) 124013} [\href{https://arxiv.org/abs/gr-qc/0110035}{{\ttfamily gr-qc/0110035}}].

\bibitem{Dittrich:2005kc}
B.~Dittrich, \emph{{Partial and complete observables for canonical general relativity}}, \href{https://doi.org/10.1088/0264-9381/23/22/006}{\emph{Class. Quant. Grav.} {\bfseries 23} (2006) 6155} [\href{https://arxiv.org/abs/gr-qc/0507106}{{\ttfamily gr-qc/0507106}}].

\bibitem{Goeller:2022rsx}
C.~Goeller, P.A.~Hoehn and J.~Kirklin, \emph{{Diffeomorphism-invariant observables and dynamical frames in gravity: reconciling bulk locality with general covariance}},  \href{https://arxiv.org/abs/2206.01193}{{\ttfamily 2206.01193}}.

\bibitem{Giesel:2012tj}
K.~Giesel and T.~Thiemann, \emph{{Scalar Material Reference Systems and Loop Quantum Gravity}},  \href{https://arxiv.org/abs/1206.3807}{{\ttfamily 1206.3807}}.

\bibitem{Bojowald:2008wn}
M.~Bojowald, \emph{{Loop Quantum Cosmology}}, {\emph{Living Rev. Relativity} {\bfseries 11} (2008) 4}.

\bibitem{Banerjee:2012fn}
K.~Banerjee, G.~Calcagni and M.~Martin-Benito, \emph{{Introduction to Loop Quantum Cosmology}}, {\emph{SIGMA} {\bfseries 8} (2012) 016} [\href{https://arxiv.org/abs/1109.6801}{{\ttfamily 1109.6801}}].

\bibitem{Kiefer2004}
C.~Kiefer, \emph{Quantum Gravity}, Oxford University Press UK (2004).

\bibitem{Bojowald:2021uqo}
M.~Bojowald, L.~Martinez and G.~Wendel, \emph{{Relational evolution with oscillating clocks}}, \href{https://doi.org/10.1103/PhysRevD.105.106020}{\emph{Phys. Rev. D} {\bfseries 105} (2022) 106020} [\href{https://arxiv.org/abs/2110.07702}{{\ttfamily 2110.07702}}].

\bibitem{Martinez:2023fsd}
L.~Martinez, M.~Bojowald and G.~Wendel, \emph{{Freeze-free cosmological evolution with a nonmonotonic internal clock}}, \href{https://doi.org/10.1103/PhysRevD.108.086001}{\emph{Phys. Rev. D} {\bfseries 108} (2023) 086001} [\href{https://arxiv.org/abs/2309.07825}{{\ttfamily 2309.07825}}].

\bibitem{Dittrich:2024awu}
B.~Dittrich, T.~Jacobson and J.~Padua-Arg\"uelles, \emph{{de Sitter horizon entropy from a simplicial Lorentzian path integral}}, \href{https://doi.org/10.1103/PhysRevD.110.046006}{\emph{Phys. Rev. D} {\bfseries 110} (2024) 046006} [\href{https://arxiv.org/abs/2403.02119}{{\ttfamily 2403.02119}}].

\bibitem{Oeckl:2003vu}
R.~Oeckl, \emph{{A 'General boundary' formulation for quantum mechanics and quantum gravity}}, \href{https://doi.org/10.1016/j.physletb.2003.08.043}{\emph{Phys. Lett. B} {\bfseries 575} (2003) 318} [\href{https://arxiv.org/abs/hep-th/0306025}{{\ttfamily hep-th/0306025}}].

\bibitem{Oeckl:2005bv}
R.~Oeckl, \emph{{General boundary quantum field theory: Foundations and probability interpretation}}, \href{https://doi.org/10.4310/ATMP.2008.v12.n2.a3}{\emph{Adv. Theor. Math. Phys.} {\bfseries 12} (2008) 319} [\href{https://arxiv.org/abs/hep-th/0509122}{{\ttfamily hep-th/0509122}}].

\bibitem{Oeckl:2006rs}
R.~Oeckl, \emph{{Probabilites in the general boundary formulation}}, \href{https://doi.org/10.1088/1742-6596/67/1/012049}{\emph{J. Phys. Conf. Ser.} {\bfseries 67} (2007) 012049} [\href{https://arxiv.org/abs/hep-th/0612076}{{\ttfamily hep-th/0612076}}].

\bibitem{Oeckl:2011qd}
R.~Oeckl, \emph{{Observables in the General Boundary Formulation}},  in \emph{{Quantum Field Theory and Gravity: Conceptual and Mathematical Advances in the Search for a Unified Framework}}, pp.~137--156, 2012, \href{https://doi.org/10.1007/978-3-0348-0043-3_8}{DOI} [\href{https://arxiv.org/abs/1101.0367}{{\ttfamily 1101.0367}}].

\bibitem{Steinhaus:2020lgb}
S.~Steinhaus, \emph{{Coarse Graining Spin Foam Quantum Gravity\textemdash{}A Review}}, \href{https://doi.org/10.3389/fphy.2020.00295}{\emph{Front. in Phys.} {\bfseries 8} (2020) 295} [\href{https://arxiv.org/abs/2007.01315}{{\ttfamily 2007.01315}}].

\bibitem{Asante:2022dnj}
S.K.~Asante, B.~Dittrich and S.~Steinhaus, \emph{{Spin foams, Refinement limit and Renormalization}},  \href{https://arxiv.org/abs/2211.09578}{{\ttfamily 2211.09578}}.

\bibitem{Freidel:2005jy}
L.~Freidel, \emph{{Group Field Theory: An Overview}}, {\emph{Int. J. Theor. Phys.} {\bfseries 44} (2005) 1769} [\href{https://arxiv.org/abs/hep-th/0505016}{{\ttfamily hep-th/0505016}}].

\bibitem{Oriti:2011jm}
D.~Oriti, \emph{{The microscopic dynamics of quantum space as a group field theory}},  in \emph{{Foundations of Space and Time: Reflections on Quantum Gravity}}, pp.~257--320, 10, 2011 [\href{https://arxiv.org/abs/1110.5606}{{\ttfamily 1110.5606}}].

\bibitem{Rovelli:2004wb}
C.~Rovelli, \emph{{Quantum Gravity}}, Cambridge University Press, Cambridge, UK (2004).

\bibitem{Sorkin:1994dt}
R.D.~Sorkin, \emph{{Quantum mechanics as quantum measure theory}}, \href{https://doi.org/10.1142/S021773239400294X}{\emph{Mod. Phys. Lett. A} {\bfseries 9} (1994) 3119} [\href{https://arxiv.org/abs/gr-qc/9401003}{{\ttfamily gr-qc/9401003}}].

\bibitem{Craig:2006ny}
D.A.~Craig, F.~Dowker, J.~Henson, S.~Major, D.~Rideout and R.D.~Sorkin, \emph{{A Bell inequality analog in quantum measure theory}}, \href{https://doi.org/10.1088/1751-8113/40/3/010}{\emph{J. Phys. A} {\bfseries 40} (2007) 501} [\href{https://arxiv.org/abs/quant-ph/0605008}{{\ttfamily quant-ph/0605008}}].

\bibitem{Frauca:2016eup}
A.M.~Frauca and R.D.~Sorkin, \emph{{How to Measure the Quantum Measure}}, \href{https://doi.org/10.1007/s10773-016-3181-x}{\emph{Int. J. Theor. Phys.} {\bfseries 56} (2017) 232} [\href{https://arxiv.org/abs/1610.02087}{{\ttfamily 1610.02087}}].

\bibitem{Dowker:2010ng}
F.~Dowker, S.~Johnston and R.D.~Sorkin, \emph{{Hilbert Spaces from Path Integrals}}, \href{https://doi.org/10.1088/1751-8113/43/27/275302}{\emph{J. Phys. A} {\bfseries 43} (2010) 275302} [\href{https://arxiv.org/abs/1002.0589}{{\ttfamily 1002.0589}}].

\bibitem{Sorkin:2019llw}
R.D.~Sorkin, \emph{{Lorentzian angles and trigonometry including lightlike vectors}},  \href{https://arxiv.org/abs/1908.10022}{{\ttfamily 1908.10022}}.

\bibitem{Wynn1956}
P.~Wynn, \emph{On a device for computing the em(sn) transformation}, {\emph{Mathematical Tables and Other Aids to Computation} {\bfseries 10} (1956) 91}.

\bibitem{Weniger2003}
E.J.~Weniger, \emph{Nonlinear sequence transformations for the acceleration of convergence and the summation of divergent series}, \href{https://doi.org/https://doi.org/10.1016/0167-7977(89)90011-7}{\emph{Computer Physics Reports} {\bfseries 10} (1989) 189} [\href{https://arxiv.org/abs/math/0306302}{{\ttfamily math/0306302}}].

\bibitem{Dona:2023myv}
P.~Don\`a and P.~Frisoni, \emph{{Summing bulk quantum numbers with Monte~Carlo in spin foam theories}}, \href{https://doi.org/10.1103/PhysRevD.107.106008}{\emph{Phys. Rev. D} {\bfseries 107} (2023) 106008} [\href{https://arxiv.org/abs/2302.00072}{{\ttfamily 2302.00072}}].

\bibitem{Schmidt1941}
R.J.~Schmidt, \emph{Xxxii. on the numerical solution of linear simultaneous equations by an iterative method}, {\emph{The London, Edinburgh, and Dublin Philosophical Magazine and Journal of Science} {\bfseries 32} (1941) 369}.

\bibitem{Shanks1955}
D.~Shanks, \emph{Non-linear transformations of divergent and slowly convergent sequences}, \href{https://doi.org/https://doi.org/10.1002/sapm19553411}{\emph{Journal of Mathematics and Physics} {\bfseries 34} (1955) 1} [\href{https://arxiv.org/abs/https://onlinelibrary.wiley.com/doi/pdf/10.1002/sapm19553411}{{\ttfamily https://onlinelibrary.wiley.com/doi/pdf/10.1002/sapm19553411}}].

\bibitem{Veltman1970}
H.~{van Dam} and M.~Veltman, \emph{Massive and mass-less yang-mills and gravitational fields}, \href{https://doi.org/https://doi.org/10.1016/0550-3213(70)90416-5}{\emph{Nuclear Physics B} {\bfseries 22} (1970) 397}.

\bibitem{Hahn:2005pf}
T.~Hahn, \emph{{The CUBA library}}, \href{https://doi.org/10.1016/j.nima.2005.11.150}{\emph{Nucl. Instrum. Meth.} {\bfseries A559} (2006) 273} [\href{https://arxiv.org/abs/hep-ph/0509016}{{\ttfamily hep-ph/0509016}}].

\bibitem{Dona:2019dkf}
P.~Don\`a, M.~Fanizza, G.~Sarno and S.~Speziale, \emph{{Numerical study of the Lorentzian Engle-Pereira-Rovelli-Livine spin foam amplitude}}, \href{https://doi.org/10.1103/PhysRevD.100.106003}{\emph{Phys. Rev. D} {\bfseries 100} (2019) 106003} [\href{https://arxiv.org/abs/1903.12624}{{\ttfamily 1903.12624}}].

\bibitem{Gozzini:2021kbt}
F.~Gozzini, \emph{{A high-performance code for EPRL spin foam amplitudes}}, \href{https://doi.org/10.1088/1361-6382/ac2b0b}{\emph{Class. Quant. Grav.} {\bfseries 38} (2021) 225010} [\href{https://arxiv.org/abs/2107.13952}{{\ttfamily 2107.13952}}].

\bibitem{Dona:2022dxs}
P.~Dona and P.~Frisoni, \emph{{How-to Compute EPRL Spin Foam Amplitudes}}, \href{https://doi.org/10.3390/universe8040208}{\emph{Universe} {\bfseries 8} (2022) 208} [\href{https://arxiv.org/abs/2202.04360}{{\ttfamily 2202.04360}}].

\bibitem{Steinhaus:2024qov}
S.~Steinhaus, \emph{{Monte~Carlo algorithm for spin foam intertwiners}}, \href{https://doi.org/10.1103/PhysRevD.110.026022}{\emph{Phys. Rev. D} {\bfseries 110} (2024) 026022} [\href{https://arxiv.org/abs/2403.04836}{{\ttfamily 2403.04836}}].

\bibitem{Asante:2024eft}
S.K.~Asante and S.~Steinhaus, \emph{{Efficient Tensor Network Algorithms for Spin Foam Models}},  \href{https://arxiv.org/abs/2406.19676}{{\ttfamily 2406.19676}}.

\bibitem{Jercher:2023nxa}
A.F.~Jercher, L.~Marchetti and A.G.A.~Pithis, \emph{{Scalar cosmological perturbations from quantum entanglement within Lorentzian quantum gravity}}, \href{https://doi.org/10.1103/PhysRevD.109.066021}{\emph{Phys. Rev. D} {\bfseries 109} (2024) 066021} [\href{https://arxiv.org/abs/2308.13261}{{\ttfamily 2308.13261}}].

\bibitem{Jercher:2023kfr}
A.F.~Jercher, L.~Marchetti and A.G.A.~Pithis, \emph{{Scalar cosmological perturbations from quantum gravitational entanglement}}, \href{https://doi.org/10.1088/1361-6382/ad6f67}{\emph{Class. Quant. Grav.} {\bfseries 41} (2024) 18LT01} [\href{https://arxiv.org/abs/2310.17549}{{\ttfamily 2310.17549}}].

\bibitem{Bargmann:1946me}
V.~Bargmann, \emph{{Irreducible unitary representations of the Lorentz group}}, \href{https://doi.org/10.2307/1969129}{\emph{Annals Math.} {\bfseries 48} (1947) 568}.

\bibitem{Lindblad:1969zz}
G.~Lindblad and B.~Nagel, \emph{{Continuous bases for unitary irreducible representations of $\mathrm{SU}(1,1)$}}, .

\bibitem{Lindblad:1970tv}
G.~Lindblad, \emph{{Eigenfunction expansions associated with unitary irreducible representations of su(1,1)}}, \href{https://doi.org/10.1088/0031-8949/1/5-6/001}{\emph{Phys. Scripta} {\bfseries 1} (1970) 201}.

\bibitem{Ruehl1970}
W.~Ruehl, \emph{Lorentz group and harmonic analysis}, W A Benjamin, Inc, United States (1970).

\bibitem{Pukanszky:1963}
L.~Pukanszky, \emph{{On the Plancherel theorem of the $2 \times 2$ real unimodular group}}, {\emph{Bulletin of the American Mathematical Society} {\bfseries 69} (1963) 504 }.

\bibitem{Takahashi:1961}
R.~Takahashi, \emph{Sur les fonctions spheriques et la formule de plancherel dans le groupe hyperbolique}, \href{https://doi.org/10.4099/jjm1924.31.0_55}{\emph{Japanese journal of mathematics :transactions and abstracts} {\bfseries 31} (1961) 55}.

\end{thebibliography}\endgroup

\end{document}